\documentclass[aps,prda4paper,twocolumn,superscriptaddress,preprintnumbers,floatfix,nofootinbib,amsmath,amssymb,table]{revtex4-2}

\usepackage{subcaption}
\usepackage{hyperref}
\usepackage{cleveref}
\usepackage{braket}
\usepackage{ragged2e}

\usepackage{amstext,amssymb}
\usepackage[section]{placeins}
\usepackage{amsmath}
\usepackage{graphicx}
\usepackage{xspace}
\usepackage{color}
\usepackage{float}
\usepackage{comment}
\usepackage{amstext,amssymb}
\usepackage{amsmath}
\usepackage[section]{placeins}
\usepackage{xspace}
\usepackage{units}
\usepackage{array}
\usepackage{amsmath,bm}
\usepackage{amssymb}
\usepackage{gensymb}
\usepackage{soul}
\usepackage{textcomp}
\usepackage{slashed} 
\usepackage{multirow}
\usepackage{hyperref}
\usepackage{mathrsfs}
\usepackage{enumitem}
\usepackage{orcidlink}

\usepackage{comment}
\usepackage{epstopdf}
\usepackage{float}
\usepackage{units}
\usepackage{feynmp}
\usepackage{slashed}
\usepackage{amsmath,bm}
\usepackage[table]{xcolor}
\usepackage[normalem]{ulem}
\definecolor{antiquewhite}{rgb}{0.98, 0.92, 0.84}
\definecolor{aqua}{rgb}{0.0, 1.0, 1.0}
\definecolor{amethyst}{rgb}{0.6, 0.4, 0.8}
\definecolor{applegreen}{rgb}{0.55, 0.71, 0.0}
\definecolor{byzantine}{rgb}{0.74, 0.2, 0.64}
\definecolor{cadetgrey}{rgb}{0.57, 0.64, 0.69}
\definecolor{candypink}{rgb}{0.89, 0.44, 0.48}
\usepackage{lipsum}
\usepackage{cals}
\usepackage{slashed} 
\newcommand{\be}{\begin{equation}}
\newcommand{\ee}{\end{equation}}
\newcommand{\bea}{\begin{eqnarray}}
\newcommand{\eea}{\end{eqnarray}}

\def\s1{\hat s}

\usepackage{xcolor}

\makeatletter

    \def\CT@@do@color{%
      \global\let\CT@do@color\relax
            \@tempdima\wd\z@
            \advance\@tempdima\@tempdimb
            \advance\@tempdima\@tempdimc
    \advance\@tempdimb\tabcolsep
    \advance\@tempdimc\tabcolsep
    \advance\@tempdima2\tabcolsep
            \kern-\@tempdimb
            \leaders\vrule
                    \hskip\@tempdima\@plus  1fill
            \kern-\@tempdimc
            \hskip-\wd\z@ \@plus -1fill }
    \makeatother

\usepackage{slashed}
\definecolor{ashgrey}{rgb}{0.7, 0.75, 0.71}
\definecolor{aureolin}{rgb}{0.99, 0.93, 0.0}
\definecolor{babypink}{rgb}{0.96, 0.76, 0.76}
\definecolor{buff}{rgb}{0.94, 0.86, 0.51}
\definecolor{chamoisee}{rgb}{0.63, 0.47, 0.35}
\definecolor{chartreuse(web)}{rgb}{0.5, 1.0, 0.0}
\definecolor{citrine}{rgb}{0.89, 0.82, 0.04}
\definecolor{emerald}{rgb}{0.31, 0.78, 0.47}
\definecolor{fawn}{rgb}{0.9, 0.67, 0.44}
\definecolor{fulvous}{rgb}{0.86, 0.52, 0.0}

\newcommand{\nua}[1]{\ensuremath{\rlap{\kern-2.5pt\ensuremath{\overset{\scriptscriptstyle(-)}{\phantom{\nu}}}}{\ensuremath{{\nu}_{#1}}}}\xspace}

\begin{document}
\title{Residual Entanglement vs. Residual Coherence: A Wave Packet Approach in Neutrino Oscillation with Matter }

\author{Lekhashri Konwar}
\email{konwar.3@iitj.ac.in}
\affiliation{Indian Institute of Technology Jodhpur, Jodhpur - 342030, India}
\author{Abhishek Kumar Jha}\email{abhiecc.jha@gmail.com}
\affiliation{School of Technology Management $\&$ Engineering, SVKM’s Narsee Monjee Institute of Management Studies (NMIMS) Deemed-to-be-University, Hyderabad- 509301, India}
\author{Bhavna Yadav}
\email{yadav.18@alumni.iitj.ac.in}
\affiliation{Wilczek Quantum Center, Shanghai Institute for Advanced Studies, Shanghai 201315, China}
\affiliation{University of Science and Technology of China, Hefei 230026, China}

\begin{abstract}
In this work, we investigate quantum correlations in three-flavor neutrino oscillations beyond the standard plane-wave (PW) description by employing a wave-packet formalism in the presence of constant matter density. We study three quantum correlation measures, namely the residual $l_1$-norm coherence, Concurrence Fill, and Three-$\pi$ entanglement, and express them in terms of transition probabilities for an initial muon-flavor state. The analysis is performed for the DUNE setup and extended beyond its baseline to study the interplay between wave-packet decoherence and matter effects over long distances. For the DUNE baseline, finite wave-packet widths suppress and smooth the oscillations relative to the PW case, with smaller widths causing stronger damping, while matter effects modify the amplitudes and positions of the oscillation features. Among the considered measures, residual $l_1$-norm coherence dominates over the entanglement measures. At very long propagation distances, wave-packet decoherence progressively washes out the oscillatory behavior, driving the oscillation probabilities and quantum correlation measures toward averaged values, with smaller wave-packet widths leading to earlier damping. Further, in the $L-\delta_{\rm CP}$ plane, residual $l_1$-norm coherence exhibits weak $\delta_{\rm CP}$ dependence, while Concurrence Fill and Three-$\pi$ entanglement show greater sensitivity at the DUNE baseline. Beyond a characteristic baseline in the extended propagation regime, all three measures approach constant values, independent of $L$ and $\delta_{\rm CP}$.

\end{abstract}

\maketitle
\flushbottom

\section{Introduction}

The quantum phenomenon of three-flavor neutrino oscillations \cite{Super-Kamiokande:1998kpq,kajita2016nobel,SNO:2002tuh} originates from the nontrivial mixing between flavor and mass eigenstates. As neutrinos propagate, the coherent superposition of mass eigenstates acquires relative phase differences, resulting in the oscillatory conversion of one flavor into another \cite{Giunti:2007ry}. Over the past few decades, the standard framework of neutrino oscillation has successfully explained results from solar, atmospheric, reactor, and accelerator experiments \cite{SNO:2001kpb,KamLAND:2002uet,DayaBay:2022eyy,T2K:2023smv,NOvA:2019cyt}. In this framework, neutrino oscillations are described within the PW approximation, both in vacuum and in matter. While the PW approach captures the essential features of flavor oscillations, it neglects the localized nature of neutrino production and detection processes \cite{Giunti:1997wq}.

 A more realistic description of neutrino propagation is provided by the wave-packet approach \citep{Nussinov:1976uw,Kiers:1995zj,Giunti:1997wq,Takeuchi:2000fz,Giunti:2003ax}, in which neutrinos are treated as finite wave packets rather than infinitely extended plane waves. Since neutrinos are produced and detected through interactions occurring within finite spatial regions, their quantum states acquire a finite spatial width. In this framework, the different mass eigenstates propagate with slightly different group velocities, causing a gradual spatial separation of the corresponding wave packets during propagation. As the overlap between these mass components decreases, a natural decoherence mechanism arises, which can suppress oscillation patterns at sufficiently long baselines. Furthermore, coherent forward scattering of neutrinos with electrons in matter generates an effective matter potential that modifies the oscillation dynamics and can influence the coherence of the propagating wave packets \citep{Wolfenstein:1977ue,Mikheyev:1985zog,Peltoniemi:2000nw}. The wave-packet approach therefore incorporates the finite spatial extent of the production and detection processes, the possible loss of coherence during propagation, and matter-induced modifications of neutrino evolution. This approach also provides a more realistic framework for studying quantum correlations in neutrino oscillations, where coherence between different mass eigenstate components plays a crucial role.

In recent years, the study of quantum correlations in neutrino oscillations has emerged as an interesting interface between neutrino physics and quantum information theory. Through the concept of single-particle mode entanglement \citep{PhysRevA.72.064306,Blasone:1995zc,Blasone_2009Entanglement,Blasone_2008,Blasone_2010}, the three-flavor neutrino system can be viewed as a multipartite quantum system in which flavor states act as different subsystems. Within this perspective, several quantum information measures, such as entanglement \cite{Blasone_2009Entanglement,banerjee2015quantum,Alok_2016,Jha:2020dav,Jha:2021itm,Jha:2022yik,Li:2021fft,Li:2022mus,Quinta:2022sgq,Blasone_2015,Konwar:2024nrd,Konwar:2025ipv,Banerjee:2024lih,Banerjee:2025gau,Bouri:2024kcl,Jha:2026jef}, coherence \cite{PhysRevA.98.050302,Dixit:2019swl,Dong:2022thu,Ravari:2022yfd,Ettefaghi:2023zsh,Dong:2022eys}, and nonlocal correlations \cite{Dixit:2018kev,sarkar2021effects,Yadav:2022grk,Yadav:2024qav,konwar2024violation} have been used to quantify the quantum nature of neutrino oscillations. These measures provide an alternative perspective on flavor conversion and highlight how neutrino flavor-transition probabilities encode information about the underlying quantum correlations. Furthermore, the study of quantum correlations in neutrino oscillations is not limited to flat spacetime but has also been extended to curved spacetime \cite{Dixit:2019lsg,Jha:2025ekn,Jha:2024gdq,Jha:2024asl}.

Among the fundamental quantum resources, entanglement \cite{Schrodinger:2008pyl,Schrodinger:2008opj,Nielsen_Chuang_2010,Sabin2008,Dur:2000zz,Cunha:2019jex,Bennett:1996gf,Guo2020,Hill:1997pfa,Wootters:1997id,Peres:1996dw,Vidal:2002zz} and quantum coherence \cite{Baumgratz:2013ecx,Singh:2015oby,Yao:2015fuw,Jiang:2020qjf} play a central role in characterizing non-classical features of quantum systems. Entanglement quantifies non-classical correlations between subsystems, whereas coherence characterizes the superposition of quantum states with respect to a chosen reference basis. Moreover, multipartite entanglement constitutes an important physical resource and plays a crucial role in large-scale quantum information processing.

In multipartite quantum systems, residual correlation measures can be used to quantify genuinely multipartite correlations that remain after accounting for pairwise contributions. In the context of three-flavor neutrino oscillations, residual entanglement measures such as the Three-$\pi$ \cite{Ou:2007ysv} and concurrence Fill \cite{Xie:2021hsy}, together with residual coherence based on the $l_1$-norm \cite{Li:2022zic}, provide a framework for investigating the distribution of quantum correlations among flavor modes during neutrino propagation.

Most previous studies of quantum correlations in neutrino oscillations have been carried out within the plane-wave approximation \cite{Blasone_2009Entanglement,banerjee2015quantum,Alok_2016,Jha:2020dav,Jha:2021itm,Jha:2022yik,Li:2021fft,Li:2022mus,Quinta:2022sgq,Blasone_2015,Konwar:2024nrd,Konwar:2025ipv,Banerjee:2024lih,Banerjee:2025gau,Bouri:2024kcl,Jha:2026jef,PhysRevA.98.050302,Dixit:2019swl,Dong:2022thu,Ravari:2022yfd,Ettefaghi:2023zsh,Dong:2022eys}. However, incorporating wave-packet effects can modify the evolution of these correlations, particularly over long propagation distances where wave-packet separation and decoherence become important. Furthermore, neutrinos propagating through matter experience an effective matter potential arising from coherent forward scattering, which modifies their oscillation dynamics and consequently influences the evolution of quantum correlations. The interplay between wave-packet decoherence and matter effects can therefore lead to nontrivial modifications of the evolution of quantum correlations \cite{Blasone:2015lya,Ettefaghi:2023zsh,Blasone:2021cau,Ravari:2022yfd}.

In this work, we present a detailed analysis of three-flavor neutrino transition probabilities, along with the residual entanglement and residual coherence measures derived from the corresponding monogamy relations, using the wave-packet formalism in the presence of matter effects. Both the residual entanglement and residual coherence measures are expressed in terms of the three-flavor transition probabilities for an initial muon-flavor neutrino state. The analysis is performed for the DUNE experimental configuration \cite{DUNE:2015lol,DUNE:2020lwj,DUNE:2021tad,DUNE:2020fgq} and is further extended to the regime beyond the DUNE baseline, in order to investigate the evolution of these quantum correlations over large propagation distances. 

The paper is structured as follows. Sec. \ref{Sec:Quantum_correlation} introduces the residual entanglement and residual coherence measures employed in this work. In Sec. \ref{Sec:Formalism}, we present the wave-packet formalism for three-flavor neutrino oscillations in matter. Sec. \ref{Sec:QC_NO} formulates the residual quantum correlation measures in terms of neutrino flavor transition probabilities. The results and their discussion are given in Sec. \ref{Sec:Result}. Finally, the conclusions are given in Sec. \ref{Sec:conclusion}.

\section{Quantum correlation measures} \label{Sec:Quantum_correlation}
\subsection{Residual Entanglement}
Residual entanglement measures the part of the total entanglement that remains after subtracting all pairwise entanglements. This remaining part represents quantum correlations shared collectively among all the particles and is therefore a signature of genuine multipartite entanglement. In a composite system consisting of three subsystems, say $A$, $B$, and $C$, one can evaluate the entanglement shared across different bipartitions, such as between $A$ and the composite subsystem $BC$, as well as the pairwise entanglements between $A$ and $B$, and $A$ and $C$. However, these bipartite contributions do not, in general, fully capture the total correlations present in the system. In particular, it is often observed that the entanglement between $A$ and the composite system $BC$ cannot be completely decomposed into the sum of pairwise entanglements shared by $A$ with $B$ and $C$. This fundamental feature is formalized through the monogamy relation introduced by Coffman, Kundu, and Wootters (CKW) \cite{Coffman:1999jd}, given by
\begin{equation}
E_{A(BC)} \geq E_{AB} + E_{AC}.
\end{equation}
The inequality expresses the restricted shareability of entanglement and naturally leads to the definition of residual entanglement as the difference between the two sides,
\begin{equation}
E^{\text{res}}_{A} = E_{A(BC)} - E_{AB} - E_{AC},
\end{equation}
which captures the amount of entanglement that is genuinely shared among all three subsystems and cannot be reduced to pairwise contributions.

To study this quantitatively, we need to use a suitable bipartite entanglement measure. In this work, we employ negativity \cite{Vidal:2002zz,Peres:1996dw,Sabin:2008naa,Miranowicz:2004ivl,Ou:2007ysv}, a widely used and computable measure based on the Peres-Horodecki criterion \cite{Peres:1996dw}. According to this criterion, a separable (i.e., unentangled) state must remain positive under partial transposition, and hence any violation of this condition signals the presence of entanglement. For a given bipartite density matrix $\rho_{AB}$, negativity quantifies the extent to which the partially transposed matrix $\rho_{AB}^{T_B}$ fails to remain positive.

It is defined as \cite{Vidal:2002zz}
\begin{equation}
N_{AB} = \frac{\|\rho_{AB}^{T_B}\| - 1}{2},
\end{equation}
where $\|\cdot\|$ denotes the trace norm. Equivalently, negativity can be interpreted as the absolute sum of the negative eigenvalues of $\rho_{AB}^{T_B}$. Thus, a non-zero value of negativity directly indicates the presence of quantum correlations, while a vanishing value implies separability of the state.

With this definition, the bipartite contributions entering the monogamy relation can be explicitly computed for a tripartite system described by the density matrix $\rho_{ABC}$. The entanglement between subsystem $A$ and the composite system $BC$ is obtained by performing a partial transpose on $\rho_{ABC}$ with respect to subsystem $A$, yielding $N_{A(BC)}$. Similarly, the pairwise negativities $N_{AB}$ and $N_{AC}$ are evaluated from the reduced density matrices $\rho_{AB} = \mathrm{Tr}_C(\rho_{ABC})$ and $\rho_{AC} = \mathrm{Tr}_B(\rho_{ABC})$, respectively.

Having established these quantities, one can now express the monogamy constraint explicitly in terms of negativity and satisfy a monogamy relation of the form \cite{Ou:2007ysv}
\begin{equation}
N_{AB}^2 + N_{AC}^2 \leq N_{A(BC)}^2,
\end{equation}
which reflects the limited shareability of entanglement across different subsystems.
In a similar manner, by considering $B$ and $C$ as the subsystems, one obtains the corresponding monogamy relations:
\begin{equation}
N_{BA}^2 + N_{BC}^2 \leq N_{B(AC)}^2,
\end{equation}
and
\begin{equation}
N_{CA}^2 + N_{CB}^2 \leq N_{C(AB)}^2.
\end{equation}

The difference between the two sides of the monogamy relation can be interpreted as the residual entanglement associated with a given focus subsystem. For instance, when subsystem $A$ is taken as the reference, the residual entanglement is defined as
\begin{equation}
\pi_A = N_{A(BC)}^2 - N_{AB}^2 - N_{AC}^2.
\end{equation}
Similarly, by choosing $B$ and $C$ as the focus subsystems, one obtains the corresponding residual entanglements
\begin{equation}
\pi_B = N_{B(AC)}^2 - N_{BA}^2 - N_{BC}^2,
\end{equation}
and
\begin{equation}
\pi_C = N_{C(AB)}^2 - N_{CA}^2 - N_{CB}^2.
\end{equation}
These residual entanglements are generally not equal, i.e., $\pi_A \neq \pi_B \neq \pi_C$. This reflects the fact that the distribution of entanglement depends on the choice of the reference subsystem and is therefore not invariant under permutations of the subsystems.

To overcome this asymmetry and obtain a permutation-invariant characterization of genuine tripartite entanglement, one introduces the quantity $\pi_{ABC}$, commonly referred to as the Three-$\pi$, defined as the average of the residual entanglements \cite{Ou:2007ysv}:
\begin{equation}\label{eq:thee-pi}
\begin{split}
\mathrm{Three}\text{-}\pi &= \frac{1}{3} \left( \pi_A + \pi_B + \pi_C \right) \\
&= \frac{1}{3} \Big[ 
N_{A(BC)}^2 + N_{B(AC)}^2 + N_{C(AB)}^2 \\
&\quad - 2N_{AB}^2 - 2N_{AC}^2 - 2N_{BC}^2 
\Big].
\end{split}
\end{equation}

This quantity symmetrizes the residual entanglement over all possible bipartitions and therefore provides a faithful characterization of genuine tripartite quantum correlations. In particular, a vanishing value of $\pi_{ABC}$ indicates that the state is at most biseparable, meaning that all correlations can be decomposed into bipartite contributions. In contrast, a non-zero value signals the presence of genuine tripartite entanglement. For instance, GHZ and W states have been identified as two important quantum information resources representing fundamentally distinct classes of genuine tripartite entanglement.

To further characterize genuine tripartite entanglement beyond the Three-$\pi$ measure discussed above, one may consider an alternative quantifier known as Concurrence Fill \cite{Xie:2021hsy}. This measure, introduced by Xie \textit{et al.} in 2021, provides a geometric perspective on multipartite entanglement by establishing a connection with Heron’s formula for the area of a triangle. The central idea is to construct a concurrence triangle, in which the squared bipartite concurrences $C_{1(23)}^2$, $C_{2(13)}^2$, and $C_{3(12)}^2$ are treated as the lengths of the three sides of a triangle corresponding to a three-qubit system. The Concurrence Fill is then defined in terms of the area of this triangle, thereby capturing the genuinely shared correlations among all three subsystems.

The explicit expression for Concurrence Fill is given by \cite{Xie:2021hsy}
\begin{equation} \label{eq:concurrence_fill}
\begin{aligned}
\text{Concurrence Fill}
=
\Bigg[
\frac{16}{3}\,
Q\,
\left(Q-C_{1(23)}^{2}\right)
\left(Q-C_{2(13)}^{2}\right) \\
\left(Q-C_{3(12)}^{2}\right)
\Bigg]^{1/4}.
\end{aligned}
\end{equation}
where
\begin{equation}
Q = \frac{1}{2} \left( C_{1(23)}^2 + C_{2(13)}^2 + C_{3(12)}^2 \right),
\end{equation}
represents the semi-perimeter of the concurrence triangle. The quantities $C_{1(23)}^2$, $C_{2(13)}^2$, and $C_{3(12)}^2$ are defined in terms of the reduced density matrices as $2\left(1 - \mathrm{Tr}(\rho_A^2)\right)$, $2\left(1 - \mathrm{Tr}(\rho_B^2)\right)$, and $2\left(1 - \mathrm{Tr}(\rho_C^2)\right)$, respectively. Here, $\rho_{A} = \mathrm{Tr}_{BC}(\rho_{ABC})$, $\rho_B=\mathrm{Tr}_{AC}(\rho_{ABC})$, and $\rho_c=\mathrm{Tr}_{AB}(\rho_{ABC})$ are three-reduced density matrices obtained from $\rho_{ABC}$.

\subsection{Residual coherence}

The $l_{1}$-norm quantum coherence \cite{Baumgratz:2013ecx,Singh:2015oby,Yao:2015fuw,Jiang:2020qjf} of a quantum state 
$\rho = \sum_{i,j} \rho_{ij}\, |i\rangle\langle j|$ 
is defined as the sum of the absolute values of its off-diagonal elements,
\begin{equation}\label{eq:l1coherence}
C_{l_{1}}(\rho) = \sum_{i \neq j} |\rho_{ij}|.
\end{equation}
The trade-off relation of the coherence is defined as in \cite{Jiang:2020qjf}
\begin{equation}
    C_{l_{1}}(\rho_{ABC}) \geq C_{l_{1}}(\rho_{AB})+C_{l_{1}}(\rho_{AC}).
\end{equation}
where the reduced density matrices are given by
\begin{equation}
\rho_{AB} = \mathrm{Tr}_{C}(\rho_{ABC}) \quad\text{and} \quad
\rho_{AC} = \mathrm{Tr}_{B}(\rho_{ABC}), 
\end{equation}
which reflects the constrained sharing of coherence among subsystems. The residual $l_1$-norm coherence ($R_{l_{1}\text{-norm}}$) is defined as 
\begin{equation} \label{coherence}
R_{l_{1}}\text{-norm}= C_{l_{1}}(\rho_{ABC}) - C_{l_{1}}(\rho_{AB}) - C_{l_{1}}(\rho_{AC}).
\end{equation}

\section{A wave packet approach: neutrino oscillation in matter }\label{Sec:Formalism}
A neutrino produced in association with a charged lepton $\ell_\alpha$ is described by a flavor state
\begin{equation}
|\nu_\alpha\rangle = \sum_i U_{\alpha i}^* |\nu_i\rangle,
\end{equation}
where $|\nu_i\rangle$ are mass eigenstates with masses $m_i$, and $U_{\alpha i}$ are elements of the leptonic mixing matrix, which is defined as 
\begin{equation}\label{PMNS}
\begin{pmatrix}
c_{13}c_{12} & c_{13}s_{12} & s_{13}e^{-i\delta} \\
- s_{12}c_{23} - c_{12}s_{23}s_{13}e^{i\delta} 
& c_{12}c_{23} - s_{12}s_{23}s_{13}e^{i\delta} 
& c_{13}s_{23} \\
s_{12}s_{23} - c_{12}c_{23}s_{13}e^{i\delta} 
& - c_{12}s_{23} - s_{12}c_{23}s_{13}e^{i\delta} 
& c_{13}c_{23}
\end{pmatrix}.
\end{equation}
Each mass eigenstate is described by a wave packet constructed as a superposition of plane waves:
\begin{equation}
|\nu_i(t,\mathbf{x})\rangle = \int \frac{dp}{(2\pi)^{1/2}} \, f_i(\mathbf{p} - \mathbf{p}_i)\, 
e^{i(\mathbf{p}\cdot \mathbf{x} - E_i(\mathbf{p}) t)} |\nu_i\rangle,
\end{equation}
where $f_i(\mathbf{p} - \mathbf{p}_i)$ is a momentum distribution function sharply peaked around the mean momentum $\mathbf{p}_i$, $\sigma_{p}$ is the momentum
width of the wave packets and $E_i(\mathbf{p}) = \sqrt{\mathbf{p}^2 + m_i^2} $.

Assuming a Gaussian momentum distribution,
\begin{equation}
f_i(\mathbf{p} - \mathbf{p}_i) = \left(\frac{1}{2\pi \sigma_p^2}\right)^{1/4}
\exp\left[-\frac{(\mathbf{p} - \mathbf{p}_i)^2}{4\sigma_p^2}\right],
\end{equation}
the corresponding spatial wave packet is also Gaussian with width
\begin{equation}
\sigma_x = \frac{1}{2\sigma_p}.
\end{equation}
Under the assumption that the momentum distribution is sharply peaked (i.e., $\sigma_p \ll E_i^2/m_i$), the energy can be expanded around the mean momentum $\mathbf{p}_i$ as
\begin{equation}
E_i(\mathbf{p}) \simeq E_i + \mathbf{v}_i \cdot (\mathbf{p} - \mathbf{p}_i),
\end{equation}
where $\mathbf{v}_i = \frac{\mathbf{p}_i}{E_i}$ is the group velocity of the $i$-th mass eigenstate.

Carrying out the Gaussian integration, the coordinate-space wave packet becomes
\begin{align}
|\nu_i(t,\mathbf{x})\rangle &=
\left(\frac{1}{2\pi \sigma_x^2}\right)^{1/4}
\exp\left[i(\mathbf{p}_i \cdot \mathbf{x} - E_i t)\right] \nonumber \\
&\quad \times
\exp\left[-\frac{(\mathbf{x} - \mathbf{v}_i t)^2}{4\sigma_x^2}\right]
|\nu_i\rangle.
\end{align}

Thus, the full flavor state at time $t$ and position $\mathbf{x}$ is given by the superposition
\begin{align} \label{eq:states}
|\nu_\alpha(t,\mathbf{x})\rangle &=
\sum_i U_{\alpha i}^*
\left(\frac{1}{2\pi \sigma_x^2}\right)^{1/4}
\exp\left[i(\mathbf{p}_i \cdot \mathbf{x} - E_i t)\right] \nonumber \\
&\quad \times
\exp\left[-\frac{(\mathbf{x} - \mathbf{v}_i t)^2}{4\sigma_x^2}\right]
|\nu_i\rangle.
\end{align}

For ultra-relativistic neutrinos in natural
units (c = 1) $L\approx t$, the following approximations can be employed:
\begin{equation}
E_i \simeq E, \qquad
\mathbf{p}_i \simeq E - \frac{m_i^2}{2E}, \qquad
\mathbf{v}_i \simeq 1 - \frac{m_i^2}{2E^2}.
\end{equation}

After performing the standard approximations, the three-flavor transition probability as a function of the baseline $L$ can be written as \cite{Giunti:1997wq}
\begin{equation}\label{PW}
P_{\alpha\gamma}(L) =
\sum_{i,j}
U_{\alpha i}^* U_{\gamma i} U_{\alpha } U_{\gamma j}^*
\exp\left[
-2\pi i \frac{L}{L^{\text{osc}}_{ij}} -\left(\frac{L}{L^{\text{coh}}_{ij}}\right)^2
\right],
\end{equation}
where
\begin{equation}
L^{\text{osc}}_{ij} = \frac{4\pi E}{\Delta m_{ij}^2},
\qquad
L^{\text{coh}}_{ij} = \frac{4\sqrt{2}\sigma_x E^2}{|\Delta m_{ij}^2|},
\end{equation}
are the oscillation wavelengths and coherence lengths, respectively.

For a medium with constant matter density, all relevant parameters can be treated as position-independent. Adopting an approach analogous to Ref.~\cite{Giunti:1997wq}, the transition probability in matter can be expressed as a generalization of Eq.~\eqref{PW} and is given by \cite{Ravari:2022yfd}:
\begin{align}
P^{m}_{\alpha\gamma}(L) \simeq
\sum_{i,j}
U^{m*}_{\alpha i} U^{m}_{\gamma i}
U^{m}_{\alpha j} U^{m*}_{\gamma j}
\exp\Bigg[-2\pi i \frac{L}{L^{m,\text{osc}}_{ij}} \nonumber \\ -\left(\frac{L}{L^{m,\text{coh}}_{ij}}\right)^2\Bigg],
\label{eq:WPm}
\end{align}
where the superscript $m$ indicates quantities evaluated in matter. The oscillation and coherence lengths in matter are defined as
\begin{equation}
L^{m,\text{osc}}_{ij} =
\frac{2\pi}{\Delta E^{m}_{ij}},
\qquad
L^{m,\text{coh}}_{ij} =
\frac{2\sqrt{2}\sigma_x}{|\Delta v^{m}_{ij}|}.
\label{eq:osc_coh_matter}
\end{equation}
With $\Delta E^{m}_{ij} = E^{m}_i - E^{m}_j$ and $\Delta v^{m}_{ij} = v^{m}_i - v^{m}_j$, takes the form
\begin{equation}\label{vdifference}
    \Delta v_{ij}^m = \frac{d(\Delta E_{ij}^m)}{dE}.
\end{equation}

$E^{m}_i$ and $v^{m}_i$ denote the eigenvalues of the effective Hamiltonian and the corresponding group velocities in matter. Their explicit expressions are presented in Appendix ~\ref{AppendixA}. 

The transition probability provides the likelihood of detecting a neutrino in a given flavor state after propagation. However, to describe the complete quantum state of the neutrino, it is convenient to employ the density matrix formalism. The resulting density matrix provides a space-time description of neutrino dynamics. In the presence of matter, the density matrix is defined as the projection operator onto the evolved flavor state,
\begin{equation}
\rho^{m}_{\alpha}(t,\mathbf{x})
=
|\nu^{m}_{\alpha}(t,\mathbf{x})\rangle
\langle\nu^{m}_{\alpha}(t,\mathbf{x})|,
\label{eq:density_matrix}
\end{equation}
where \(\alpha\) denotes the flavor of the neutrino at the production point, and \(|\nu^{m}_{\alpha}(t,\mathbf{x})\rangle\) represents the corresponding flavor state after propagating to the space-time point \((t,\mathbf{x})\) in matter. Since the neutrino is described by a pure quantum state, the density matrix satisfies \(\mathrm{Tr}(\rho^{m}_{\alpha})=1\) and contains the complete information about its quantum evolution.

The final density matrix is obtained by Gaussian averaging of $\rho^{m}_{\alpha}(t,\mathbf{x})$ over time as,
\begin{equation}
\rho^{m}_{\alpha}(x)
=
\sum_{i,j}
U^{m}_{\alpha i} U^{m*}_{\alpha j}\,
f^{m}_{ij}(x)\,
|\nu^{m}_i\rangle\langle\nu^{m}_j|,
\label{eq:density_matrix_wp}
\end{equation}

\begin{equation}
f^{m}_{ij}(x)
=
\exp\left[
-i{\Delta E^{m}_{ij}}\,x
-
\left(
\frac{|\Delta v^{m}_{ij}|\,x}
{2\sqrt{2}\,\sigma_x}
\right)^{2}
\right].
\label{eq:fjk}
\end{equation}

\section{Residual Entanglement and Residual Coherence in matter induced neutrino oscillation}\label{Sec:QC_NO}
The neutrino flavor states can be represented in the three-mode (occupation-number) basis. In this representation, the presence (absence) of a neutrino in a given flavor mode is denoted by 1 (0). The flavor states are therefore expressed as
\begin{align}
|\nu_e\rangle &= |1\rangle_e \otimes |0\rangle_\mu \otimes |0\rangle_\tau
\equiv |100\rangle, \nonumber\\
|\nu_\mu\rangle &= |0\rangle_e \otimes |1\rangle_\mu \otimes |0\rangle_\tau
\equiv |010\rangle, \\
|\nu_\tau\rangle &= |0\rangle_e \otimes |0\rangle_\mu \otimes |1\rangle_\tau
\equiv |001\rangle. \nonumber
\label{eq:occupation_basis}
\end{align}

The mass eigenstates can be expressed in terms of the flavor eigenstates through the relation
\begin{equation}
|\nu_i\rangle=\sum_{\alpha=e,\mu,\tau}U_{\alpha i}\,|\nu_\alpha\rangle,
\end{equation}
where \(i=1,2,3\). Using this transformation, the density matrix given in Eq.~\eqref{eq:density_matrix_wp} can be written in the flavor basis as
\begin{equation}
\rho^{m}_\alpha(x)=\sum_{\alpha,\gamma=e,\mu,\tau}
F_{\alpha\gamma}^{m}(x)
|\nu_\alpha\rangle\langle\nu_\gamma|,
\end{equation}
where the coefficients \(F_{\alpha\gamma}^{m}(x)\) are defined by
\begin{equation} \label{38}
F_{\alpha\gamma}^{\,m}(x)=
\sum_{i,j=1}^{3}
U_{\alpha i}^{m*}\,
U_{\gamma j}^{m}\,
f^{m}_{ij}(x)\,
U_{\alpha i}^{m}\,
U_{\gamma j}^{m*}.
\end{equation}

From Eq.~\eqref{38}, the density matrix corresponding to the neutrino state can be written as
\begin{equation}
\rho_{e\mu\tau}^{m}(x)=
\begin{pmatrix}
0 & 0 & 0 & 0 & 0 & 0 & 0 & 0\\
0 & 0 & 0 & 0 & 0 & 0 & 0 & 0\\
0 & 0 & 0 & 0 & 0 & 0 & 0 & 0\\
0 & 0 & 0 & F_{ee}^{m}(x) & 0 & F_{e\mu}^{m}(x) & F_{e\tau}^{m}(x) & 0\\
0 & 0 & 0 & 0 & 0 & 0 & 0 & 0\\
0 & 0 & 0 & F_{\mu e}^{m}(x) & 0 & F_{\mu\mu}^{m}(x) & F_{\mu\tau}^{m}(x) & 0\\
0 & 0 & 0 & F_{\tau e}^{m}(x) & 0 & F_{\tau\mu}^{m}(x) & F_{\tau\tau}^{m}(x) & 0\\
0 & 0 & 0 & 0 & 0 & 0 & 0 & 0
\end{pmatrix}.
\label{eq:density_matrix_fock}
\end{equation}
The probability that a neutrino initially produced in the flavor state \(\nu_\alpha\) is detected as a flavor state \(\nu_\gamma\) at a distance \(x\) is obtained from the density matrix \(\rho^m_{\alpha}(x)\) as
\begin{equation}
P(\nu_\alpha \rightarrow \nu_\gamma)(x)
= \mathrm{Tr}\!\left[\,|\nu_\gamma\rangle\langle\nu_\gamma|\,
\rho^{m}_{\alpha}(x)\,\right]
= F_{\gamma\gamma}^{(m)}(x)=P^{m}_{\alpha \gamma}.
\label{eq:transition_probability}
\end{equation}
The flavor evolution of an initially produced muon-neutrino state, $|\nu_\mu\rangle$, is considered as it propagates through matter. The three-flavor transition probabilities such as $P(\nu_\mu \rightarrow \nu_e)$ and $P(\nu_\mu \rightarrow \nu_\tau)$, denoted by $P^{m}_{\mu e}$ and $P^{m}_{\mu\tau}$, respectively, represent the oscillation probabilities, while $P(\nu_\mu \rightarrow \nu_\mu)$, denoted by $P^{m}_{\mu\mu}$, represents the survival probability. By substituting the neutrino density matrix given in Eq.~\eqref{eq:density_matrix_fock} into the definitions of the Three-$\pi$ entanglement measure as in Eq.~\eqref{eq:thee-pi} and the Concurrence Fill as in Eq.~\eqref{eq:concurrence_fill}, their expressions in terms of the oscillation and survival probabilities in matter can be obtained as in \cite{Li:2022mus}
\begin{equation}
\begin{aligned}
\text{Three-}\pi
= \frac{4}{3}\Bigg[
&- \left(P_{\mu e}^{m}\right)^2
- \left(P_{\mu\mu}^{m}\right)^2
- \left(P_{\mu\tau}^{m}\right)^2  \\
&+ P_{\mu e}^{m}
\sqrt{\left(P_{\mu e}^{m}\right)^2
+4P_{\mu\mu}^{m}P_{\mu\tau}^{m}} \\
&+ P_{\mu\mu}^{m}
\sqrt{\left(P_{\mu\mu}^{m}\right)^2
+4P_{\mu e}^{m}P_{\mu\tau}^{m}} \\
&+ P_{\mu\tau}^{m}
\sqrt{\left(P_{\mu\tau}^{m}\right)^2
+4P_{\mu e}^{m}P_{\mu\mu}^{m}}
\Bigg],
\end{aligned}
\label{eq:three_pi_probability}
\end{equation}
and
\begin{equation}
\begin{aligned}
\text{Concurrence Fill}
&=
\frac{8}{3}
\Bigg\{
\left(
P_{\mu e}^{m}
P_{\mu\mu}^{m}
P_{\mu\tau}^{m}
\right)^2 \times   \\
&\qquad
\Big[
P_{\mu\tau}^{m}P_{\mu\mu}^{m}
+
P_{\mu e}^{m}
\left(
P_{\mu\mu}^{m}
+
P_{\mu\tau}^{m}
\right)
\Big]
\Bigg\}^{1/4},
\end{aligned}
\label{eq:CF_probability}
\end{equation}
respectively.
The $l_1$-norm coherence, expressed in terms of the three-flavor transition probabilities, can be written as \cite{Dixit:2019swl,Jha:2020dav}
\begin{equation}
    C_{l_{1}}(\rho_{ABC})= 2(\sqrt{P_{\mu e}^{m}P_{\mu \mu}^{m}}+\sqrt{P_{\mu e}^{m}P_{\mu \tau}^{m}}+\sqrt{P_{\mu \mu}^{m}P_{\mu \tau}^{m}}),
\end{equation}
similarly,
\begin{equation}
C_{l_{1}}(\rho_{AB}) = 2(\sqrt{P_{\mu e}^{m} P_{\mu \mu}^{m}}),
\quad
C_{l_{1}}(\rho_{AC}) =  2(\sqrt{P_{\mu e}^{m} P_{\mu \mu}^{m}}).
\end{equation}

Therefore, the residual $l_1$-norm coherence as in Eq. \eqref{coherence} for the neutrino system can be written as
\begin{equation}
R_{l_{1}\text{-norm}}
=2\sqrt{P_{\mu\mu}^{m}P_{\mu\tau}^{m}}.
\label{eq:l1norm_probability}
\end{equation}

\section{Result and discussions} \label{Sec:Result}
\begin{table}[ht]
\centering
\begin{tabular}{|l|c|}
\hline
\textbf{Oscillation Parameters} & \textbf{ Best-fit value} \\
\hline
$\Delta m_{21}^{2}\,(\mathrm{eV}^{2})$ & $7.537\times10^{-5}$ \\
\hline
$\Delta m_{3\ell}^{2}\,(\mathrm{eV}^{2})$ & $2.511\times10^{-3}$ \\
\hline
$\sin^{2}\theta_{12}$ & 0.3088 \\
\hline
$\sin^{2}\theta_{13}$ & 0.02248 \\
\hline
$\sin^{2}\theta_{23}$ & 0.470 \\
\hline
$\delta_{\mathrm{CP}}$ & $212^\circ$ \\
\hline
\end{tabular}
\caption{Best-fit values of the standard three-neutrino oscillation parameters used in our analysis \cite{Esteban_2024}.}
\label{del1}
\end{table}

\begin{figure*}[htb]
\centering
     \begin{subfigure}{0.32\textwidth}
        \centering
 \includegraphics[height=45mm, width=53mm]{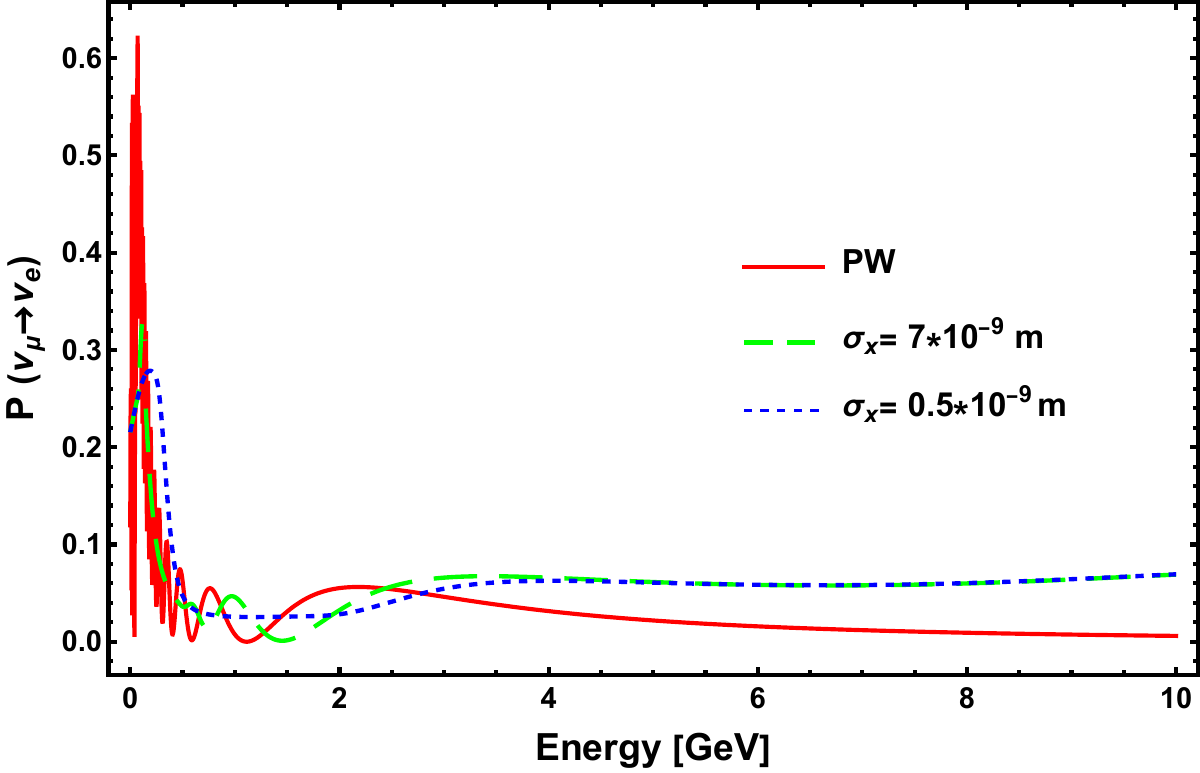}
        \caption{DUNE}
        \label{1a_fig:sub1}
    \end{subfigure}
    \hfill
    \begin{subfigure}{0.32\textwidth}
        \centering
      \includegraphics[height=45mm, width=53mm]{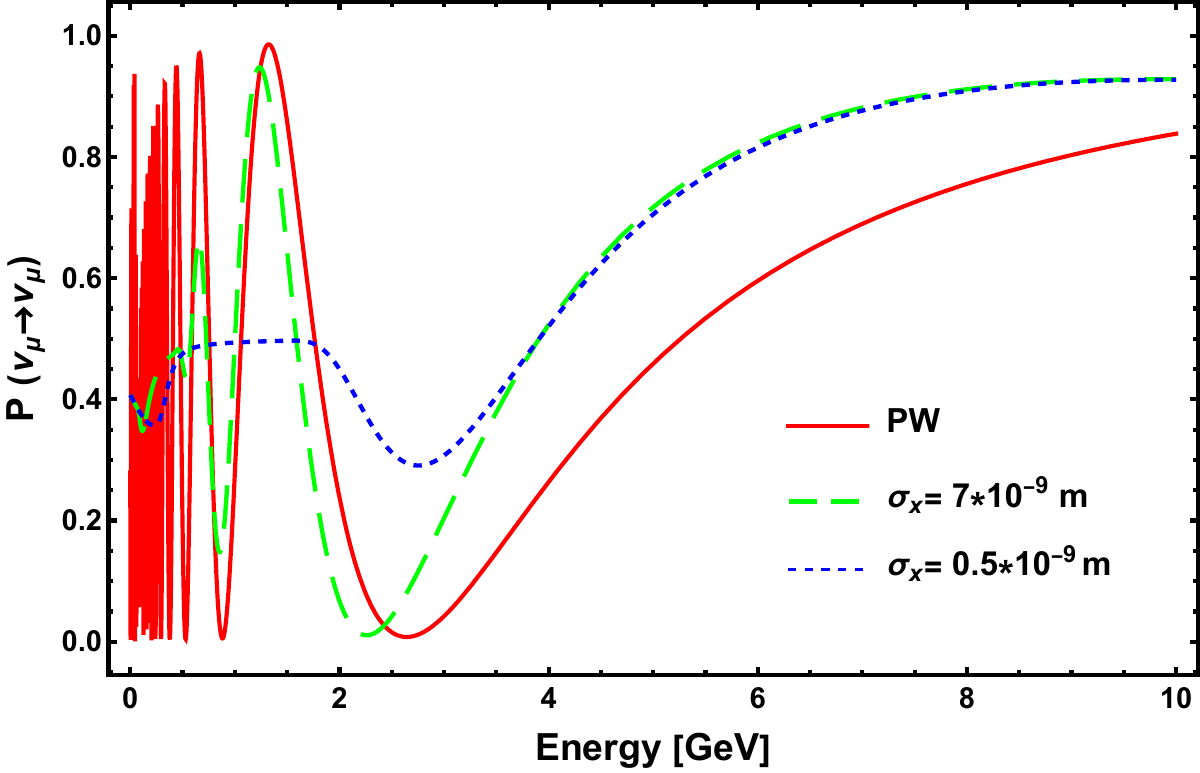}
        \caption{DUNE}
        \label{1a_fig:sub2}
    \end{subfigure}
     \hfill
    \begin{subfigure}{0.32\textwidth}
        \centering
       \includegraphics[height=45mm, width=53mm]{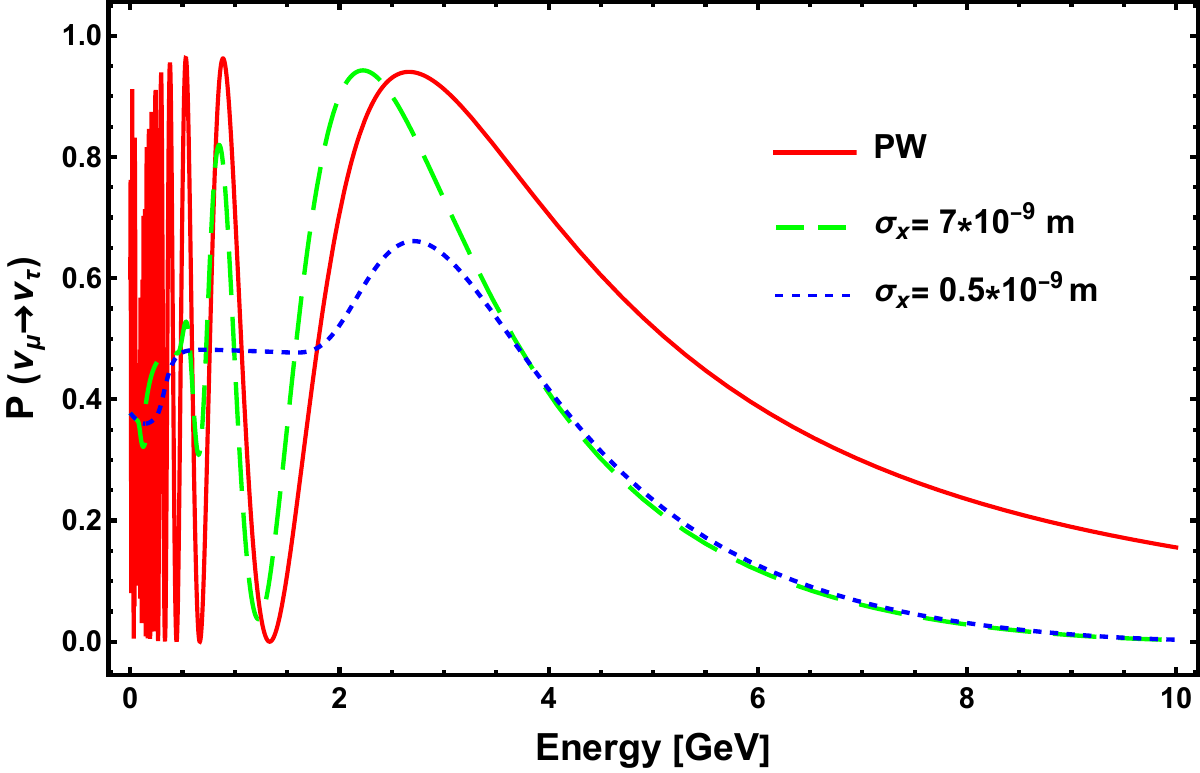}

        \caption{DUNE}
        \label{1a_fig:sub3}
    \end{subfigure}
    
    \medskip

    \begin{subfigure}{0.32\textwidth}
        \centering
   \includegraphics[height=45mm, width=53mm]{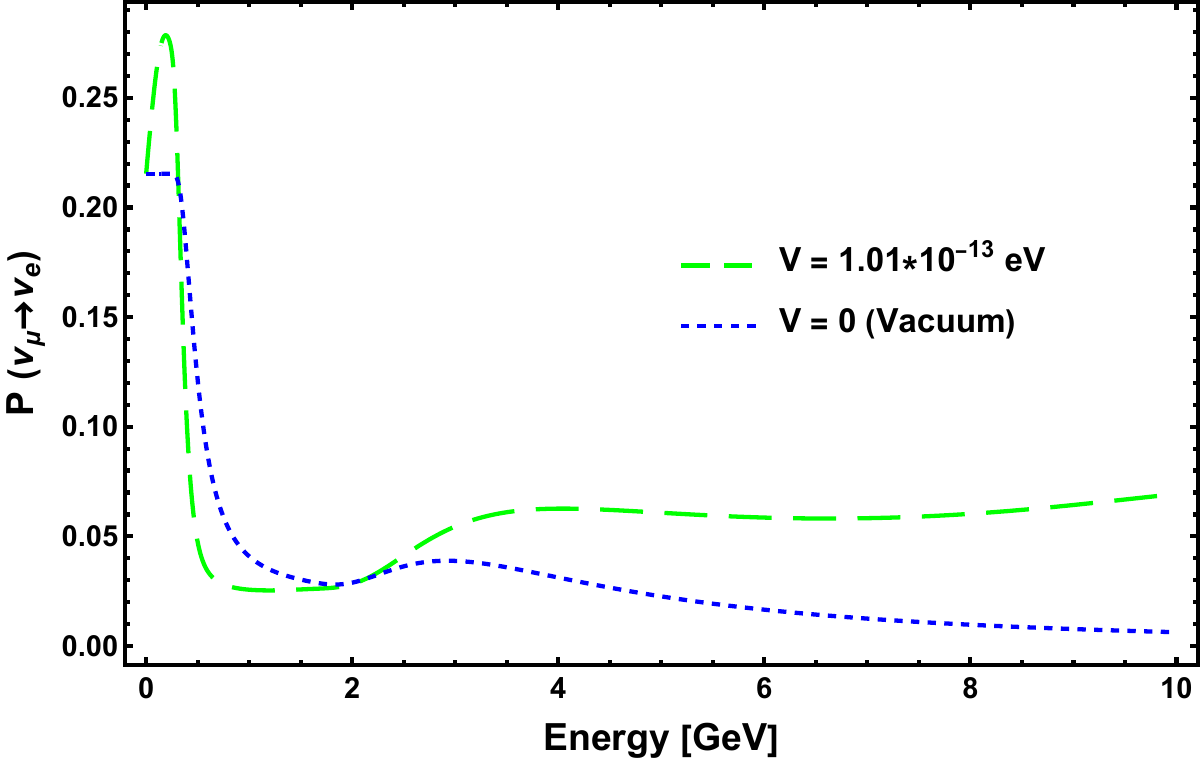}
        \caption{DUNE}
        \label{1a_fig:sub4}
    \end{subfigure}
    \hfill
    \begin{subfigure}{0.32\textwidth}
        \centering
     \includegraphics[height=45mm, width=53mm]{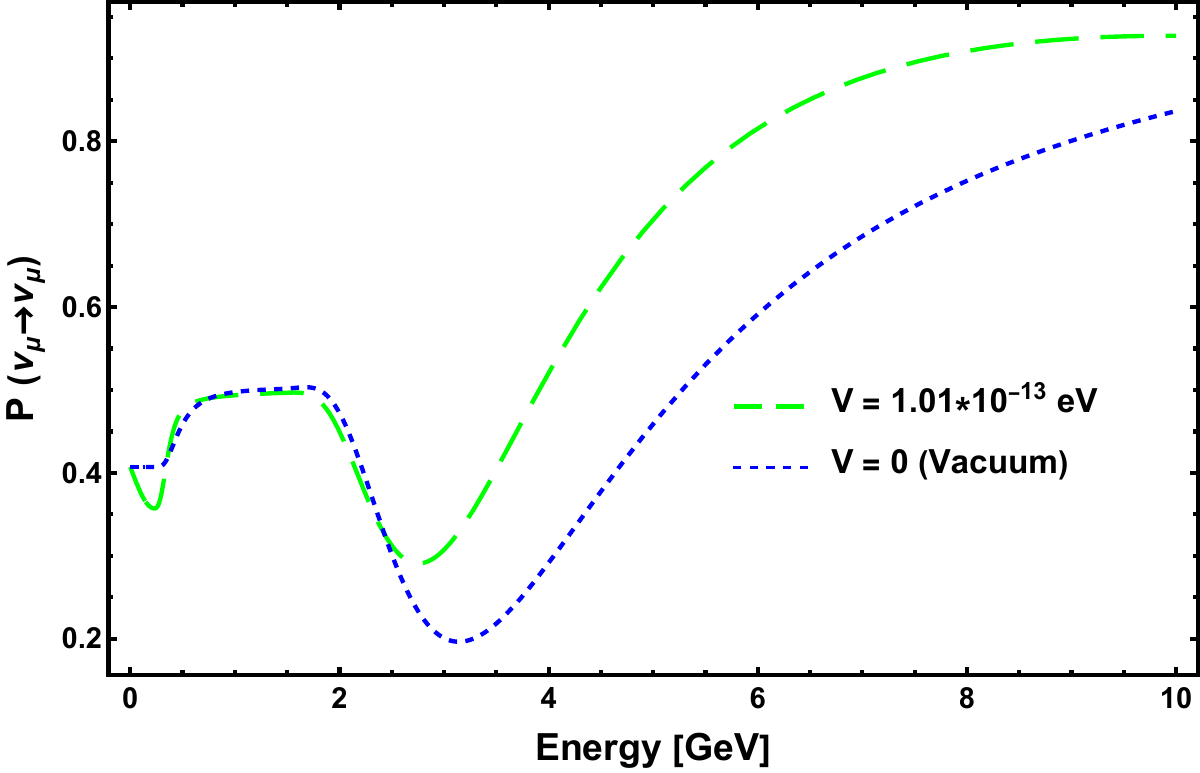}
        \caption{DUNE }
        \label{1a_fig:sub5}
    \end{subfigure}
     \hfill
    \begin{subfigure}{0.32\textwidth}
        \centering
  \includegraphics[height=45mm, width=53mm]{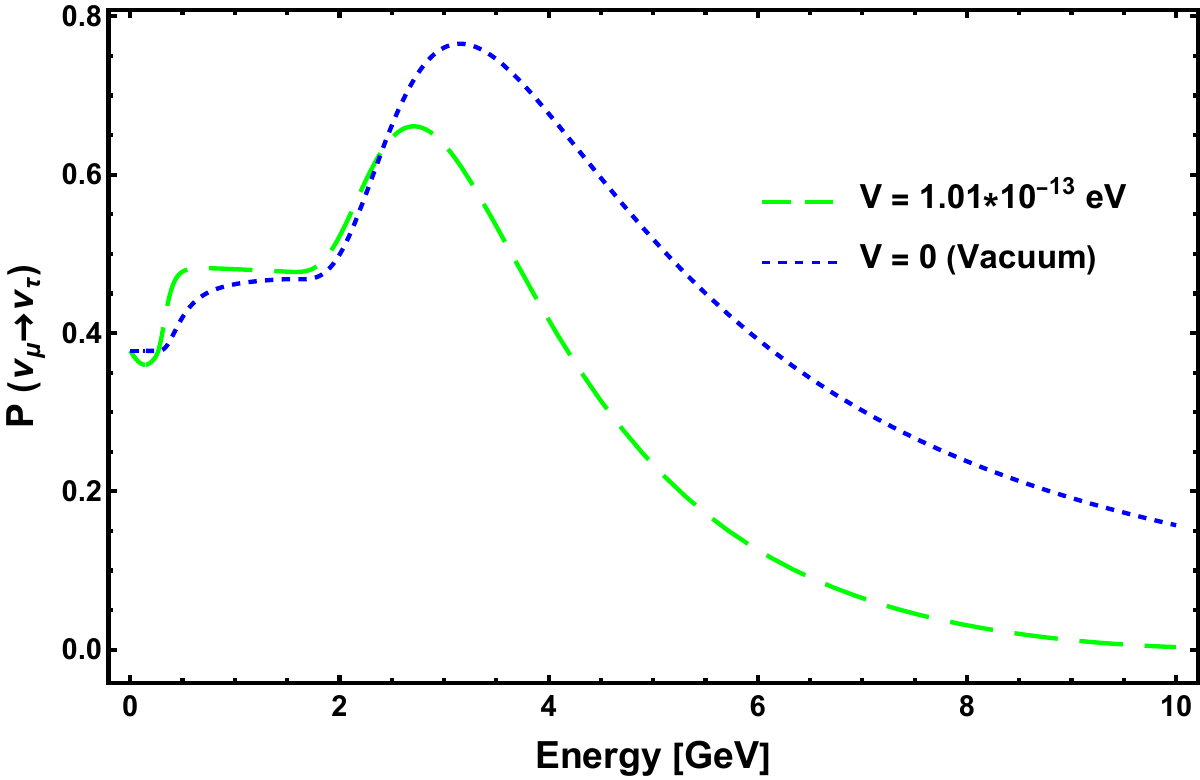}
        \caption{DUNE}
        \label{1a_fig:sub6}
    \end{subfigure}

\caption{\justifying{The oscillation ($P(\nu_\mu \rightarrow \nu_e)$ \& $P(\nu_\mu \rightarrow \nu_\tau)$) and survival ($P(\nu_\mu \rightarrow \nu_\mu)$) probabilities of an initial state $\ket{\nu_\mu}$ are shown as functions of neutrino energy\,$[\text{GeV}]$ for the DUNE experimental configuration with a baseline of $L=1300$ km. The upper panels present a comparison between PW and wave-packet formalisms for two different wave-packet widths, with the matter potential fixed at $V=1.01\times10^{-13}\,\mathrm{eV}$, shown in Figs. 1(a)-1(c). The lower panels illustrate the effect of matter by comparing neutrino propagation in matter and vacuum for a fixed wave-packet width of $\sigma_x=0.5\times10^{-9}\,\mathrm{m}$, shown in Figs. 1(d)-1(f).}}
\label{fig1}
\end{figure*}

This section presents the results obtained for the DUNE experimental configuration \cite{DUNE:2015lol} in Sec.~\ref{sec:7.3.1}, followed by the analysis for an extended propagation distance beyond the DUNE baseline in Sec.~\ref{sec:7.3.2}. Throughout the analysis, the standard neutrino oscillation parameters are taken at their best-fit values listed in Table~\ref{del1}, where the neutrino mass ordering is assumed to be normal ordering (NO). To investigate the combined influence of matter effects and wave-packet decoherence on the evolution of oscillation probabilities as well as the quantum correlation measures, we consider two cases: one in which the matter potential $V$ is held constant while the wave-packet width $\sigma_x$ is varied, and another in which $\sigma_x$ is fixed while the matter potential is changed.

\subsection{Analysis for the DUNE configuration} \label{sec:7.3.1}
The Deep Underground Neutrino Experiment (DUNE) \cite{DUNE:2015lol} is a next-generation long-baseline neutrino oscillation experiment that aims to address several fundamental questions in neutrino physics. The experiment utilizes an intense wide-band neutrino beam generated at the Long-Baseline Neutrino Facility (LBNF) at Fermilab, which propagates over a baseline of 1300 km to a far detector located at the Sanford Underground Research Facility (SURF) in South Dakota. The combination of a long propagation distance and a broad neutrino energy spectrum, extending from approximately 0.5 to 8 GeV with a flux peak around 2.5 GeV, significantly enhances matter effects during neutrino propagation. These features provide excellent sensitivity to the determination of the neutrino mass ordering, precise measurements of the leptonic CP-violating phase $\delta_{\rm CP}$, and the exploration of possible signatures of physics beyond the Standard Model, including non-standard neutrino interactions.

Fig. \ref{fig1} presents the neutrino oscillation ($P(\nu_\mu \rightarrow \nu_e)$ [\ref{fig1}(\subref{1a_fig:sub1})] \& $P(\nu_\mu \rightarrow \nu_\tau)$ [\ref{fig1}(\subref{1a_fig:sub2})]) and survival ($P(\nu_\mu \rightarrow \nu_\mu)$)[\ref{fig1}(\subref{1a_fig:sub3})] probabilities of the initial state $\ket{\nu_\mu}$ as functions of neutrino energy \,[\text{GeV}] for the DUNE baseline ($L = 1300~\rm km$) within the wave-packet formalism, incorporating matter effects. For the DUNE configuration, the matter potential is fixed at $V = 1.01 \times 10^{-13}~\rm eV$, which corresponds to the Earth's matter potential \cite{Bouri:2024kcl}. The upper panels of Fig.~\ref{fig1} display the energy dependence of the oscillation and survival probabilities for a constant matter potential of $V = 1.01 \times 10^{-13}~\rm eV$, while considering different values of the wave-packet width $\sigma_x$ together with the PW scenario. The PW result is shown by the solid red curve, whereas the wave-packet cases with $\sigma_x = 7 \times 10^{-9}~\rm m$ and $\sigma_x = 0.5 \times 10^{-9}~\rm m$ are represented by the green dashed and blue dotted curves, respectively. On the other hand, the lower panels of Fig.~\ref{fig1} illustrate the effect of matter by comparing the probabilities in matter and vacuum for a fixed wave-packet width of $\sigma_x = 0.5 \times 10^{-9}~\rm m$. The green dashed curve corresponds to propagation in matter, while the blue dotted curve represents propagation in vacuum.

 \begin{figure*}[htb]
 \centering
     \begin{subfigure}{0.49\textwidth}
        \centering
 \includegraphics[height=55mm, width=65mm]{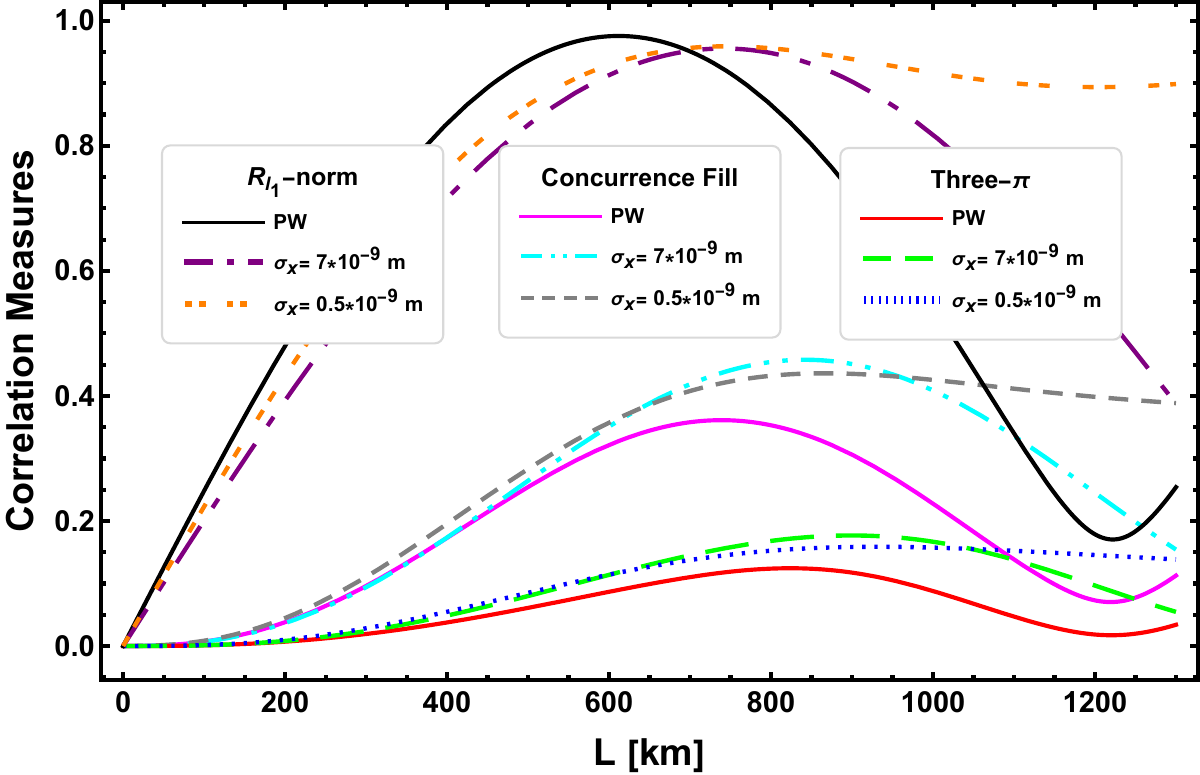}
        \caption{DUNE}
        \label{2a_fig:sub1}
    \end{subfigure}
    \hfill
    \begin{subfigure}{0.49\textwidth}
        \centering
      \includegraphics[height=55mm, width=65mm]{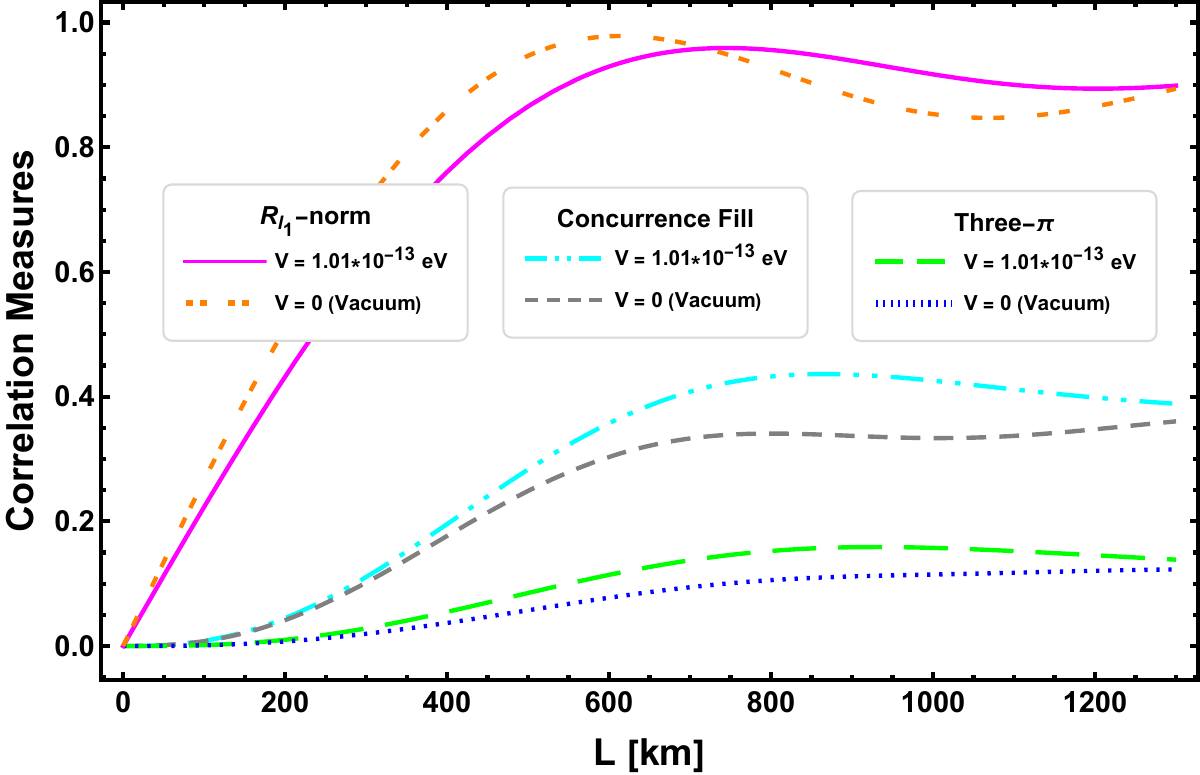}
        \caption{DUNE}
        \label{2a_fig:sub2}
    \end{subfigure}

\caption{\justifying{For an initial state $\ket{\nu_\mu}$, various residual quantum correlation measures, including the $R_{l_{1}\text{-norm}}$, Concurrence Fill, and Three-$\pi$, are shown as functions of the propagation length $L\,[\text{km}]$ for the DUNE experimental configuration at $E = 2.5$ GeV. The left plot illustrates these measures for different values of the wave-packet width $\sigma_x$, with the matter potential fixed at $V = 1.01 \times 10^{-13}$ eV [see Fig. 2(a)]. The right plot shows the results for a fixed wave-packet width of $\sigma_x = 0.5 \times 10^{-9}$ m and different values of the matter potential [see Fig. 2(b)].}}
\label{fig2}
\end{figure*}

The oscillation probability $P(\nu_\mu \rightarrow \nu_e)$, shown in Fig. \ref{fig1}(\subref{1a_fig:sub1}) in the upper left panel, exhibits strong oscillatory behavior in the PW scenario at low energies, whereas the probability decreases smoothly as the energy increases. Incorporating the wave-packet description modifies this behavior considerably. For $\sigma_x = 7 \times 10^{-9}\,\rm m$, the oscillation probability closely follows the PW prediction at lower energies, but as the energy increases, wave-packet effects reduce the oscillation amplitude and shift the positions of the oscillation maxima and minima relative to the PW case. Consequently, the oscillation pattern becomes smoother at higher energies. With a further decrease in the wave-packet width to $\sigma_x = 0.5 \times 10^{-9}\,\rm m$, the oscillatory structure is further reduced, and a much smoother energy dependence of the appearance probability is observed in a constant matter throughout the considered energy range.

The upper middle panel in Fig. \ref{fig1}(\subref{1a_fig:sub2}) shows the survival probability, $P(\nu_\mu \rightarrow \nu_\mu)$, as a function of neutrino energy. In the PW scenarios, the survival probability displays pronounced oscillatory behavior in the low-energy region, with the first oscillation maximum occurring around $E \simeq 2.5$ - $3$ GeV. At higher energies, the survival probability gradually increases. When wave-packet effects are taken into account with $\sigma_x = 7 \times 10^{-9}\,\rm m$, the probability remains close to the PW prediction at low energies. However, the position of the first oscillation maximum shifts toward the lower-energy range of $E \simeq 2$ - $2.5$ GeV, and noticeable deviations from the PW result emerge at higher energies. For the smaller wave-packet width, $\sigma_x = 0.5 \times 10^{-9}\,\rm m$, the oscillatory features are suppressed more strongly, resulting in a smoother energy dependence of the survival probability. This behavior highlights the impact of wave-packet decoherence on neutrino survival probabilities during propagation in constant matter.

The upper right panel as in Fig. \ref{fig1}(\subref{1a_fig:sub3}) depicts the oscillation probability $P(\nu_\mu \rightarrow \nu_\tau)$. In the PW case, the oscillation probability undergoes rapid oscillations in the low-energy region, with a maximum appearing at approximately $E \simeq 0.5$ GeV and the probability decreases gradually with increasing energy. For a wave-packet width of $\sigma_x = 7 \times 10^{-9}\,\rm m$, the probability follows the overall behavior of the PW prediction at lower energies, although the location of the maximum is shifted toward $E \simeq 2.5$ GeV. This deviation from the PW result becomes more evident as the energy increases. When the wave-packet width is reduced to $\sigma_x = 0.5 \times 10^{-9}\,\rm m$, the oscillatory features are substantially damped, yielding a considerably smoother energy dependence. Moreover, for $E > 3.5~\rm GeV$, the results for $\sigma_x = 0.5 \times 10^{-9}\,\rm m$ and $\sigma_x = 7 \times 10^{-9}\,\rm m$ become nearly indistinguishable, while the PW prediction remains clearly distinguishable for all three probabilities. It is evident that the smaller wave-packet widths modify the interference pattern of neutrino oscillations compared to the conventional PW description in constant matter.

The lower panels shown in Figs. \ref{fig1}(\subref{1a_fig:sub4})-\ref{fig1}(\subref{1a_fig:sub6}) examine the impact of the matter potential on neutrino oscillation and survival probabilities by comparing propagation in matter, with $V = 1.01 \times 10^{-13}~\rm eV$, to the vacuum case, $V=0$. In this comparison, the wave-packet width is fixed at $\sigma_x = 0.5 \times 10^{-9}~\rm m$. The lower left panel as in Fig. \ref{fig1}(\subref{1a_fig:sub4}) shows the oscillation probability $P(\nu_\mu \rightarrow \nu_e)$ as functions of energy\,[\text{GeV}]. At lower neutrino energies, the probability in matter remains below the corresponding vacuum prediction. As the energy increases, however, this behavior changes, and for $E \gtrsim 2.5~\rm GeV$, the matter-induced probability exceeds the vacuum result. This enhancement becomes increasingly evident in the higher-energy region, demonstrating the significant role of matter effects in modifying the $\nu_\mu \rightarrow \nu_e$ oscillation probability. The matter effect also modifies the overall shape of the oscillation curve, leading to visible differences in the oscillation pattern compared to vacuum propagation.

In the lower middle panel shown in Fig. \ref{fig1}(\subref{1a_fig:sub5}), which represents the survival probability $P(\nu_\mu \rightarrow \nu_\mu)$, the inclusion of matter effects changes both the depth and the position of the oscillation minimum. At higher energies, the survival probability in matter reaches larger values compared to the vacuum case. For the lower right panel shown in Fig. \ref{fig1}(\subref{1a_fig:sub6}), corresponding to the oscillation probability $P(\nu_\mu \rightarrow \nu_\tau)$, matter effects significantly alter the oscillation profile. The peak structure and the damping behavior differ noticeably from the vacuum case, particularly at energies $\geq$ 2.5 GeV.

Overall, Fig.~\ref{fig1} shows that both wave-packet effects and matter effects strongly influence neutrino flavor transitions for the DUNE baseline. While wave-packet effects tend to smooth and suppress the oscillatory structures, matter effects modify the amplitudes and positions of the oscillation features, leading to clear deviations from the vacuum scenario.

Figure~\ref{fig2} displays the dependence of the residual quantum correlation measures on the propagation distance $L$ (km) for the DUNE configuration, considering a fixed neutrino energy of $E = 2.5~\rm GeV$. The analysis considers three residual quantum correlation measures: the residual entanglement measures, Three-$\pi$ and Concurrence Fill, along with the residual $l_1$-norm coherence ($R_{l_{1}\text{-norm}}$).

The left panel shown in Fig. \ref{fig2}(\subref{2a_fig:sub1}) illustrates the dependence of the residual quantum correlation measures on the wave-packet width $\sigma_x$, while keeping the matter potential fixed at $V = 1.01 \times 10^{-13}~\rm eV$. For the $R_{l_{1}\text{-norm}}$, the black solid, magenta dashed, and orange dotted curves denote the PW scenario, $\sigma_x = 7 \times 10^{-9}~\rm m$, and $\sigma_x = 0.5 \times 10^{-9}~\rm m$, respectively. In the case of Concurrence Fill, the corresponding results are shown by the pink solid, cyan dash-dotted, and gray dashed curves. For the Three-$\pi$ measure, the PW result is represented by the red solid curve, while the green dashed and blue dotted curves correspond to $\sigma_x = 7 \times 10^{-9}~\rm m$ and $\sigma_x = 0.5 \times 10^{-9}~\rm m$, respectively. It is observed that all three correlation measures initially increase with the propagation distance, indicating the gradual generation of quantum correlations during neutrino flavor evolution. The PW case generally exhibits a stronger oscillatory nature, whereas the inclusion of finite wave-packet widths smooths the evolution profiles. For the smaller wave-packet width, $\sigma_x = 0.5 \times 10^{-9}~\rm m$, the correlation measures attain higher values towards the end of the baseline. Among the three measures, the $R_{l_{1}\text{-norm}}$ remains dominant over the entire propagation range, while the Three-$\pi$ entanglement has comparatively smaller values. The Concurrence Fill shows an intermediate behavior between these two measures.

\begin{figure*}[htb]
    \centering
     \begin{subfigure}{0.32\textwidth}
        \centering
       \includegraphics[width=40mm]{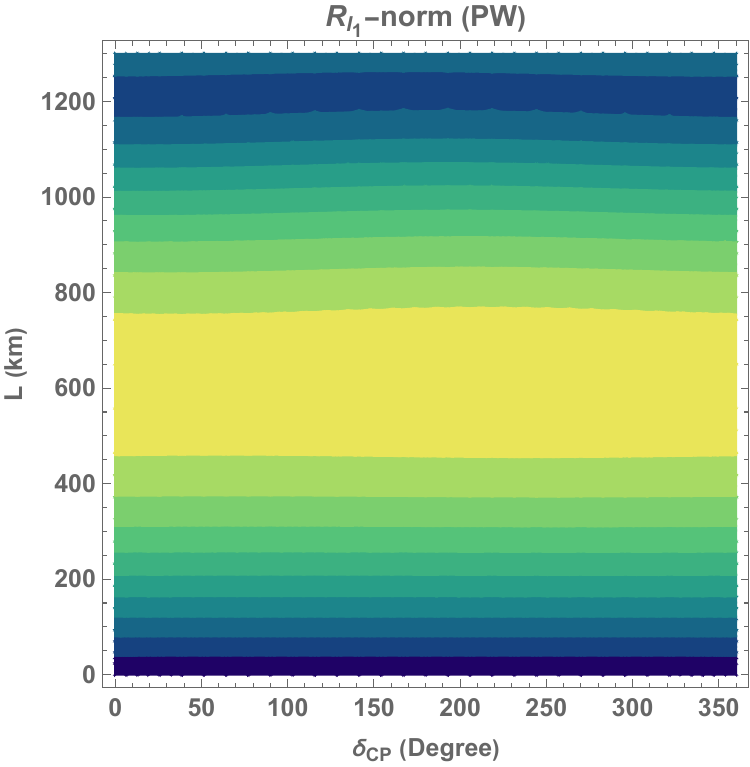}
       \includegraphics[width=7.4mm]{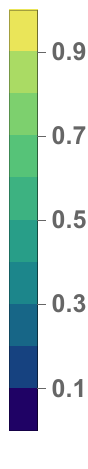}
        \caption{PW with $V=1.01\times10^{-13}$ eV}
        \label{5a_fig:sub1}
    \end{subfigure}
    \hfill
    \begin{subfigure}{0.32\textwidth}
        \centering
       \includegraphics[width=40mm]{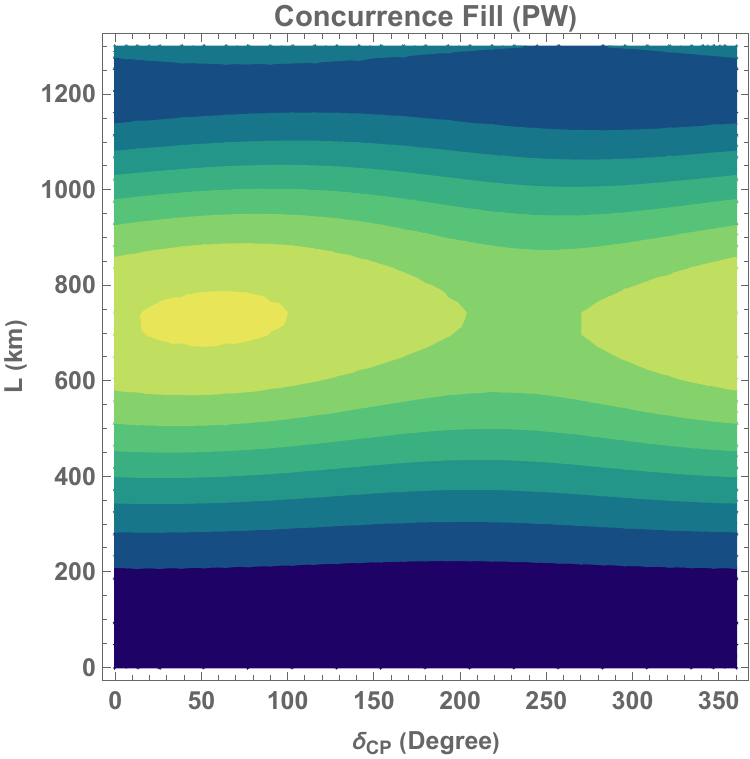}
       \includegraphics[width=8.4mm]{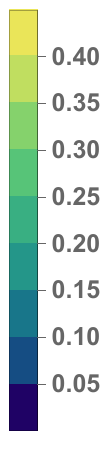}
        \caption{PW with $V=1.01\times10^{-13}$ eV}
        \label{5a_fig:sub2}
    \end{subfigure}
     \hfill
    \begin{subfigure}{0.32\textwidth}
        \centering
      \includegraphics[width=40mm]{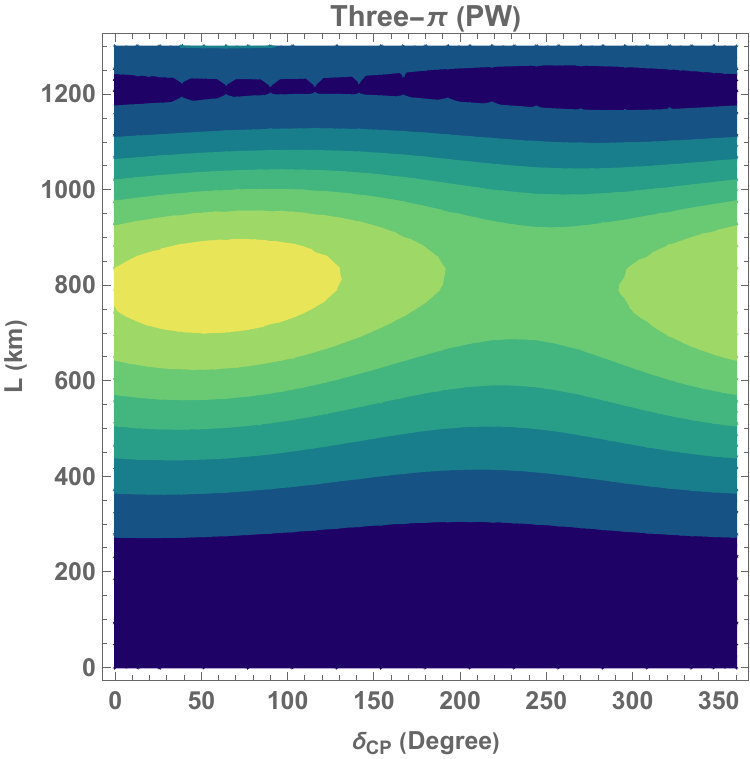}
      \includegraphics[width=8.4mm]{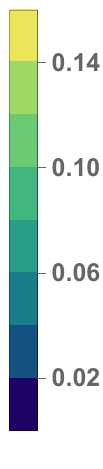}
        \caption{PW with $V=1.01\times10^{-13}$ eV}
        \label{5a_fig:sub3}
    \end{subfigure}
    
    \medskip

    \begin{subfigure}{0.32\textwidth}
        \centering
       \includegraphics[width=40mm]{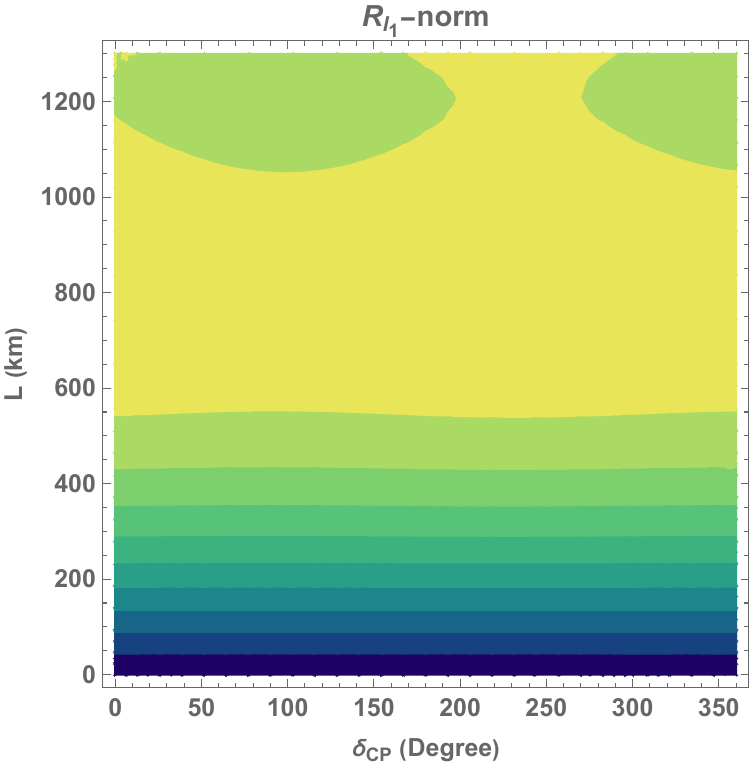}
       \includegraphics[width=7.4mm]{contour_plots/con_mark_rl1.pdf}
        \caption{$\sigma_x=0.5\times10^{-9}$m and $V=1.01\times10^{-13}$ eV}
        \label{5a_fig:sub4}
    \end{subfigure}
    \hfill
    \begin{subfigure}{0.32\textwidth}
        \centering
      \includegraphics[width=40mm]{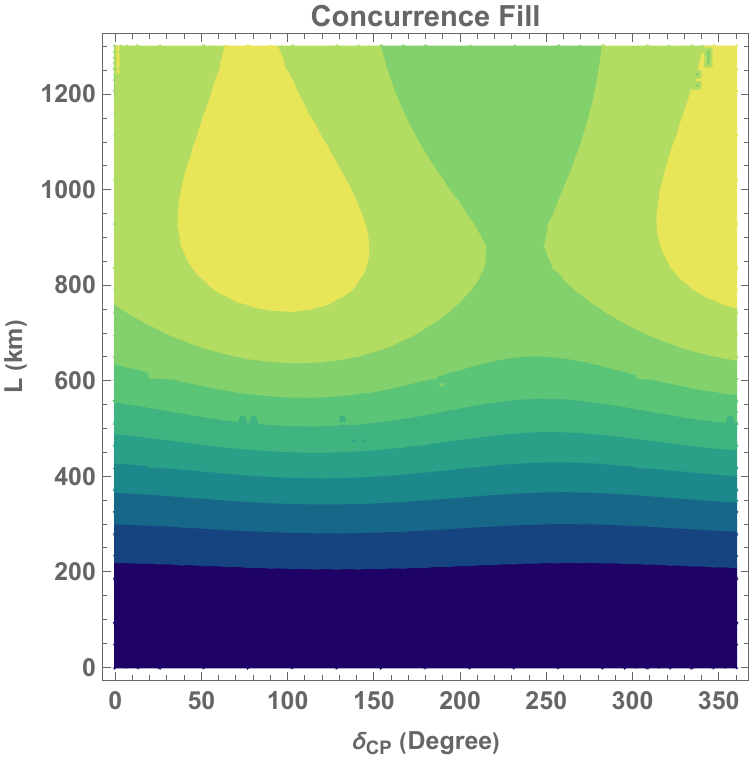}
\includegraphics[width=8.6mm]{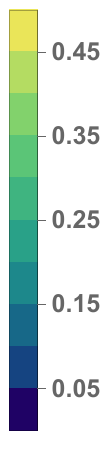}
        \caption{$\sigma_x=0.5\times10^{-9}$ m and $V=1.01\times10^{-13}$ eV }
        \label{5a_fig:sub5}
    \end{subfigure}
     \hfill
    \begin{subfigure}{0.32\textwidth}
        \centering
      \includegraphics[width=40mm]{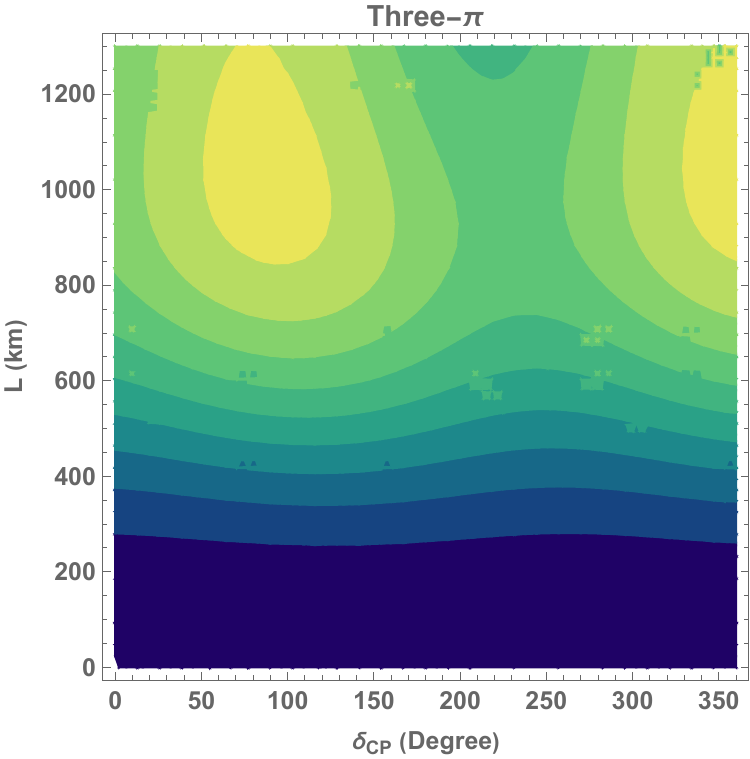}
 \includegraphics[width=8.6mm]{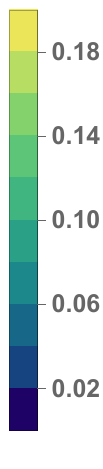}
        \caption{$\sigma_x=0.5\times10^{-9}$ m and $V=1.01\times10^{-13}$ eV}
        \label{5a_fig:sub6}
    \end{subfigure}

\medskip

    \begin{subfigure}{0.32\textwidth}
        \centering
       \includegraphics[width=40mm]{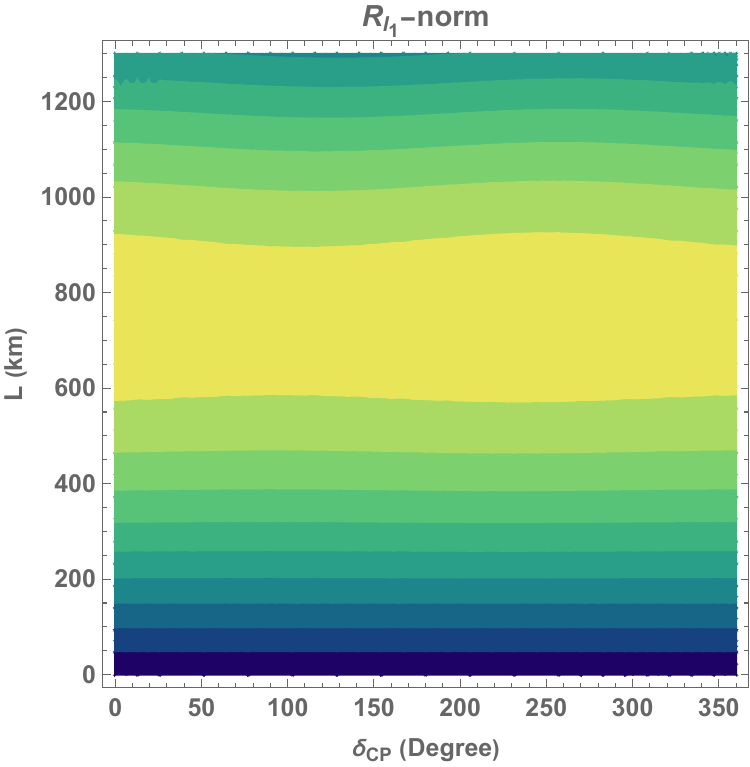}
\includegraphics[width=7.4mm]{contour_plots/con_mark_rl1.pdf}
        \caption{$\sigma_x=7\times10^{-9}$ m and $V=1.01\times10^{-13}$ eV}
        \label{5a_fig:sub7}
    \end{subfigure}
    \hfill
    \begin{subfigure}{0.32\textwidth}
        \centering
      \includegraphics[width=40mm]{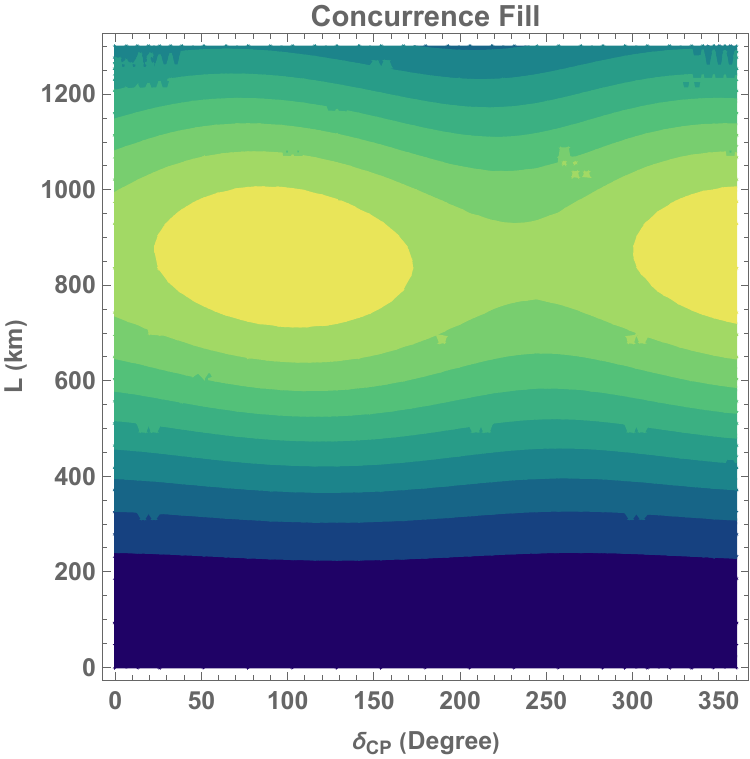}
\includegraphics[width=8.6mm]{contour_plots/con_mark_confill.pdf}
        \caption{$\sigma_x=7\times10^{-9}$ m and $V=1.01\times10^{-13}$ eV}
        \label{5a_fig:sub8}
    \end{subfigure}
     \hfill
    \begin{subfigure}{0.32\textwidth}
        \centering
      \includegraphics[width=40mm]{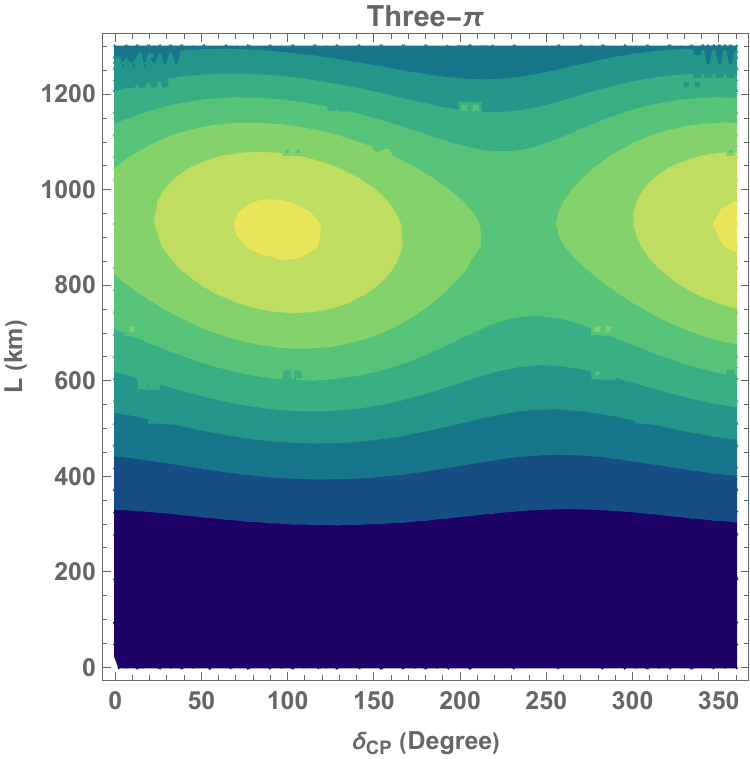}
 \includegraphics[width=9.6mm]{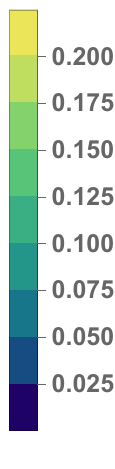}

        \caption{$\sigma_x=7\times10^{-9}$ m and $V=1.01\times10^{-13}$ eV}
        \label{5a_fig:sub9}
    \end{subfigure}

\medskip

    \begin{subfigure}{0.32\textwidth}
        \centering
       \includegraphics[width=40mm]{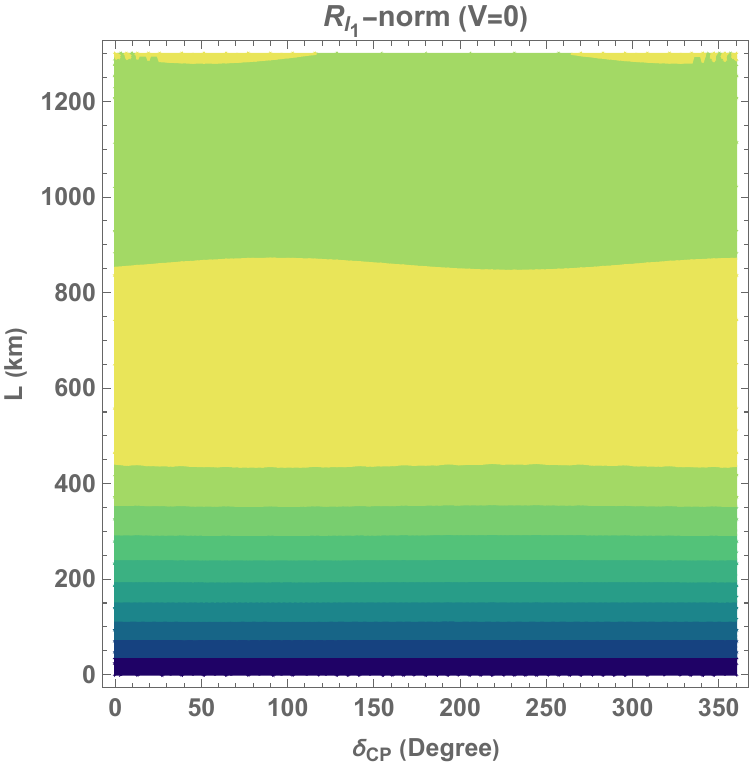}
\includegraphics[width=7.4mm]{contour_plots/con_mark_rl1.pdf}
        \caption{V=0 and $\sigma_x=0.5\times10^{-9}$ m }
        \label{5a_fig:sub10}
    \end{subfigure}
    \hfill
    \begin{subfigure}{0.32\textwidth}
        \centering
      \includegraphics[width=40mm]{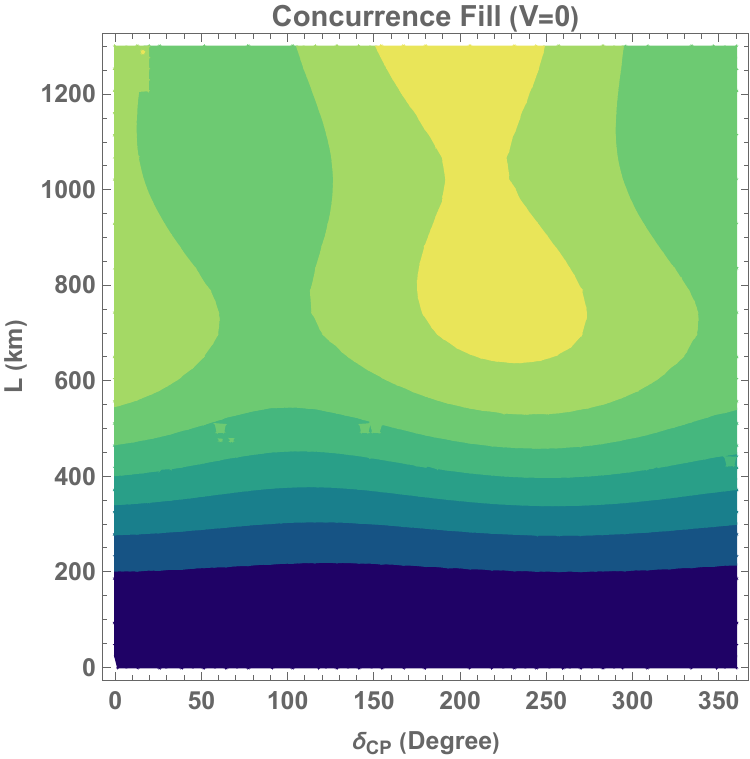}
\includegraphics[width=8.6mm]{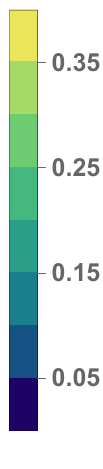}
        \caption{V=0 and $\sigma_x=0.5\times10^{-9}$ m}
        \label{5a_fig:sub11}
    \end{subfigure}
     \hfill
    \begin{subfigure}{0.32\textwidth}
        \centering
      \includegraphics[width=40mm]{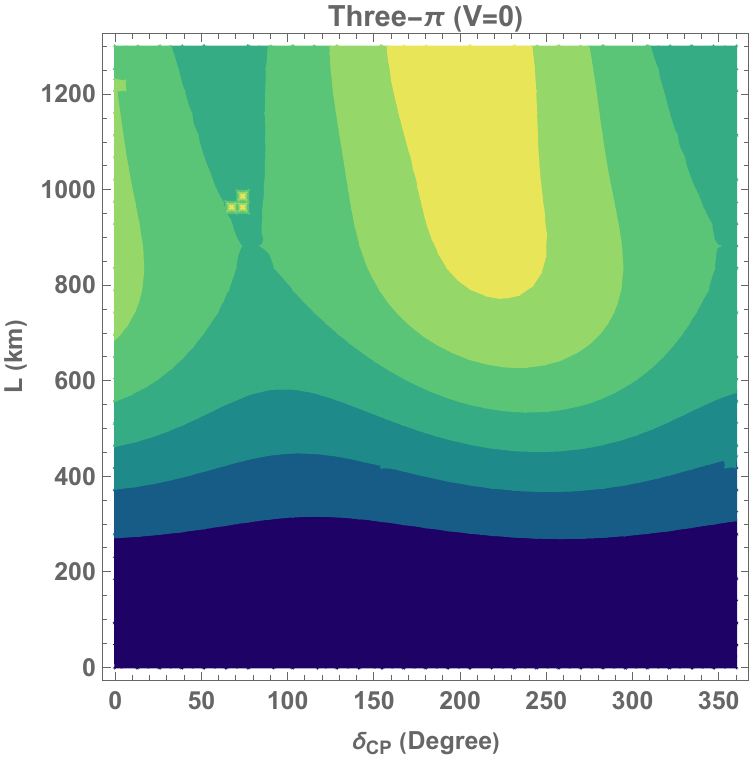}
 \includegraphics[width=8.6mm]{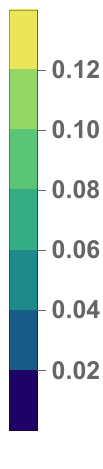}
      \caption{V=0 and $\sigma_x=0.5\times10^{-9}$ m}
        \label{5a_fig:sub12}
    \end{subfigure}
\caption{\justifying{Oscillogram plots of the $R_{l_{1}\text{-norm}}$, Concurrence Fill, and Three-$\pi $ entanglement in the $L-\delta_{\rm CP}$ plane, showing their variation with the baseline length L (km) and the CP-violating phase $\delta_{\rm CP}$ (degree), over the full range $0^\circ \leq \delta_{\rm CP} \leq 360^\circ$ with energy E=2.5 GeV. From left to right, the columns represent the $R_{l_{1}\text{-norm}}$ coherence, Concurrence Fill, and Three-$\pi$ entanglement, respectively. From top to bottom, the rows correspond to: (i) PW with $V=1.01\times10^{-13}$ eV, shown in Figs. 3(a)-3(c); (ii) $\sigma_x=0.5\times10^{-9}$ m and $V=1.01\times10^{-13}$ eV, shown in Figs. 3(d)-3(f); (iii) $\sigma_x=7\times10^{-9}$ m and $V=1.01\times10^{-13}$ eV, shown in Figs. 3(g)-3(i); and (iv) $V=0$ and $\sigma_x=0.5\times10^{-9}$ m, shown in Figs. 3(j)-3(l).}}
\label{fig3}
\end{figure*}

The Fig. \ref{fig2}(\subref{2a_fig:sub2}) in the right panel of Fig.~\ref{fig2} shows the influence of the matter potential by comparing neutrino propagation in matter and vacuum for a fixed wave-packet width of $\sigma_x = 0.5 \times 10^{-9}~\rm m$. For the $R_{l_{1}\text{-norm}}$, the pink solid and orange dotted curves represent the matter case, with $V = 1.01 \times 10^{-13}~\rm eV$, and the vacuum case, $V=0$, respectively. The cyan dash-dotted and gray dashed curves show the corresponding Concurrence Fill results in matter and vacuum, while the green dashed and blue dotted curves denote the Three-$\pi$ entanglement measure for the two propagation scenarios. For the $R_{l_{1}\text{-norm}}$, the vacuum case exhibits slightly larger values than the matter case up to a baseline of approx $700~\rm km$. Beyond this distance, the coherence in matter becomes dominant and remains higher than the corresponding vacuum result over the subsequent propagation range. At the end of the baseline, however, the two curves gradually converge and become nearly indistinguishable. For the Concurrence Fill measure, the correlation in matter remains consistently larger than the corresponding vacuum case throughout the entire baseline range. A similar trend is observed for the Three-$\pi$ entanglement measure, where the matter case also exhibits larger values than the vacuum case over the full propagation range. However, the values of the Three-$\pi$ measure remain comparatively smaller than those of the Concurrence Fill.

To further understand the behavior of \(R_{l_1\text{-norm}}\) as in Fig. \ref{fig2}(\subref{2a_fig:sub1}), we examine its relationship with the corresponding neutrino transition probabilities, particularly how variations in the survival and oscillation probabilities influence the resulting \(R_{l_1\text{-norm}}\) in different wave-packet scenarios for a fixed matter potential $V = 1.01 \times 10^{-13}~\rm eV$. For the DUNE experimental configuration with $E=2.5\,\text{GeV}$ and at L=1300 km, the values of the matter-modified survival and oscillation probabilities in the PW scenario are $P_{\mu\mu}^{m}=0.016$ and $P_{\mu\tau}^{m}=0.930$, respectively. Substituting these values into the definition of $R_{l_1\text{-norm}}$ as in Eq. \eqref{eq:l1norm_probability}, we obtain $R_{l_1\text{-norm}}\approx0.242$. The relatively small coherence value reflects the strong dominance of the $\nu_\mu\rightarrow\nu_\tau$ oscillation probability over the $\nu_\mu\rightarrow\nu_\mu$ survival probability. For $\sigma_x=7\times10^{-9}\,\mathrm{m}$, the corresponding probabilities are $P_{\mu\mu}^{m}=0.041$ and $P_{\mu\tau}^{m}=0.901$, giving $R_{l_1\text{-norm}}\approx0.387$. Compared with the PW case, the increase in the survival probability, accompanied by a slight reduction in the oscillation probability, results in enhanced coherence. A more pronounced change is observed for $\sigma_x=0.5\times10^{-9}\,\mathrm{m}$, where $P_{\mu\mu}^{m}=0.312$ and $P_{\mu\tau}^{m}=0.646$, yielding $R_{l_1\text{-norm}}\approx0.898$. In this case, the two probabilities become more comparable, leading to a higher coherence. Thus, the variation of $R_{l_1\text{-norm}}$ across the three scenarios can be understood through the changes in the relative contributions of the survival and oscillation probabilities. In particular, a more balanced distribution between these probabilities results in a larger value of $R_{l_1\text{-norm}}$, whereas a strong dominance of one probability over the other suppresses the coherence. To further investigate the influence of matter effects on $R_{l_1\text{-norm}}$ for different matter potentials while keeping the wave-packet width fixed at $\sigma_x=0.5\times10^{-9}\,\mathrm{m}$ as in Fig. \ref{fig2}(\subref{2a_fig:sub2}). When the matter potential is $V=1.01\times10^{-13}\,\mathrm{eV}$, the corresponding survival and oscillation probabilities are $P_{\mu\mu}^{m}=0.312$ and $P_{\mu\tau}^{m}=0.646$, respectively, yielding $R_{l_1\text{-norm}}=0.898$. In comparison, for the $V=0$, the probabilities are $P_{\mu\mu}^{m}=0.302$ and $P_{\mu\tau}^{m}=0.662$, resulting in $R_{l_1\text{-norm}}=0.893$. The slight increase in the survival probability and decrease in the oscillation probability in the presence of matter lead to a marginal enhancement of the coherence. However, the small difference between the two coherence values indicates that the matter potential has a relatively weak influence on $R_{l_1\text{-norm}}$ at the considered energy and baseline. Similarly, the behavior of the Concurrence Fill and Three-\(\pi\) entanglement measures can be examined by relating their variations to the corresponding neutrino transition probabilities.

Overall, Fig.~\ref{fig2} demonstrates that the combined interplay of wave-packet decoherence and matter effects plays a significant role in governing the evolution of residual quantum correlations in neutrino oscillations. Wave-packet decoherence generally leads to the suppression and smoothing of the correlation measures over long propagation distances, whereas matter effects modify their magnitude and evolution patterns. These results reveal that, within the DUNE experimental setup, genuine tripartite entanglement persists in neutrino oscillations not only in the PW scenario but also in the wave-packet scenario.

\begin{figure*}[htb]
    \centering

\begin{subfigure}{0.32\textwidth}
        \centering
       \includegraphics[height=45mm, width=53mm]{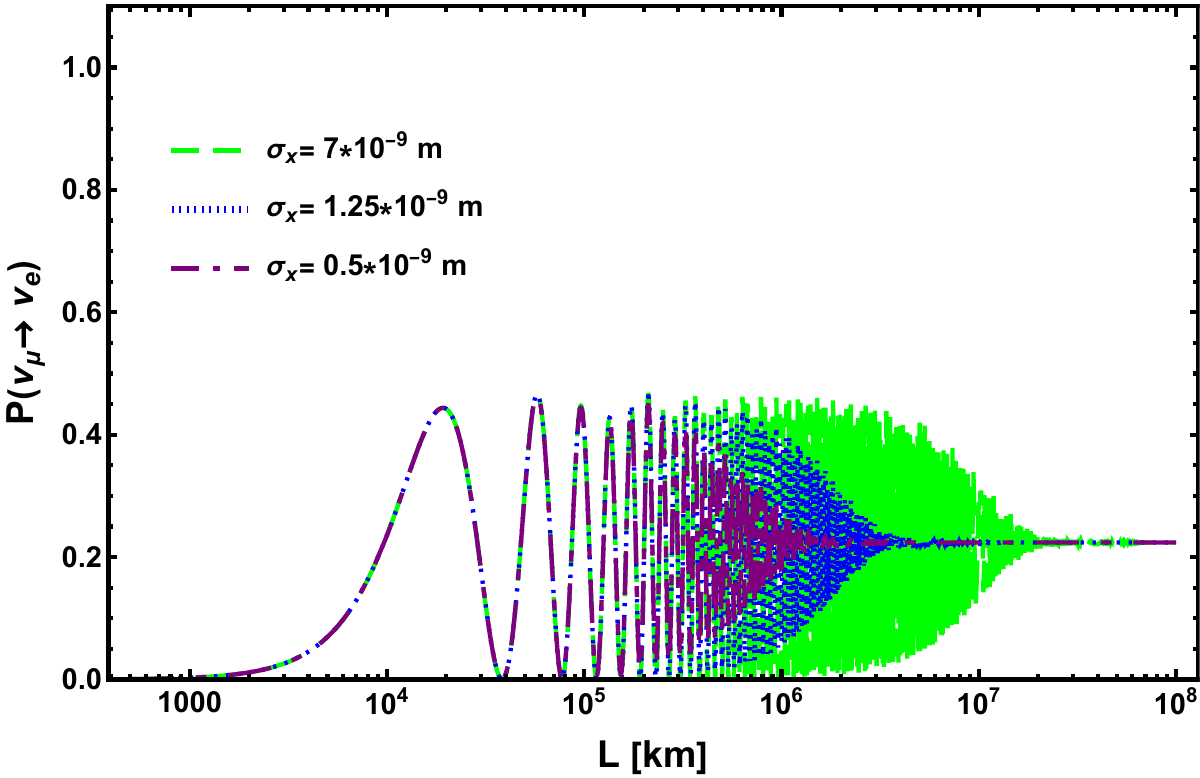}
        \caption{$V = 2.824 \times 10^{-14}~\rm eV$}
        \label{4a_fig:sub1}
    \end{subfigure}
 .   \hfill
    \begin{subfigure}{0.32\textwidth}
        \centering
       \includegraphics[height=45mm, width=53mm]{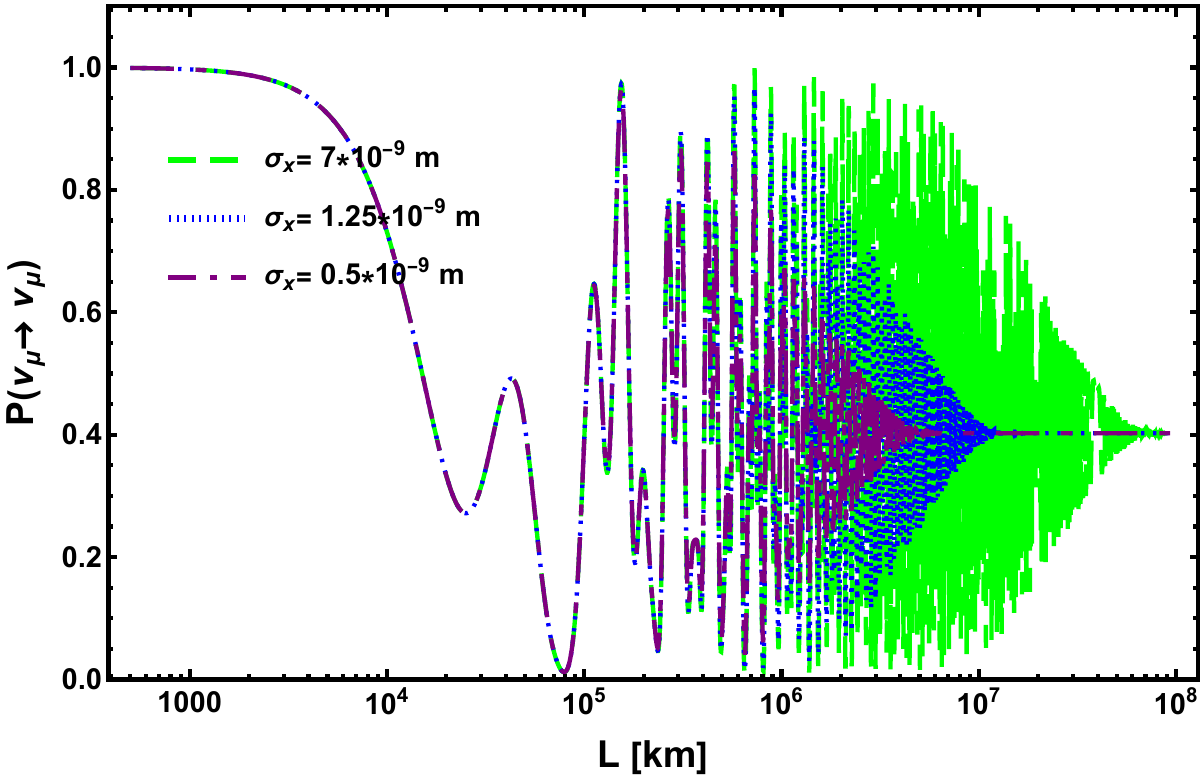}
        \caption{$V = 2.824 \times 10^{-14}~\rm eV$}
        \label{4a_fig:sub2}
    \end{subfigure}
     \hfill
    \begin{subfigure}{0.32\textwidth}
        \centering
      \includegraphics[height=45mm, width=53mm]{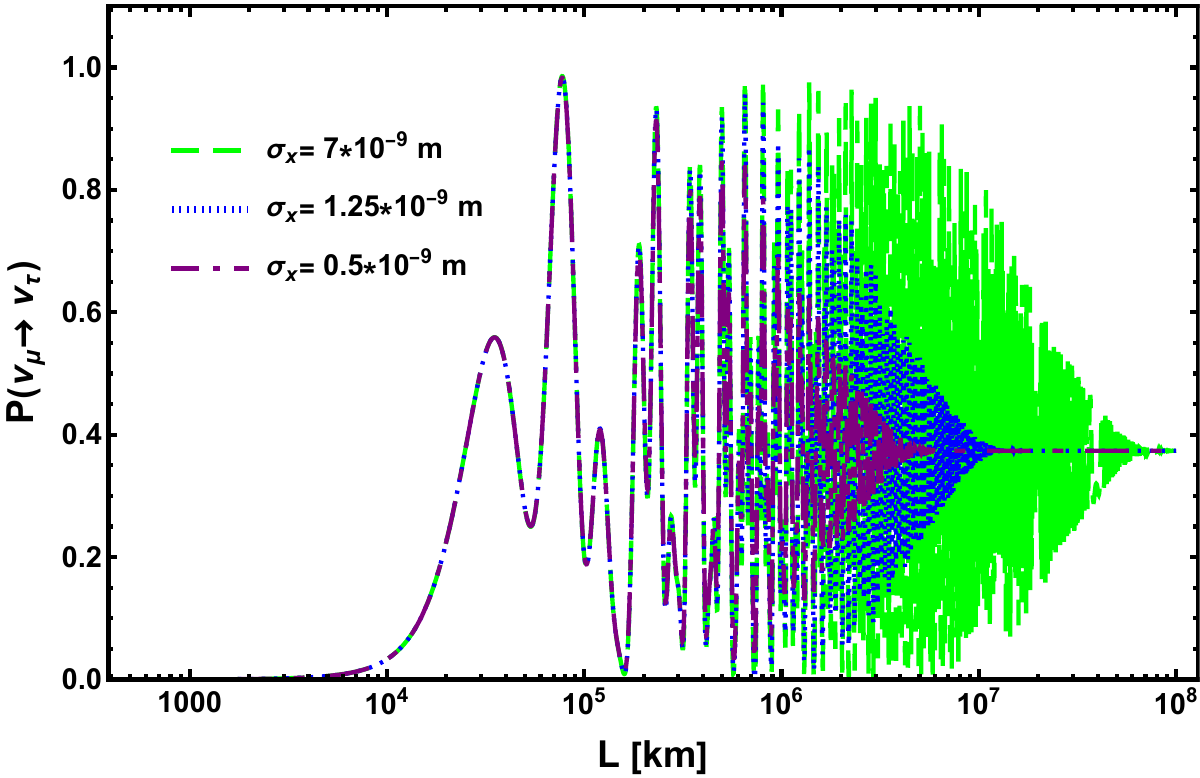}
        \caption{$V = 2.824 \times 10^{-14}~\rm eV$}
        \label{4a_fig:sub3}
    \end{subfigure}
    
    \medskip

    \begin{subfigure}{0.32\textwidth}
        \centering
      \includegraphics[height=45mm, width=53mm]{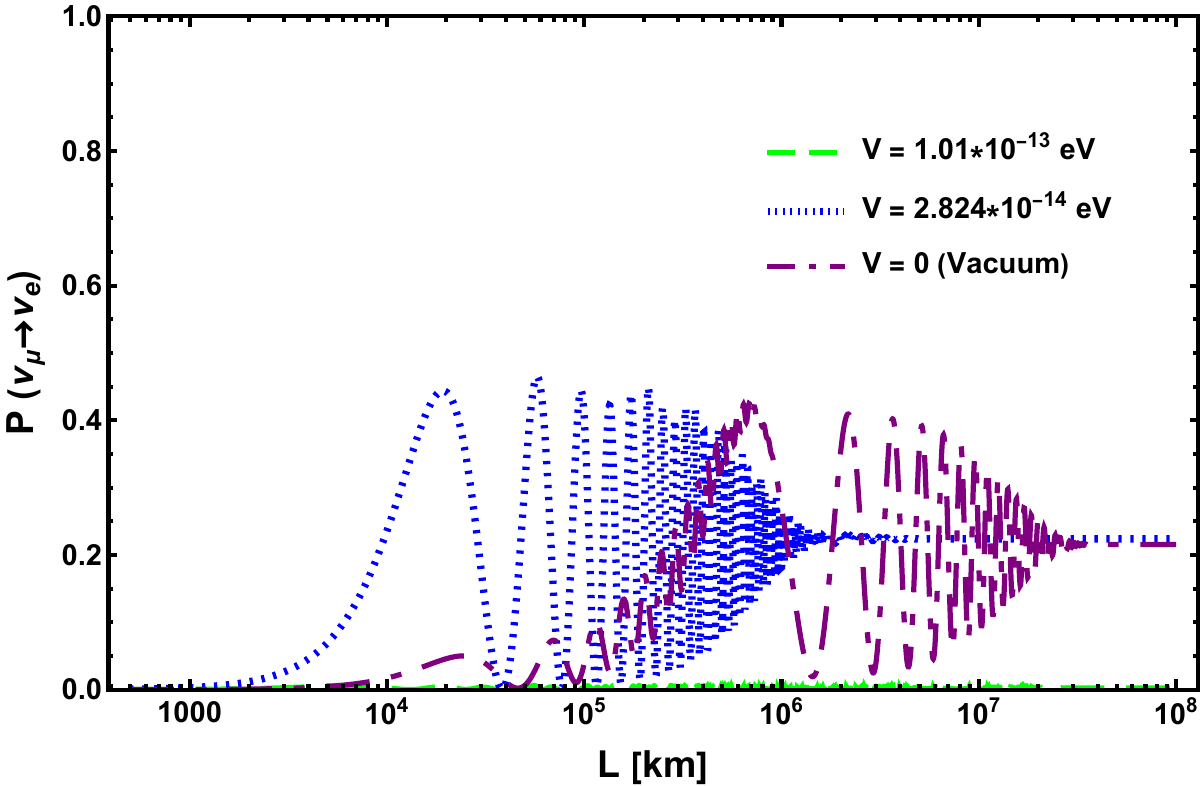}
        \caption{$\sigma_x=0.5\times10^{-9}$m }
        \label{4a_fig:sub4}
    \end{subfigure}
    \hfill
    \begin{subfigure}{0.32\textwidth}
        \centering
 \includegraphics[height=45mm, width=53mm]{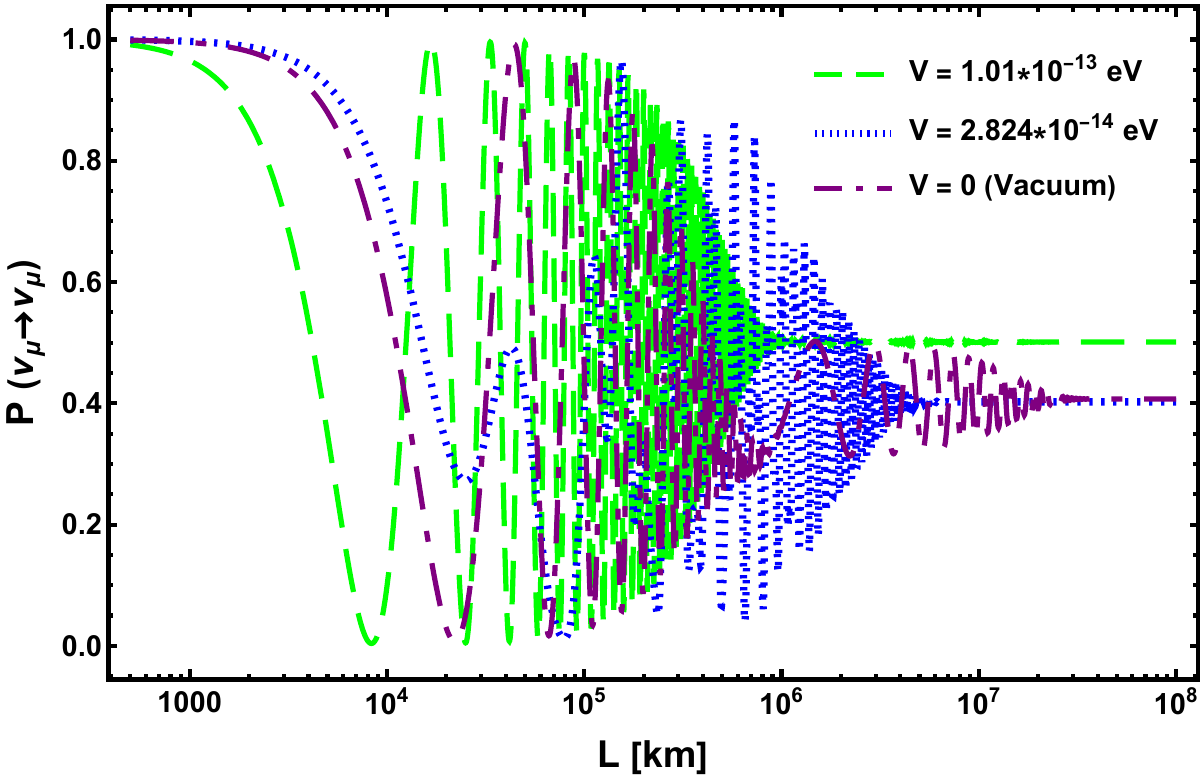}
        \caption{$\sigma_x=0.5\times10^{-9}$ m}
        \label{4a_fig:sub5}
    \end{subfigure}
     \hfill
    \begin{subfigure}{0.32\textwidth}
        \centering
\includegraphics[height=45mm, width=53mm]{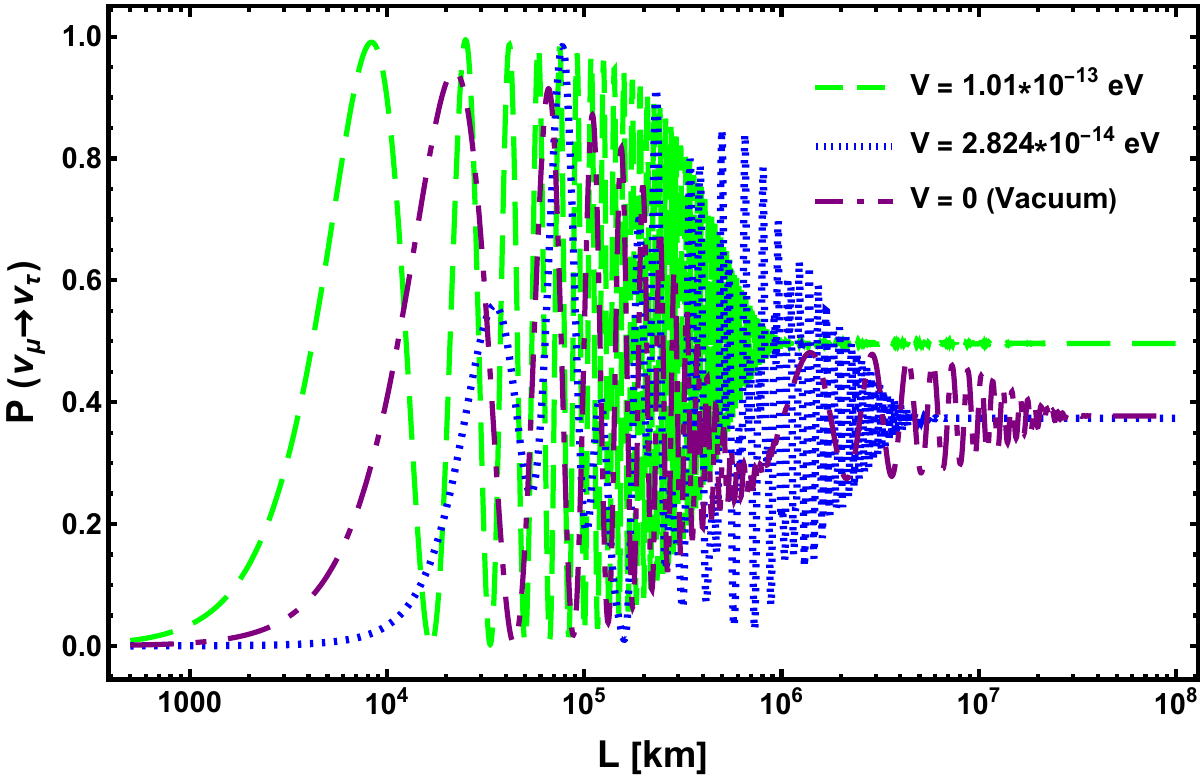}
        \caption{$\sigma_x=0.5\times10^{-9}$ m }
        \label{4a_fig:sub6}
    \end{subfigure}

\caption{\justifying{The oscillation ($P(\nu_\mu \rightarrow \nu_e)$ \& $P(\nu_\mu \rightarrow \nu_\tau)$) and survival ($P(\nu_\mu \rightarrow \nu_\mu)$) probabilities of an initial state $\ket{\nu_\mu}$ are shown as functions of the long-range propagation beyond the DUNE baseline, at $E = 45\,\text{GeV}$ \cite{Ravari:2022yfd}. The upper panels present the results for a fixed matter potential of $V = 2.824 \times 10^{-14}\,\text{eV}$ and different values of the wave-packet width $\sigma_x$, shown in Figs. 4(a)-4(c), while the lower panels show the results for a fixed wave-packet width of $\sigma_x = 0.5 \times 10^{-9}\,\text{m}$ and different values of the matter potential, shown in Figs. 4(d)-4(f).}}
\label{fig4}
\end{figure*}

Fig. \ref{fig3} illustrates the variation of the $R_{l_{1}\text{-norm}}$, Concurrence Fill, and Three-$\pi $ entanglement in the $L-\delta_{\rm CP}$ plane for the DUNE baseline. Figs. \ref{fig3}(\subref{5a_fig:sub1}), \ref{fig3}(\subref{5a_fig:sub2}) and \ref{fig3}(\subref{5a_fig:sub3}) correspond to the PW scenario in the presence of matter, with $V=1.01\times10^{-13}\,\mathrm{eV}$. As shown in Fig. \ref{fig3}(\subref{5a_fig:sub1}), the $R_{l_{1}\text{-norm}}$ exhibits a predominantly baseline-dependent behavior, with only a weak dependence on the CP-violating phase $\delta_{\rm CP}$, and the maximum value is observed in the baseline range of approximately 450-750 km. In contrast, both Concurrence Fill, shown in Fig. \ref{fig3}(\subref{5a_fig:sub2}) and Three-$\pi$ entanglement as in Fig. \ref{fig3}(\subref{5a_fig:sub3}) exhibit a more pronounced dependence on the $\delta_{\rm CP}$, particularly at intermediate baseline. The maximum values of Concurrence Fill and Three-$\pi$ entanglement occur in the CP-phase intervals of ($25^\circ\leq \delta_{\rm CP} \leq100^\circ$) and ($0^\circ\leq \delta_{\rm CP} \leq150^\circ$), respectively. Figs. \ref{fig3}(\subref{5a_fig:sub4}), \ref{fig3}(\subref{5a_fig:sub5}), and \ref{fig3}(\subref{5a_fig:sub6}) correspond to $\sigma_x=0.5\times10^{-9}$ m and $V=1.01\times10^{-13}$ eV. In Fig. \ref{fig3}(\subref{5a_fig:sub4}), at the end of the baseline, it is observed that the maximum value of $R_{l_1\text{-norm}}$ occurs within the $\delta_{\rm CP}$ interval of approximately $180^\circ\leq\delta_{\rm CP}\leq280^\circ$. Similarly, for the Concurrence Fill, as shown in Fig. \ref{fig3}(\subref{5a_fig:sub5}), the maximum is observed within the $\delta_{\rm CP}$ interval of approximately $50^\circ$ to $150^\circ$ and $330^\circ$ to $360^\circ$ for a baseline length ranging from 800 to 1300 km. For the Three-$\pi$ as shown in Fig. \ref{fig3}(\subref{5a_fig:sub6}), the behaviour is similar to that observed for the Concurrence Fill, but over a narrower range of $\delta_{\rm CP}$. The results for the three residual measures are presented in Figs. \ref{fig3}(\subref{5a_fig:sub7}), \ref{fig3}(\subref{5a_fig:sub8}), and \ref{fig3}(\subref{5a_fig:sub9}), corresponding to $\sigma_x=7\times10^{-9}$ m and $V=1.01\times10^{-13}$ eV. In Fig. \ref{fig3}(\subref{5a_fig:sub7}), the maximum of $R_{l_1\text{-norm}}$ is observed over the whole range of $\delta_{\rm CP}$ for baseline lengths of approximately 700-1100 km. In Fig. \ref{fig3}(\subref{5a_fig:sub8}), the maximum of the Concurrence Fill is observed in the $\delta_{\rm CP}$ intervals of approximately $40^\circ$-$90^\circ$ and $270^\circ$-$330^\circ$, corresponding to baseline lengths of around 800-1000 km. The behaviour of the Three-$\pi$ entanglement in Fig. \ref{fig3}(\subref{5a_fig:sub7}) is similar to that of the Concurrence Fill, with the maxima appearing in nearly the same regions of the $\delta_{\rm CP}$-baseline plane. For $ V=0$, as shown in Figs. \ref{fig3}(\subref{5a_fig:sub10}), \ref{fig3}(\subref{5a_fig:sub11}), and \ref{fig3}(\subref{5a_fig:sub12}), the $R_{l_1\text{-norm}}$ as shown in Fig. \ref{fig3}(\subref{5a_fig:sub10}) attains its maximum values over a broad range of $\delta_{\rm CP}$, particularly for baseline lengths above approximately 500 km to 800 km. In Fig. \ref{fig3}(\subref{5a_fig:sub11}), the Concurrence Fill exhibits its maximum values within a relatively narrow $\delta_{\rm CP}$ interval of approximately $150^\circ$-$250^\circ$, mainly for baseline lengths of 700-900 km. A similar pattern is observed for the Three-$\pi$ entanglement in Fig. \ref{fig3}(\subref{5a_fig:sub12}), although its maximum region is shifted towards longer baseline lengths, approximately 900-1100 km, within a comparable range of $\delta_{\rm CP}$.

\subsection{Analysis for baselines beyond DUNE} \label{sec:7.3.2}
Fig.~\ref{fig4} shows the neutrino oscillation and survival probabilities with the propagation distance $L$, considering the range beyond the DUNE baseline and $E = 45~\rm GeV$ \cite{Ravari:2022yfd}. The analysis is performed within the wave-packet formalism, taking matter effects into account. In the upper panels, the wave-packet width $\sigma_x$ is varied while the matter potential is held constant at $V = 2.824 \times 10^{-14}~\rm eV$. The magenta dash-dotted, blue dotted, and green dashed curves correspond to $\sigma_x=0.5\times10^{-9}\,\mathrm{m}$, $\sigma_x=1.25\times10^{-9}\,\mathrm{m}$, and $\sigma_x=7\times10^{-9}\,\mathrm{m}$, respectively. The lower panels examine the dependence on the matter potential for a fixed wave-packet width of $\sigma_x = 0.5 \times 10^{-9}~\rm m$. The magenta dash-dotted, blue dotted, and green dashed curves correspond to matter potential values of V = 0, $V=1.01\times10^{-13}\,\mathrm{eV}$, and $V=2.824\times10^{-14}\,\mathrm{eV}$, respectively. From left to right, the three columns display the oscillation probability $P(\nu_\mu \rightarrow \nu_e)$, the survival probability $P(\nu_\mu \rightarrow \nu_\mu)$, and $P(\nu_\mu \rightarrow \nu_\tau)$, respectively.

\begin{figure*}[htb]
    \centering
  
\begin{subfigure}{0.32\textwidth}
        \centering
       \includegraphics[height=45mm, width=53mm]{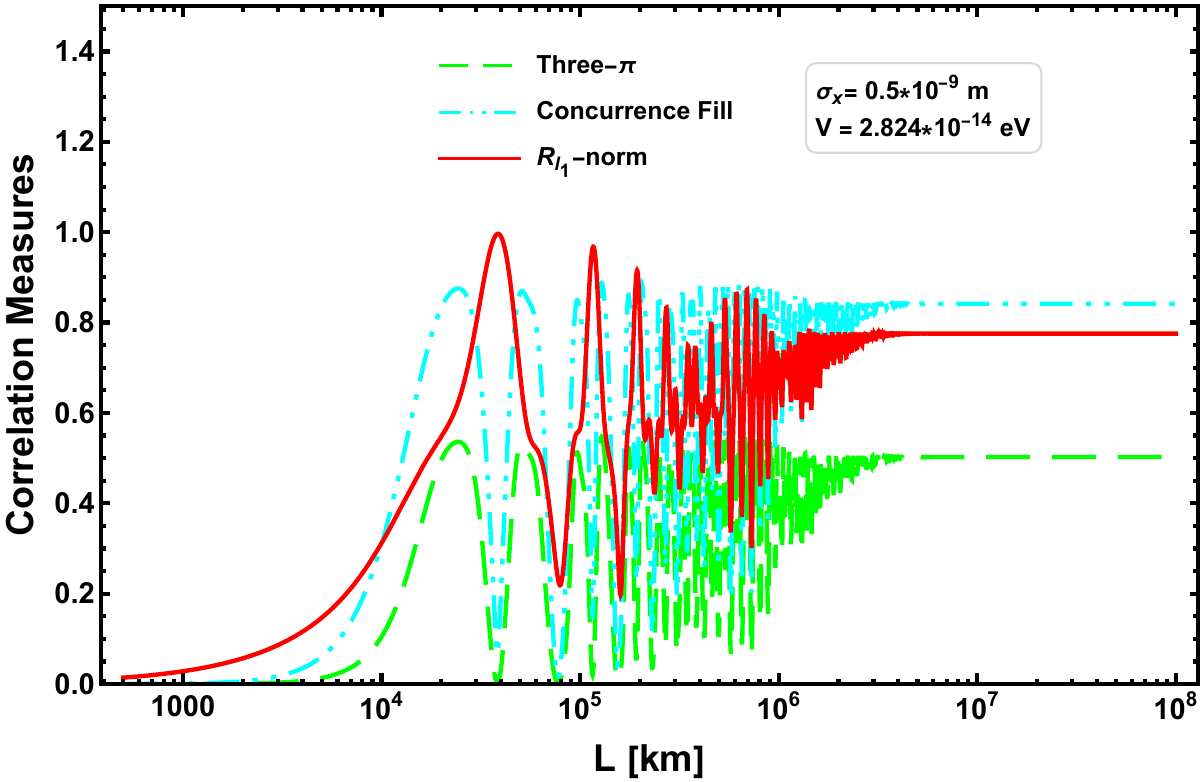}
        \caption{ $\sigma_x=0.5\times10^{-9}$ m and $V = 2.824 \times 10^{-14}~\rm eV$}
        \label{55a_fig:sub1}
    \end{subfigure}
 .   \hfill
    \begin{subfigure}{0.32\textwidth}
        \centering
     \includegraphics[height=45mm, width=53mm]{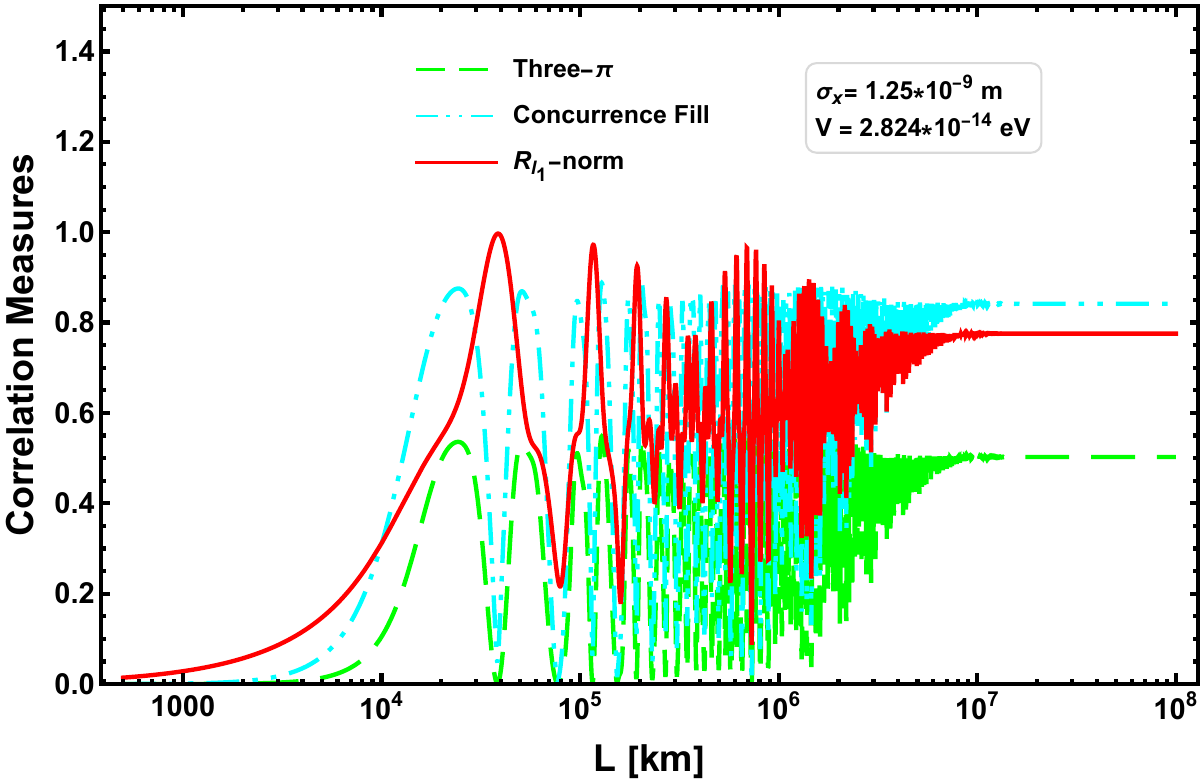}
        \caption{ $\sigma_x=1.25\times10^{-9}$ m and $V = 2.824 \times 10^{-14}~\rm eV$}
        \label{55a_fig:sub2}
    \end{subfigure}
     \hfill
    \begin{subfigure}{0.32\textwidth}
        \centering
      \includegraphics[height=45mm, width=53mm]{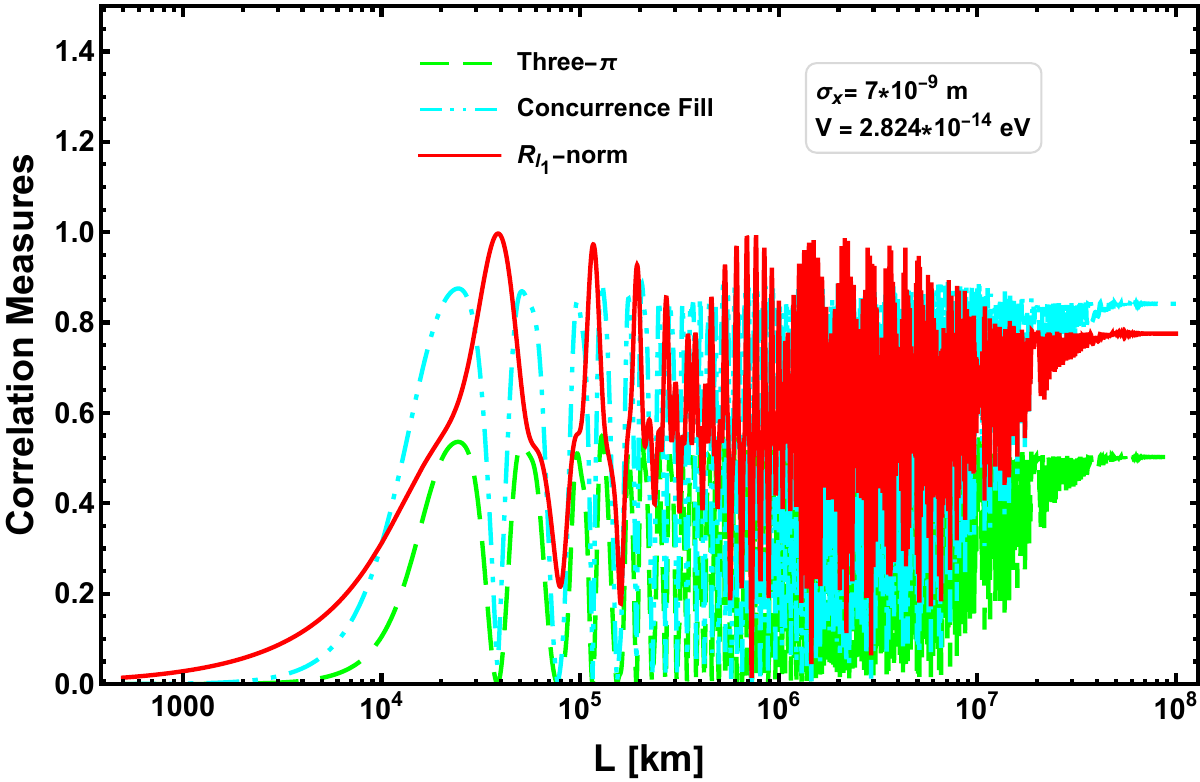}
        \caption{ $\sigma_x=7\times10^{-9}$ m and $V = 2.824 \times 10^{-14}~\rm eV$}
        \label{55a_fig:sub3}
    \end{subfigure}
    
    \medskip

    \begin{subfigure}{0.32\textwidth}
        \centering
      \includegraphics[height=45mm, width=53mm]{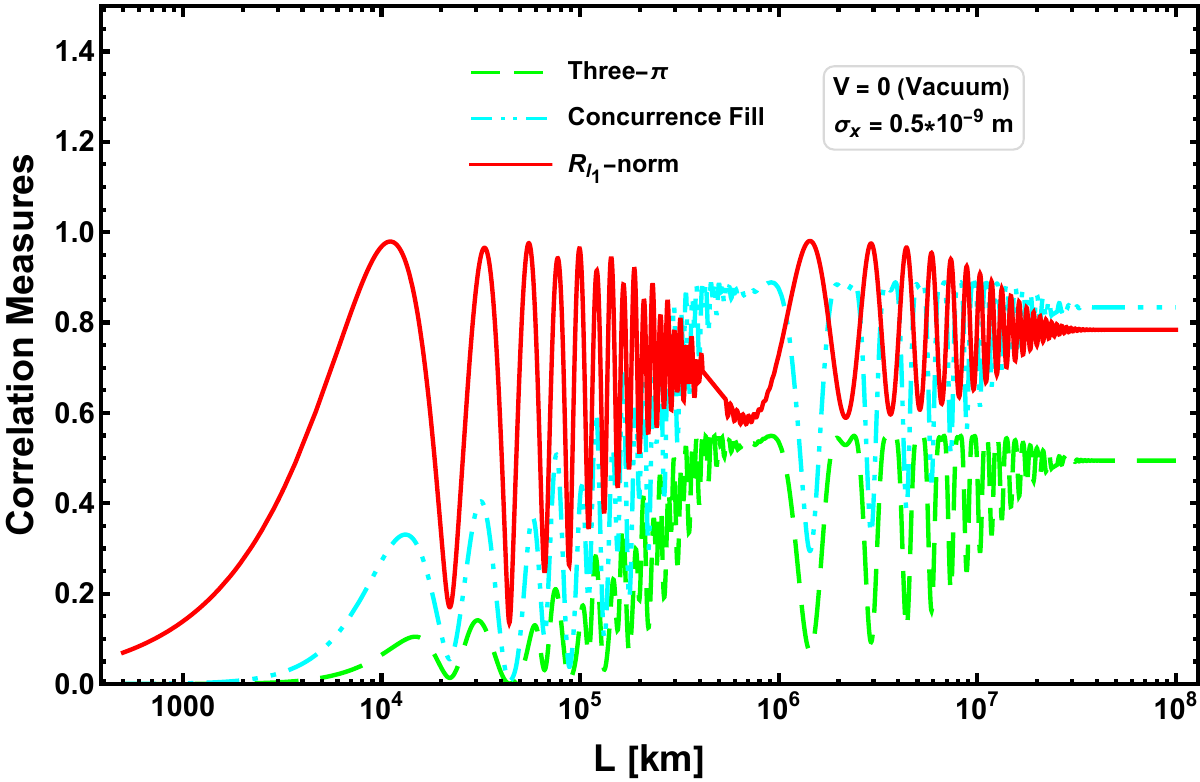}
        \caption{V = 0 (vacuum) and $\sigma_x=0.5\times10^{-9}$m }
        \label{55a_fig:sub4}
    \end{subfigure}
    \hfill
    \begin{subfigure}{0.32\textwidth}
        \centering
\includegraphics[height=45mm, width=53mm]{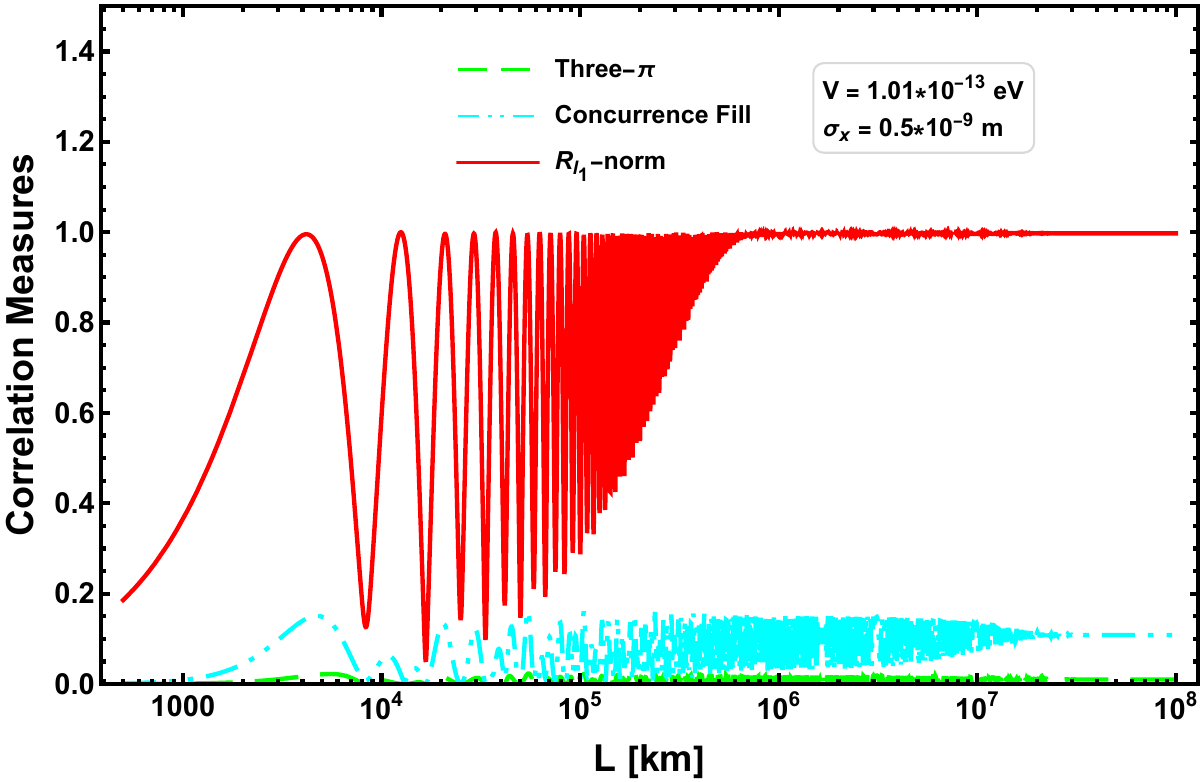}
        \caption{$V = 1.01 \times 10^{-13}~\rm eV$ and $\sigma_x=0.5\times10^{-9}$ m}
        \label{55a_fig:sub5}
    \end{subfigure}
     \hfill
    \begin{subfigure}{0.32\textwidth}
        \centering
\includegraphics[height=45mm, width=53mm]{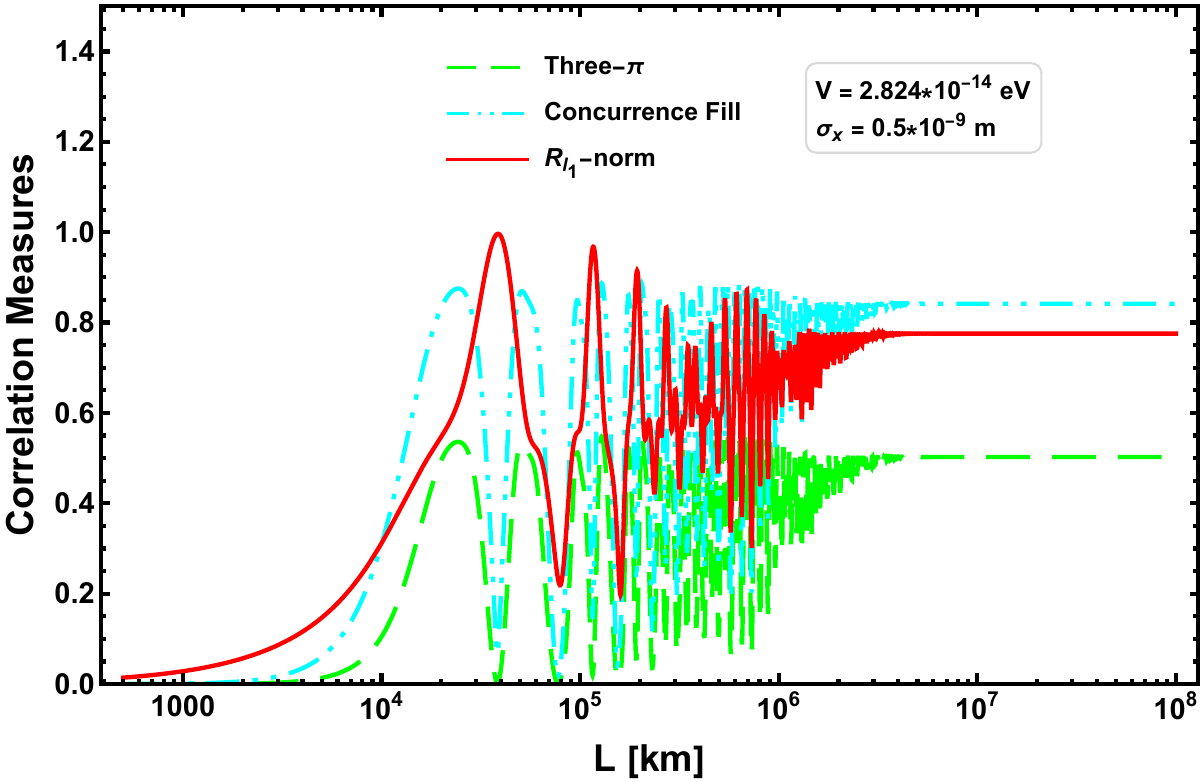}

        \caption{$V = 2.824 \times 10^{-14}~\rm eV$ and $\sigma_x=0.5\times10^{-9}$ m }
        \label{55a_fig:sub6}
    \end{subfigure}

\caption{\justifying{For an initial state $\ket{\nu_\mu}$, various residual quantum correlation measures, including the $R_{l_{1}\text{-norm}}$, Concurrence Fill, and Three-$\pi$, are shown as functions of the long-range propagation length, at $E = 45\,\text{GeV}$ \cite{Ravari:2022yfd}. The upper panels illustrate these measures for different values of the wave-packet width $\sigma_x$, with the matter potential fixed at $V = 2.824 \times 10^{-14}\,\text{eV}$, as shown in Figs. 5(a)-5(c). The lower panels show the results for a fixed wave-packet width of $\sigma_x = 0.5 \times 10^{-9}\,\text{m}$ and different values of the matter potential, shown in Figs. 5(d)-5(f).}}
  \label{fig5}
\end{figure*}

The upper panels shown in Figs. \ref{fig4}(\subref{4a_fig:sub1})-\ref{fig4}(\subref{4a_fig:sub3}) highlight the effect of wave-packet decoherence on neutrino oscillations over extremely long propagation distances. In the upper left panel as in Fig. \ref{fig4}(\subref{4a_fig:sub1}), the oscillation probability $P(\nu_\mu \rightarrow \nu_e)$ initially increases with distance and develops pronounced oscillatory structures. For the larger wave-packet width, $\sigma_x = 7 \times 10^{-9}~\rm m$, these oscillations remain visible over a comparatively longer distance, with noticeable damping becoming apparent only beyond $L \sim 10^{7}~\rm km$. On the other hand, for the smaller wave-packet widths, $\sigma_x = 1.25 \times 10^{-9}~\rm m$ and $\sigma_x = 0.5 \times 10^{-9}~\rm m$, damping begins at an earlier distance of approximately $L \approx 10^{6}~\rm km$, leading to a more rapid reduction in the oscillation amplitudes. At sufficiently large propagation distances, all three cases gradually converge toward nearly constant averaged values, reflecting the gradual loss of coherence between the propagating mass-eigenstate wave packets. The upper middle panel shown in Fig. \ref{fig4}(\subref{4a_fig:sub2}) depicts the survival probability $P(\nu_\mu \rightarrow \nu_\mu)$. At short distances, the survival probability remains close to unity, after which it develops rapid oscillatory variations as the baseline increases. With further increase in the propagation distance, these oscillations become increasingly suppressed, and the survival probability eventually approaches an averaged value. The extent of this damping is strongly dependent on the wave-packet width $\sigma_x$, highlighting the significant role of wave-packet separation in driving the decoherence process. The upper right panel shown in Fig. \ref{fig4}(\subref{4a_fig:sub3}), corresponding to the oscillation probability $P(\nu_\mu \rightarrow \nu_\tau)$, shows a similar oscillatory behavior followed by damping at large distances. The probability first increases from zero, reaches significant oscillation amplitudes, and later evolves toward a stable averaged value due to decoherence effects. Compared to the other channels, the $\nu_\tau$ oscillation probability exhibits a broader oscillation structure over the considered distance range.

In the lower left panel as in Fig. \ref{fig4}(\subref{4a_fig:sub4}), the oscillation probability $P(\nu_\mu \rightarrow \nu_e)$ exhibits a dependence on the matter potential. For $V = 1.01 \times 10^{-13}~\rm eV$, the oscillation probability is highly suppressed over the entire baseline range and rapidly approaches zero at large distances. When the matter potential is reduced to $V = 2.824 \times 10^{-14}~\rm eV$, pronounced oscillatory features emerge at intermediate baselines, followed by a gradual suppression of the oscillations due to decoherence at longer distances. In the vacuum case, however, the oscillations remain visible over a considerably broader baseline range and become significantly averaged only for $L \gtrsim 10^{7}~\rm km$. At these large distances, $P(\nu_\mu \rightarrow \nu_e)$ approaches an approximately constant, nonzero value as a result of decoherence. The lower middle panel, as in Fig. \ref{fig4}(\subref{4a_fig:sub5}) shows the survival probability $P(\nu_\mu \rightarrow \nu_\mu)$ and demonstrates a clear dependence on the strength of the matter potential. For $V = 1.01 \times 10^{-13}~\rm eV$, rapid oscillatory variations are observed at intermediate baselines, followed by a relatively faster decoherence at larger distances. With the smaller potential, $V = 2.824 \times 10^{-14}~\rm eV$, $P(\nu_\mu \rightarrow \nu_\mu)$ remains pronounced over a larger baseline interval, while the transition toward the decoherent regime occurs more gradually. By comparison, the vacuum scenario maintains its oscillatory behavior up to substantially larger distances, with $P(\nu_\mu \rightarrow \nu_\mu)$ becoming averaged only beyond $L \gtrsim 10^{7}~\rm km$. The lower right panel as in Fig. \ref{fig4}(\subref{4a_fig:sub6}) presents the oscillation probability $P(\nu_\mu \rightarrow \nu_\tau)$. In the case of $V = 1.01 \times 10^{-13}~\rm eV$, relatively large oscillation amplitudes appear at shorter baselines, followed by rapid damping as the propagation distance increases. For $V = 2.824 \times 10^{-14}~\rm eV$, the oscillation peaks are displaced toward larger baselines and occur with comparatively reduced amplitudes. In the absence of matter effects, the oscillatory pattern extends across a much wider baseline range and is gradually washed out at sufficiently large distances as decoherence becomes dominant.

Fig.~\ref{fig5} presents the dependence of the residual quantum correlation measures on the propagation distance $L$ (km) in the long-distance regime, considering the wave-packet framework with matter effects. The analysis includes three quantum correlation measures: $R_{l_{1}\text{-norm}}$, Concurrence Fill, and Three-$\pi$ entanglement represented by red solid, cyan dashed-dotted, and green dashed lines, respectively. In the upper panels shown in Figs. \ref{fig5}(\subref{55a_fig:sub1})-\ref{fig5}(\subref{55a_fig:sub3}), the impact of different wave-packet widths $\sigma_x$ is examined while keeping the matter potential fixed at $V = 2.824 \times 10^{-14}~\rm eV$. The lower panels, as shown in Figs. \ref{fig5}(\subref{55a_fig:sub4})-\ref{fig5}(\subref{55a_fig:sub5}), on the other hand, demonstrate the effect of varying the matter potential for a fixed wave-packet width of $\sigma_x = 0.5 \times 10^{-9}~\rm m$.

For the upper left panel as in Fig. \ref{fig5}(\subref{55a_fig:sub1}), corresponding to $\sigma_x = 0.5 \times 10^{-9}~\rm m$, all three measures $R_{l_{1}\text{-norm}}$, Concurrence Fill, and Three-$\pi$ entanglement show an initial increase with the propagation distance, followed by damped oscillations and, eventually, convergence toward nearly stable values at sufficiently large distances. Throughout the propagation, the $R_{l_{1}\text{-norm}}$ remains the dominant contribution among the three measures. For the upper middle panel shown in Fig. \ref{fig5}(\subref{55a_fig:sub2}), with $\sigma_x = 1.25 \times 10^{-9}~\rm m$, the oscillatory features extend over a comparatively larger distance range and exhibit enhanced amplitudes. In this case, both the Concurrence Fill and Three-$\pi$ entanglement retain appreciable values over longer propagation distances, suggesting that decoherence is less pronounced than for the smaller wave-packet width. The upper right panel, as in Fig. \ref{fig5}(\subref{55a_fig:sub3}), corresponding to $\sigma_x = 7 \times 10^{-9}~\rm m$, displays the most pronounced and extended oscillatory behavior among the three cases. The increased wave-packet width allows coherence and entanglement to survive over substantially longer propagation distances, thereby leading to a more gradual suppression of the oscillatory patterns induced by decoherence.

The dependence of \(R_{l_1\text{-norm}}\) on the survival probability \(P_{\mu\mu}^{m}\) and oscillation probability \(P_{\mu\tau}^{m}\) is examined over extended propagation distances for a fixed matter potential of \(V=2.824\times10^{-14}\,\mathrm{eV}\) and different wave-packet widths \(\sigma_x\). For \(\sigma_x=0.5\times10^{-9}\,\mathrm{m}\), at \(L=4\times10^4\) km, the survival and oscillation probabilities are \(P_{\mu\mu}^{m}=0.48\) and \(P_{\mu\tau}^{m}=0.51\), respectively, yielding \(R_{l_1\text{-norm}}\approx0.99\) as shown in Fig. \ref{fig5}(\subref{55a_fig:sub1}). At a larger baseline of \(L=3\times10^6\) km, these probabilities decrease to \(0.34\) and \(0.43\), respectively, resulting in \(R_{l_1\text{-norm}}=0.77\). Similarly, as in Fig. \ref{fig5}(\subref{55a_fig:sub2}) for \(\sigma_x=1.25\times10^{-9}\,\mathrm{m}\), at \(L=4\times10^4\) km, the corresponding survival and oscillation probabilities are \(P_{\mu\mu}^{m}=0.47\) and \(P_{\mu\tau}^{m}=0.51\), respectively, giving \(R_{l_1\text{-norm}}\approx0.99\). However, at \(L=7\times10^6\) km, the survival probability increases to \(0.48\), while the oscillation probability decreases to \(0.29\), leading to a reduced coherence. For Fig. \ref{fig5}(\subref{55a_fig:sub3}), the larger wave-packet width of \(\sigma_x=7\times10^{-9}\,\mathrm{m}\), at \(L=4\times10^4\) km, we obtain \(P_{\mu\mu}^{m}=0.47\) and \(P_{\mu\tau}^{m}=0.51\), corresponding to \(R_{l_1\text{-norm}}\approx0.99\). At \(L=4\times10^7\) km, these probabilities become \(P_{\mu\mu}^{m}=0.39\) and \(P_{\mu\tau}^{m}=0.38\), yielding \(R_{l_1\text{-norm}}\approx0.77\). The results demonstrate that the evolution of \(R_{l_1\text{-norm}}\) is closely related to the interplay between the survival and oscillation probabilities. When these two probabilities have comparable magnitudes, the  \(R_{l_1\text{-norm}}\) approaches its maximum value. However, as the baseline increases, changes in their relative contributions lead to a reduction in  \(R_{l_1\text{-norm}}\). This behavior highlights how the wave-packet width influences the evolution of neutrino survival and oscillation probabilities and, consequently, the residual quantum coherence over extended propagation distances.

The lower left panel as in Fig. \ref{fig5}(\subref{55a_fig:sub4}) represents the vacuum scenario ($V=0$) with a fixed wave-packet width of $\sigma_x = 0.5 \times 10^{-9}~\rm m$. Here, the $R_{l_{1}\text{-norm}}$ undergoes pronounced oscillations across a broad range of propagation distances and gradually settles toward a stable behavior. The Concurrence Fill increases progressively and approaches an approximately constant value at sufficiently large distances, while the Three-$\pi$ measure exhibits relatively smaller amplitudes but continues to oscillate over an extended propagation range. Without matter effects, the oscillatory structures therefore remain comparatively uniform over long propagation distances. For the lower middle panel shown in Fig. \ref{fig5}(\subref{55a_fig:sub5}), corresponding to $V = 1.01 \times 10^{-13}~\rm eV$, the correlation measures exhibit a substantially different behavior. The $R_{l_{1}\text{-norm}}$ rapidly rises toward unity and remains nearly constant across most of the propagation range. In contrast, both the Concurrence Fill and Three-$\pi$ entanglement are strongly suppressed, retaining values close to zero. This behavior demonstrates that strong matter effects can enhance the residual coherence while simultaneously reducing the entanglement-based correlations. The lower right panel as in Fig. \ref{fig5}(\subref{55a_fig:sub6}) corresponds to the smaller matter potential, $V = 2.824 \times 10^{-14}~\rm eV$. In this case, all three quantum correlation measures exhibit appreciable oscillations before gradually approaching stable values at large propagation distances. Relative to the vacuum scenario, the presence of matter modifies the amplitudes as well as the damping characteristics of the correlations. The $R_{l_{1}\text{-norm}}$ remains the dominant measure, whereas the Concurrence Fill and Three-$\pi$ entanglement retain moderate values throughout the considered propagation range. Thus, genuine tripartite entanglement in neutrino oscillations persists within the wave-packet approximation over baselines extending well beyond the DUNE baseline.  

To further examine the influence of matter effects on the evolution of $R_{l_1\text{-norm}}$, we consider different matter potentials while keeping the wave-packet width fixed at $\sigma_x=0.5\times10^{-9}\,\mathrm{m}$ over extended propagation distances. For the vacuum case, V = 0 as in Fig. \ref{fig5}(\subref{55a_fig:sub4}). At $L=10^4$ km, the survival and oscillation probabilities are $P_{\mu\mu}^{m}=0.57$ and $P_{\mu\tau}^{m}=0.41$, respectively, resulting in $R_{l_1\text{-norm}}\approx0.97$. As the baseline increases to $7\times10^5$ km, these probabilities decrease to 0.31 and 0.27, respectively, leading to a reduction in coherence to approximately 0.59. At a further increased baseline of $L=2.6\times10^7$ km, the survival and transition probabilities become 0.40 and 0.37, respectively, and the $R_{l_1\text{-norm}}$ increases to $\approx$ 0.77. This indicates that the coherence does not decrease monotonically with propagation distance but follows the changes in the survival and transition probabilities. For $V=2.824\times10^{-14}\,\mathrm{eV}$, as shown in Fig. \ref{fig5}(\subref{55a_fig:sub6}) at $L=4\times10^4$ km, the corresponding probabilities are $P_{\mu\mu}^{m}=0.48$ and $P_{\mu\tau}^{m}=0.51$, yielding $R_{l_1\text{-norm}}\approx0.99$. At $L=3\times10^6$ km, these probabilities become 0.34 and 0.43, respectively, resulting in $R_{l_1\text{-norm}}=0.77$. Similarly, for $V=1.01\times10^{-13}\,\mathrm{eV}$, as in Fig. \ref{fig5}(\subref{55a_fig:sub5}) at $L=7\times10^5$ km, the survival and transition probabilities are approximately 0.58 and 0.44, respectively, corresponding to $R_{l_1\text{-norm}}$ close to unity. The comparison demonstrates that matter effects modify the evolution of the survival and transition probabilities and, consequently, influence the residual quantum coherence over extended propagation distances.

\begin{figure*}[htb]
    \centering
     \begin{subfigure}{0.32\textwidth}
        \centering
       \includegraphics[width=41mm]{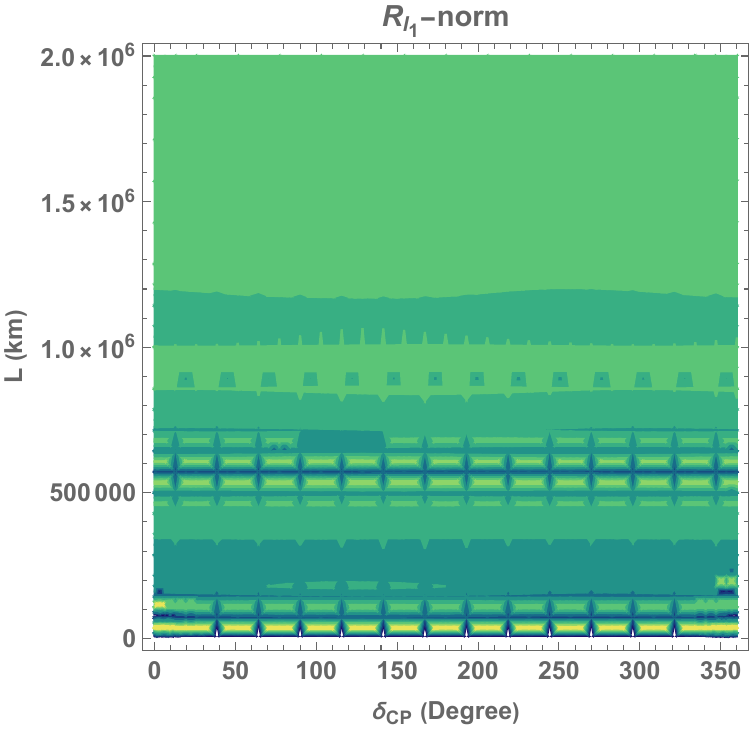}
\includegraphics[width=7.2mm]{contour_plots/con_mark_rl1.pdf}
        \caption{$\sigma_x=0.5\times10^{-9}$ m and $V = 2.824 \times 10^{-14}~\rm eV$}
        \label{6a_fig:sub1}
    \end{subfigure}
    \hfill
    \begin{subfigure}{0.32\textwidth}
        \centering
      \includegraphics[width=41mm]{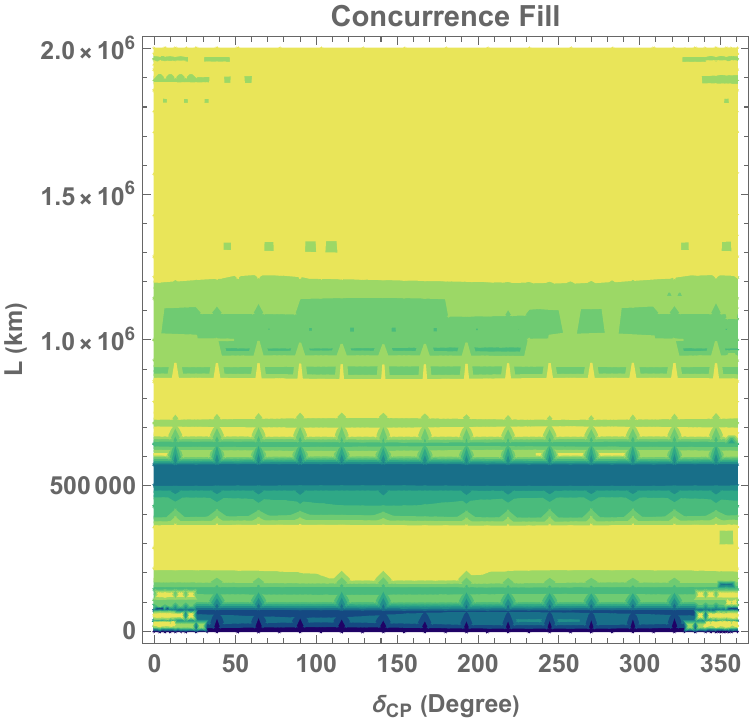}
 \includegraphics[width=7.2mm]{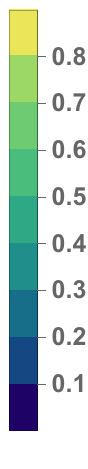}
        \caption{$\sigma_x=0.5\times10^{-9}$ m and $V = 2.824 \times 10^{-14}~\rm eV$}
        \label{6a_fig:sub2}
    \end{subfigure}
     \hfill
    \begin{subfigure}{0.32\textwidth}
        \centering
      \includegraphics[width=41mm]{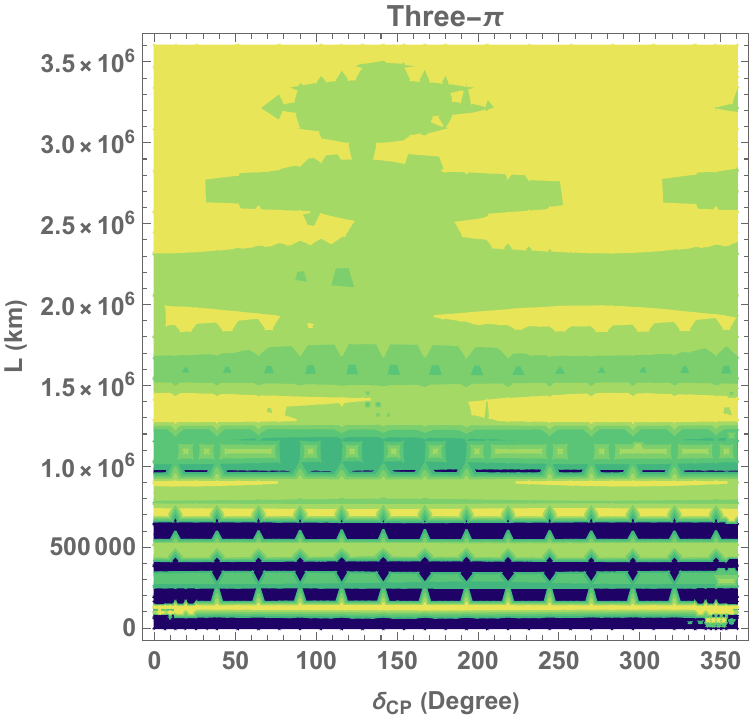}
\includegraphics[width=8.2mm]{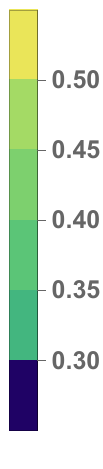}
        \caption{$\sigma_x=0.5\times10^{-9}$ m and $V = 2.824 \times 10^{-14}~\rm eV$}
        \label{6a_fig:sub3}
    \end{subfigure}
    
    \medskip

    \begin{subfigure}{0.32\textwidth}
        \centering
     \includegraphics[width=41mm]{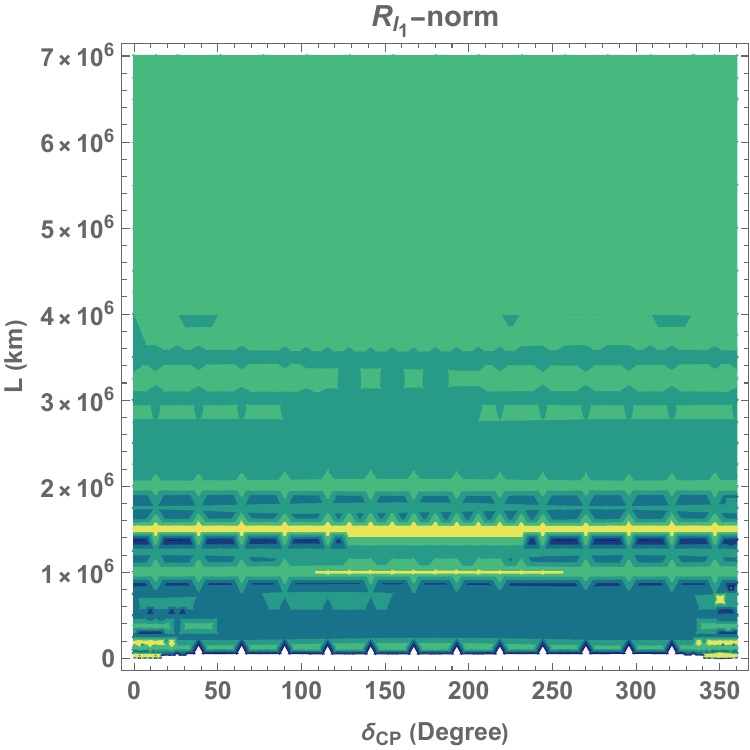}
\includegraphics[width=7.4mm]{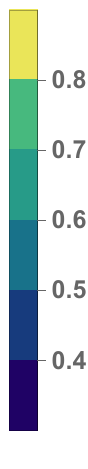}
        \caption{$\sigma_x=1.25\times10^{-9}$ m and $V = 2.824 \times 10^{-14}~\rm eV$}
        \label{6a_fig:sub4}
    \end{subfigure}
    \hfill
    \begin{subfigure}{0.32\textwidth}
        \centering
      \includegraphics[width=41mm]{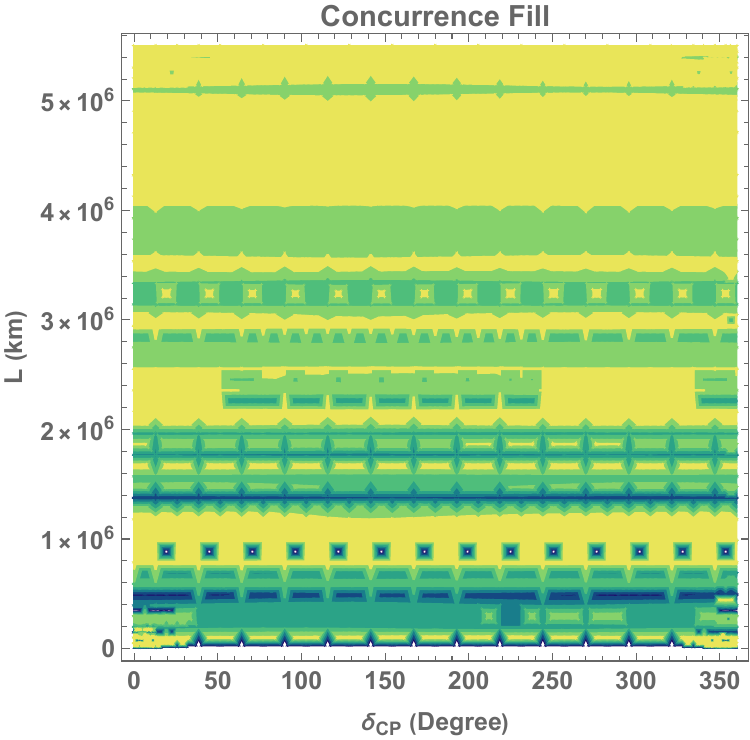}
 \includegraphics[width=7.4mm]{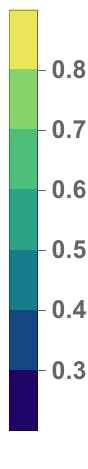}
        \caption{$\sigma_x=1.25\times10^{-9}$ m and $V = 2.824 \times 10^{-14}~\rm eV$ }
        \label{6a_fig:sub5}
    \end{subfigure}
     \hfill
    \begin{subfigure}{0.32\textwidth}
        \centering
  \includegraphics[width=41mm]{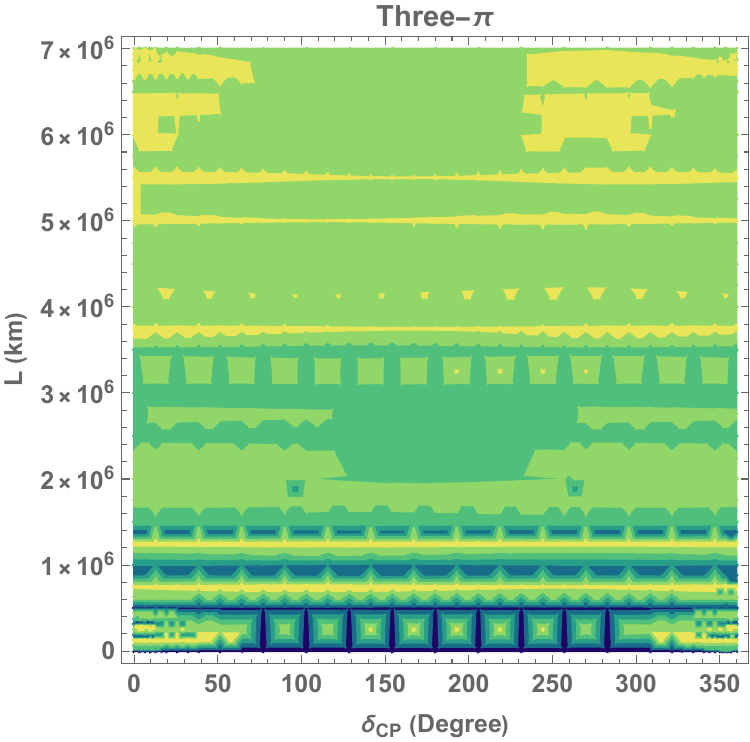}
\includegraphics[width=7.4mm]{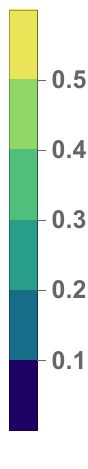}
        \caption{$\sigma_x=1.25\times10^{-9}$ m and $V = 2.824 \times 10^{-14}~\rm eV$}
        \label{6a_fig:sub6}
    \end{subfigure}

\medskip

    \begin{subfigure}{0.32\textwidth}
        \centering
     \includegraphics[width=43mm]{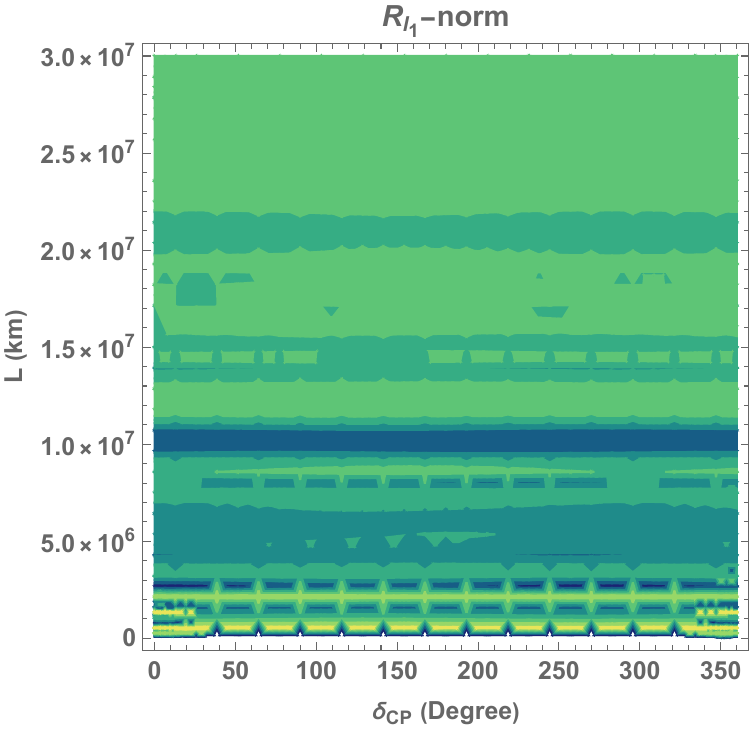}
     \includegraphics[width=7.2mm]{contour_plots/con_mark_rl1.pdf}
        \caption{$\sigma_x=7\times10^{-9}$ m and $V = 2.824 \times 10^{-14}~\rm eV$}
        \label{6a_fig:sub7}
    \end{subfigure}
    \hfill
    \begin{subfigure}{0.32\textwidth}
        \centering
   \includegraphics[width=41mm]{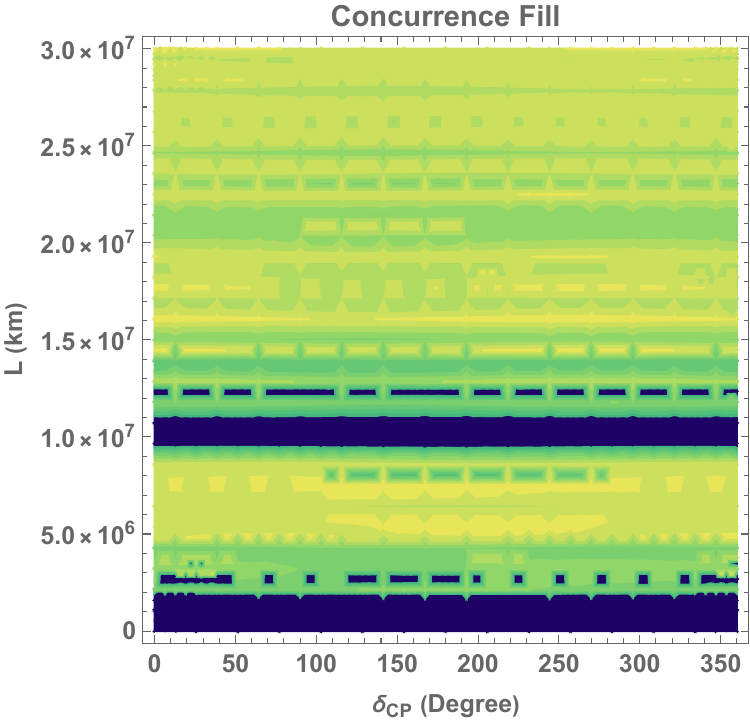}
 \includegraphics[width=8.4mm]{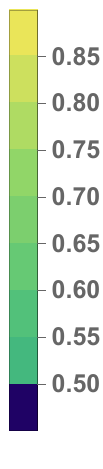}
        \caption{$\sigma_x=7\times10^{-9}$ m and $V = 2.824 \times 10^{-14}~\rm eV$}
        \label{6a_fig:sub8}
    \end{subfigure}
     \hfill
    \begin{subfigure}{0.32\textwidth}
        \centering
      \includegraphics[width=41mm]{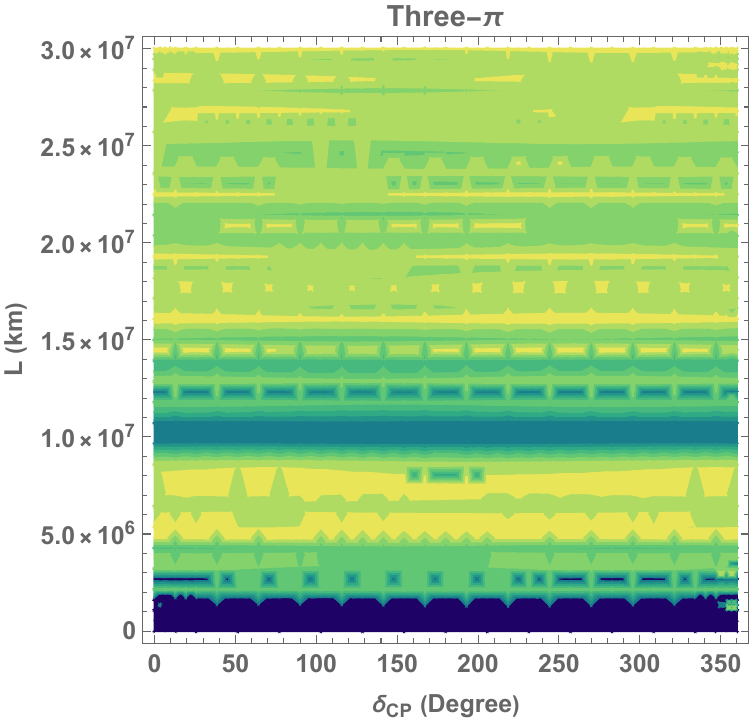}
      \includegraphics[width=7.4mm]{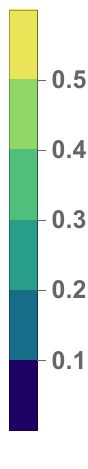}
        \caption{$\sigma_x=7\times10^{-9}$ m and $V = 2.824 \times 10^{-14}~\rm eV$}
        \label{6a_fig:sub9}
    \end{subfigure}

\medskip

    \begin{subfigure}{0.32\textwidth}
        \centering
       \includegraphics[width=41mm]{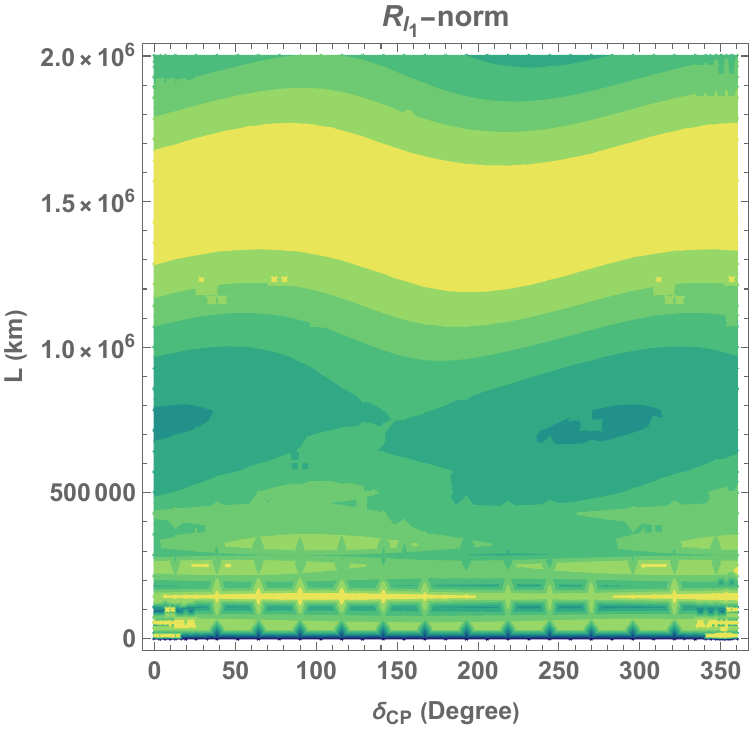}
       \includegraphics[width=7.3mm]{contour_plots/con_mark_rl1.pdf}
        \caption{V=0 and $\sigma_x=0.5\times10^{-9}$ m }
        \label{6a_fig:sub10}
    \end{subfigure}
    \hfill
    \begin{subfigure}{0.32\textwidth}
        \centering
      \includegraphics[width=41mm]{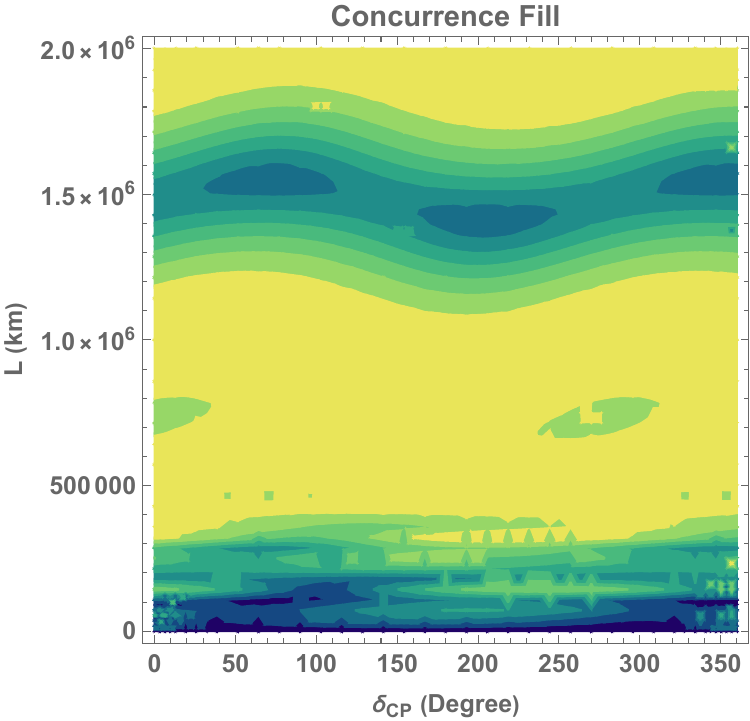}
       \includegraphics[width=7.2mm]{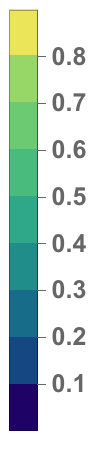}
        \caption{V=0 and $\sigma_x=0.5\times10^{-9}$ m}
        \label{6a_fig:sub11}
    \end{subfigure}
     \hfill
    \begin{subfigure}{0.32\textwidth}
        \centering
      \includegraphics[width=41mm]{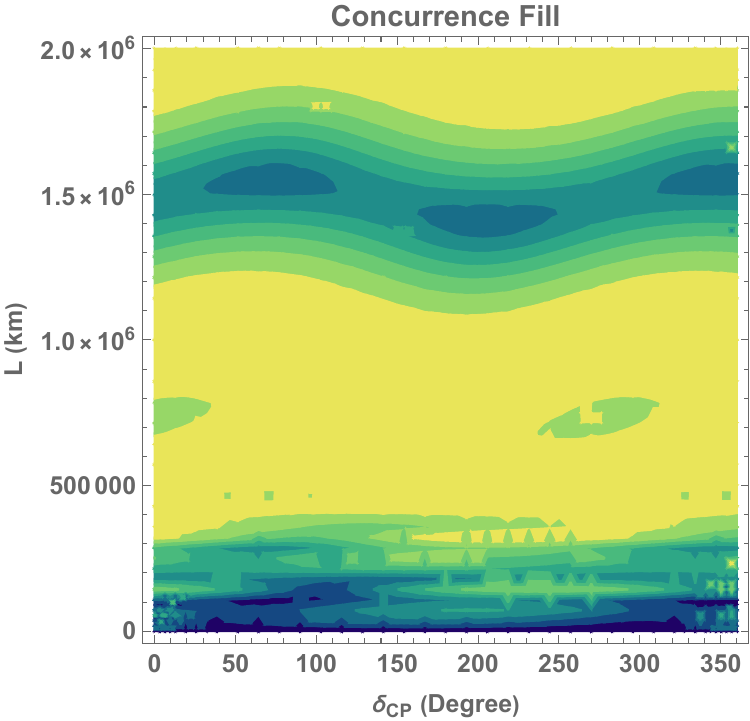}
\includegraphics[width=7.3mm]{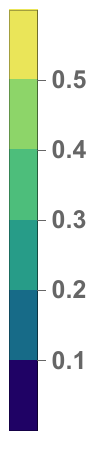}
      \caption{V=0 and $\sigma_x=0.5\times10^{-9}$ m}
        \label{6a_fig:sub12}
    \end{subfigure}
\caption{\justifying{Oscillogram plots of the $R_{l_{1}\text{-norm}}$, Concurrence Fill, and Three-$\pi $ entanglement in the $L-\delta_{\rm CP}$ plane, showing their variation with the baseline length L (km) and the CP-violating phase $\delta_{\rm CP}$ (degree), over the full range $0^\circ \leq \delta_{\rm CP} \leq 360^\circ$  with energy E=45 GeV. From left to right, the columns represent the $R_{l_{1}\text{-norm}}$, Concurrence Fill, and Three-$\pi$ entanglement, respectively. From top to bottom, the rows correspond to: (i)$\sigma_x=0.5\times10^{-9}$ m and $V = 2.824 \times 10^{-14}~\rm eV$ shown in Figs. 6(a)-6(c); (ii) $\sigma_x=0.5\times10^{-9}$ m and $V = 2.824 \times 10^{-14}~\rm eV$ shown in Figs. 6(d)-6(f); (iii) $\sigma_x=7\times10^{-9}$ m and $V = 2.824 \times 10^{-14}~\rm eV$; shown in Figs. 6(g)-6(i)and (iv) $ V=0$ and $\sigma_x=0.5\times10^{-9}$ m shown in Figs. 6(j)-6(l).}}
\label{fig6}
\end{figure*}

Fig. \ref{fig6} illustrates the variation of the $R_{l_1\text{-norm}}$, Concurrence Fill, and Three-$\pi$ entanglement in the $L-\delta_{\rm CP}$ plane for extended propagation distances beyond the DUNE baseline. Figs. \ref{fig6}(\subref{6a_fig:sub1})-\ref{fig6}(\subref{6a_fig:sub3}) illustrate the variation of the three residual correlation measures in the \(L\)-\(\delta_{\rm CP}\) plane for  \(\sigma_x=0.5\times10^{-9}\,\mathrm{m}\) and \(V=2.824\times10^{-14}\,\mathrm{eV}\). In Fig. \ref{fig6}(\subref{6a_fig:sub1}), $R_{l_1\text{-norm}}$ is observed to exhibit higher values in several distinct patches corresponding to different values of the CP-violating phase, $\delta_{\rm CP}$. Furthermore, at \(L \geq 1.2\times10^6\) km, the $R_{l_1\text{-norm}}$ value remains constant, irrespective of variations in \(L\) and \(\delta_{\rm CP}\). This character directly reflects the corresponding behaviour of the transition probabilities. A similar character is observed for Concurrence Fill shown in Fig. \ref{fig6}(\subref{6a_fig:sub2}), while Three-\(\pi\) shown in Fig. \ref{fig6}(\subref{6a_fig:sub3}) becomes independent of both \(L\) and \(\delta_{\rm CP}\) for \(L>3.5\times10^{6}\,\mathrm{km}\). The dependence of the three residual correlation measures on the baseline length \(L\) and the CP-violating phase \(\delta_{\rm CP}\) is presented in Figs. \ref{fig6}(\subref{6a_fig:sub4})-\ref{fig6}(\subref{6a_fig:sub6}) for \(\sigma_x=1.25\times10^{-9}\,\mathrm{m}\) and \(V=2.824\times10^{-14}\,\mathrm{eV}\). The $R_{l_1\text{-norm}}$ in Fig. \ref{fig6}(\subref{6a_fig:sub4}) follows a similar pattern to that in Fig. \ref{fig6}(\subref{6a_fig:sub1}), exhibiting negligible sensitivity to $\delta_{\rm CP}$ for baseline lengths exceeding 4$\times10^6$ km. Likewise, for baseline lengths exceeding this value, the maximum Concurrence Fill as in Fig. \ref{fig6}(\subref{6a_fig:sub5}) remains nearly constant, exhibiting negligible dependence on variations in both the baseline length L and the CP-violating phase $\delta_{\rm CP}$. For the Three-\(\pi\) entanglement as in Fig. \ref{fig6}(\subref{6a_fig:sub6}), a similar behaviour is observed for baseline lengths exceeding \(7\times10^{6}\,\mathrm{km}\). The variations of the three residual correlation measures with respect to the baseline length L and the CP-violating phase $\delta_{\rm CP}$ are illustrated in Figs. \ref{fig6}(\subref{6a_fig:sub7})-\ref{fig6}(\subref{6a_fig:sub9}), for a $\sigma_x=7\times10^{-9}\,\mathrm{m}$ and $V=2.824\times10^{-14}\,\mathrm{eV}$. The overall behaviour is similar to that observed in Figs. \ref{fig6}(\subref{6a_fig:sub1})-\ref{fig6}(\subref{6a_fig:sub3}) and Figs. \ref{fig6}(\subref{6a_fig:sub4})-\ref{fig6}(\subref{6a_fig:sub6}), with the difference being the baseline length beyond which the three residual correlation measures approach constant values, exhibiting negligible dependence on both L and $\delta_{\rm CP}$. This behaviour is consistent with the corresponding changes in the transition probabilities, indicating that an increase in the wave-packet width extends the propagation distance over which coherence is maintained before decoherence becomes significant. Figs. \ref{fig6}(\subref{6a_fig:sub10})-\ref{fig6}(\subref{6a_fig:sub12}) illustrate the variation of the three residual quantum correlation measures in the \(L\)-\(\delta_{\rm CP}\) plane for the vacuum scenario (\(V=0\)) with a fixed wave-packet width of \(\sigma_x=0.5\times10^{-9}\,\mathrm{m}\). As shown in Fig. \ref{fig6}(\subref{6a_fig:sub10}), the $R_{l_1\text{-norm}}$ attains its maximum values over a broad range of baseline lengths, approximately $1.3\times10^6$ km to $2.0\times10^6$ km, covering the entire range of $\delta_{\rm CP}$. For the Concurrence Fill shown in Fig. \ref{fig6}(\subref{6a_fig:sub11}), the maximum values are observed over the baseline length range of \(5\times10^{5}\)-\(1.2\times10^{6}\,\mathrm{km}\) across the entire range of \(\delta_{\rm CP}\). In contrast, at \(L\approx1.5\times10^{6}\,\mathrm{km}\), three distinct regions exhibiting near-minimum values of the Concurrence Fill are observed, corresponding to \(\delta_{\rm CP}=30^\circ\)-\(120^\circ\), \(180^\circ\)-\(280^\circ\), and \(320^\circ\)-\(360^\circ\). A similar behaviour is also observed for the Three-\(\pi\) entanglement as shown in Fig. \ref{fig6}(\subref{6a_fig:sub12}).

\section{Conclusion}\label{Sec:conclusion}
In this work, we investigate the three-flavor transition probabilities of an initial muon-flavor neutrino state and its residual quantum correlations using the wave-packet formalism while incorporating matter effects. The dependence of these quantities on the matter potential $V$ and the wave-packet width $\sigma_x$ is examined to understand how matter-induced interactions and wave-packet decoherence jointly influence neutrino flavor evolution. The study is performed for the DUNE experimental configuration as well as for an extended propagation distance.

For the DUNE baseline, the results show that finite wave-packet widths suppress and smooth the oscillation patterns compared to the PW case. The suppression becomes more pronounced as $\sigma_x$ is reduced, resulting in stronger damping of both oscillation and survival probabilities. The presence of matter additionally alters the oscillation amplitudes and shifts the positions of the extrema, leading to appreciable deviations from the corresponding vacuum results. The residual quantum correlation measures, namely Three-$\pi$ entanglement, Concurrence Fill, and residual $l_1$-norm coherence , also respond significantly to both wave-packet decoherence and matter effects. Among these measures, the residual $l_1$-norm coherence maintains the largest magnitude over the propagation range, whereas the entanglement measures exhibit comparatively stronger suppression. This behaviour can be primarily ascribed to the survival and oscillation probabilities of the initial muon-flavor neutrino state. Furthermore, we observe that the residual $l_1$-norm coherence exhibits a relatively weak dependence on $\delta_{\rm CP}$ in the $L-\delta_{\rm CP}$ plane, whereas the Concurrence Fill and Three-$\pi$ entanglement exhibit a stronger sensitivity to the CP-violating phase, $\delta_{\rm CP}$ for the DUNE baseline. Moreover, changes in the wave-packet width and the inclusion or omission of matter effects significantly influence the positions and extent of the regions corresponding to their maximum values in the \(L\)-\(\delta_{\rm CP}\) parameter space. These results indicate that genuine tripartite entanglement in neutrino oscillations is not restricted to the PW approximation but also persists within the wave-packet approximation.

In the extended propagation regime, wave-packet decoherence becomes increasingly important as the baseline increases. With increasing propagation distance, the oscillatory features of the neutrino flavor transition probabilities are progressively washed out, eventually approaching nearly constant averaged values at very large baselines as the neutrino wave packets become spatially separated. The extent to which neutrino flavor transitions and quantum correlations persist is strongly dependent on the wave-packet width: larger values of $\sigma_x$ allow these features to survive over greater distances, whereas smaller values result in more rapid damping. The matter potential also substantially modifies the distance-dependent evolution of both transition probabilities and quantum correlations. In particular, increasing the matter potential tends to enhance the residual coherence while producing a pronounced suppression of the entanglement measures. In the extended propagation regime, the results in the $L-\delta_{\rm CP}$ plane show that, beyond a certain propagation distance, all three residual quantum correlation measures approach constant values, becoming independent of both L and $\delta_{\rm CP}$. Thus, the wave-packet analysis reveals that residual quantum correlations persist over baselines far exceeding the DUNE baseline, highlighting the robustness of genuine tripartite entanglement and the long-range quantum correlations underlying neutrino oscillations.

\section{Acknowledgement}
L. K. acknowledges the Ministry of Education (MoE), Government of India, for providing financial support. B. Y. gratefully acknowledge the support of the National Natural Science Foundation of China under Grant No. 12075059, as well as the start-up fund of the University of Science and Technology of China (USTC).

\appendix 
\section{Neutrino flavor oscillations: wave-packet approach in matter}
\label{AppendixA}

The time evolution of a neutrino flavor state is governed by the Schrödinger equation
\begin{equation}
i\frac{d}{dt} |\nu_\alpha(t)\rangle = H |\nu_\alpha(t)\rangle.
\end{equation}
Where the effective Hamiltonian is in the mass basis and takes the form
\begin{equation}
  H =\frac{1}{2E}\big[U\mbox{diag}(0,\Delta m_{21}^{2},\Delta m_{31}^{2})U^{\dagger}+\mbox{diag}(V,0,0)\big],
\label{eq:effH}
\end{equation}
$V$ is the matter potential and takes the form $V=\pm2\sqrt{2}G_{F} N_{e} E$, where $G_F$ is the Fermi constant and $N_{e}$ is the electron number density. The positive sign is for neutrinos, whereas the negative sign is used for antineutrinos.

The Hamiltonian can be diagonalized through successive rotations. First, a rotation in the $(1-3)$ sector introduces the effective mixing angle in matter:
\begin{equation}
\theta_{13}^m = \frac{1}{2}\tan^{-1}\left(\frac{\epsilon \sin 2\theta_{13}}{\epsilon \cos 2\theta_{13} - 2EV}\right),
\end{equation}
where $\epsilon = \Delta m^2_{31} - \Delta m^2_{21}\sin^2\theta_{12}.$ Next, a rotation in the $(1-2)$ sector yields
\begin{equation}
\theta_{12}^m = \frac{1}{2}\tan^{-1}\left(\frac{\Delta m^2_{21}\sin 2\theta_{12}}{l_2 - l_1}\cos(\theta_{13}-\theta_{13}^m)\right).
\end{equation}

The matter mixing matrix $U^m_{\alpha i}$ has the same form as the PMNS matrix defined in Eq. \eqref{PMNS}, with the vacuum mixing angles $\theta_{13}$ and $\theta_{12}$ replaced by their matter-modified counterparts $\theta_{13}^{m}$ and $\theta_{12}^{m}$, respectively. After these rotations, the effective Hamiltonian becomes approximately diagonal, with eigenvalues
\begin{equation}
\begin{aligned}
E_{1,2}^{m}
&=
\frac{1}{4E}
\Bigg[
(l_{1}+l_{2})\, \mp \\
&\qquad
\sqrt{
(l_{1}-l_{2})^{2}+
\Delta m_{21}^{4}
\sin^{2}2\theta_{12}
\cos^{2}\!\left(\theta_{13}-\theta_{13}^{m}\right)
}
\Bigg],
\label{eq:E12_matter}
\end{aligned}
\end{equation}
and
\begin{equation}
E_3^m=\frac{l_3}{2E}.
\end{equation}
The quantities $l_i$ appearing in the matter-modified eigenvalues are defined as
\begin{equation}
\label{l13}
\begin{aligned}
l_{1,3}
&=
\frac{1}{2}
\Bigg[
\Delta m_{31}^{2}
+2EV
+\Delta m_{21}^{2}\sin^{2}\theta_{12} \\
&\qquad\qquad
\mp
\sqrt{
(2EV)^{2}
+\epsilon^{2}
-4EV\epsilon\cos2\theta_{13}
}
\Bigg].
\end{aligned}
\end{equation}
and
\begin{equation}
l_2=\Delta m_{21}^{2}\cos^{2}\theta_{12}.
\end{equation}

To determine the coherence lengths, it is necessary to evaluate the differences between the group velocities of the matter eigenstates. These velocity differences are obtained from the energy splittings through Eq. \eqref{vdifference}. In the three-flavor framework, two independent velocity differences can be identified:
\begin{equation}
v_{21}^{m}=\frac{1}{8E}\left(-8\Delta E_{21}^{m}+\xi\right),
\end{equation}
\begin{equation}
v_{32}^{m}=\frac{1}{8E}\left(6\frac{dl_3}{dE}-4V-8\Delta E_{32}^{m}-\frac{\xi}{2}\right),
\end{equation}

with 
\begin{equation}
    \xi=\frac{1}{2}\,
\frac{1}{4E\,\Delta E_{21}^{m}}
\left[
8(l_1-l_2)\frac{dl_1}{dE}
+\frac{\zeta}{(\epsilon\cos2\theta_{13}-2EV)^2}
\right],
\end{equation}
where
\begin{equation}
    \zeta=
4\Delta m_{21}^{4}\epsilon V
\cos^{2}2\theta_{13}^{m}
\sin^{2}2\theta_{12}
\sin2\theta_{13}
\sin2(\theta_{13}-\theta_{13}^{m}).
\end{equation}

Differentiating Eq. \eqref{l13} with respect to the neutrino energy yields
\begin{equation}
\frac{dl_{1,3}}{dE}
=V\left[
1\mp
\frac{2EV-\epsilon\cos2\theta_{13}}
{\sqrt{\epsilon^{2}+(2EV)^{2}-4EV\epsilon\cos2\theta_{13}}}
\right].
\end{equation}
\bibliography{main}

@article{PhysRevA.72.064306,
  title = {Single-particle entanglement},
  author = {van Enk, S. J.},
  journal = {Phys. Rev. A},
  volume = {72},
  issue = {6},
  pages = {064306},
  numpages = {3},
  year = {2005},
  month = {Dec},
  publisher = {American Physical Society},
  doi = {10.1103/PhysRevA.72.064306},
  url = {https://link.aps.org/doi/10.1103/PhysRevA.72.064306}
}

@article{Blasone:2021cau,
    author = "Blasone, Massimo and De Siena, Silvio and Matrella, Cristina",
    title = "{Wave packet approach to quantum correlations in neutrino oscillations}",
    eprint = "2104.03166",
    archivePrefix = "arXiv",
    primaryClass = "quant-ph",
    doi = "10.1140/epjc/s10052-021-09471-4",
    journal = "Eur. Phys. J. C",
    volume = "81",
    number = "7",
    pages = "660",
    year = "2021"
}

@article{Peltoniemi:2000nw,
    author = {Peltoniemi, J. T. and Sipilainen, V.},
    title = {Neutrino propagation in matter using the wave packet approach},
    eprint = {hep-ph/0004162},
    archivePrefix = {arXiv},
    reportNumber = {UO-REPORT-130-2000, UO-130-2000},
    doi = {10.1088/1126-6708/2000/06/011},
    journal = {JHEP},
    volume = {06},
    number = {06},
    pages = {011},
    year = {2000}
}

@article{Ravari:2022yfd,
    author = "Ravari, Z. Askaripour and Ettefaghi, M. M. and Miraboutalebi, S.",
    title = "{Quantum coherence in neutrino oscillation in matter}",
    eprint = "2204.12332",
    archivePrefix = "arXiv",
    primaryClass = "quant-ph",
    doi = "10.1140/epjp/s13360-022-02717-4",
    journal = "Eur. Phys. J. Plus",
    volume = "137",
    number = "4",
    pages = "488",
    year = "2022"
}

@article{Ettefaghi:2023zsh,
    author = "Ettefaghi, M. M. and Ravari, Z. Askaripour",
    title = "{Quantum coherence and entanglement in neutral-current neutrino oscillation in matter}",
    eprint = "2305.12194",
    archivePrefix = "arXiv",
    primaryClass = "hep-ph",
    doi = "10.1140/epjc/s10052-023-11505-y",
    journal = "Eur. Phys. J. C",
    volume = "83",
    number = "5",
    pages = "417",
    year = "2023"
}

@article{Blasone:2015lya,
    author = "Blasone, Massimo and Dell'Anno, Fabio and De Siena, Silvio and Illuminati, Fabrizio",
    title = "{Flavor entanglement in neutrino oscillations in the wave packet description}",
    eprint = "1510.06761",
    archivePrefix = "arXiv",
    primaryClass = "quant-ph",
    doi = "10.1209/0295-5075/112/20007",
    journal = "EPL",
    volume = "112",
    number = "2",
    pages = "20007",
    year = "2015"
}

@article{Esteban_2024,
    author = {Esteban, Ivan and Gonzalez-Garcia, M. C. and Maltoni, Michele and Martinez-Soler, Ivan and Pinheiro, Jo{\~a}o Paulo and Schwetz, Thomas},
    title = {NuFit-6.0: updated global analysis of three-flavor neutrino oscillations},
    eprint = {2410.05380},
    archivePrefix = {arXiv},
    primaryClass = {hep-ph},
    reportNumber = {IFT-UAM/CSIC-24-140, YITP-SB-2024-24, IPPP/24/64},
    doi = {10.1007/JHEP12(2024)216},
    journal = {JHEP},
    volume = {12},
    number = {12},
    pages = {216},
    year = {2024}
}

@article{Jha:2020dav,
    author = "Jha, Abhishek Kumar and Mukherjee, Supratik and Bambah, Bindu A.",
    title = "{Tri-Partite entanglement in Neutrino Oscillations}",
    eprint = "2004.14853",
    archivePrefix = "arXiv",
    primaryClass = "hep-ph",
    doi = "10.1142/S0217732321500565",
    journal = "Mod. Phys. Lett. A",
    volume = "36",
    number = "09",
    pages = "2150056",
    year = "2021"
}

@article{Jiang:2020qjf,
    author = "Jiang, Zhengmin and Zhang, Tinggui and Huang, Xiaofen and Fei, Shao-Ming",
    title = "{Trade-off relations of {\$}l{\_}1{\textbackslash})-norm coherence for multipartite systems}",
    eprint = "2002.00165",
    archivePrefix = "arXiv",
    primaryClass = "quant-ph",
    doi = "10.1007/s11128-020-2588-8",
    journal = "Quant. Inf. Proc.",
    volume = "19",
    number = "3",
    pages = "92",
    year = "2020"
}

@article{Li:2022zic,
    author = "Li, Yu-Wen and Li, Li-Juan and Song, Xue-Ke and Wang, Dong",
    title = "{Trade-off relations of quantum resource theory in neutrino oscillations}",
    eprint = "2212.09320",
    archivePrefix = "arXiv",
    primaryClass = "quant-ph",
    doi = "10.1140/epjp/s13360-022-03459-z",
    journal = "Eur. Phys. J. Plus",
    volume = "137",
    number = "11",
    pages = "1267",
    year = "2022"
}

@article{Yao:2015fuw,
    author = "Yao, Yao and Xiao, Xing and Ge, Li and Sun, C. P.",
    title = "{Quantum coherence in multipartite systems}",
    eprint = "1506.01773",
    archivePrefix = "arXiv",
    primaryClass = "quant-ph",
    doi = "10.1103/PhysRevA.92.022112",
    journal = "Phys. Rev. A",
    volume = "92",
    number = "2",
    pages = "022112",
    year = "2015"
}

@article{Giunti:1997wq,
    author = "Giunti, C. and Kim, C. W.",
    title = "{Coherence of neutrino oscillations in the wave packet approach}",
    eprint = "hep-ph/9711363",
    archivePrefix = "arXiv",
    reportNumber = "KIAS-P97007, JHU-TIPAC-97019, DFTT-71-97",
    doi = "10.1103/PhysRevD.58.017301",
    journal = "Phys. Rev. D",
    volume = "58",
    pages = "017301",
    year = "1998"
}

@article{Kiers:1995zj,
    author = "Kiers, Ken and Nussinov, Shmuel and Weiss, Nathan",
    title = "{Coherence effects in neutrino oscillations}",
    eprint = "hep-ph/9506271",
    archivePrefix = "arXiv",
    doi = "10.1103/PhysRevD.53.537",
    journal = "Phys. Rev. D",
    volume = "53",
    pages = "537--547",
    year = "1996"
}

@article{Takeuchi:2000fz,
    author = "Takeuchi, Yoshihiro and Tazaki, Yuichi and Tsai, S. Y. and Yamazaki, Takashi",
    title = "{How do neutrinos propagate? Wave packet treatment of neutrino oscillation}",
    eprint = "hep-ph/0006334",
    archivePrefix = "arXiv",
    reportNumber = "NUP-A-2000-9",
    doi = "10.1143/PTP.105.471",
    journal = "Prog. Theor. Phys.",
    volume = "105",
    pages = "471--482",
    year = "2001"
}

@article{Coffman:1999jd,
    author = "Coffman, Valerie and Kundu, Joydip and Wootters, William K.",
    title = "{Distributed entanglement}",
    eprint = "quant-ph/9907047",
    archivePrefix = "arXiv",
    doi = "10.1103/PhysRevA.61.052306",
    journal = "Phys. Rev. A",
    volume = "61",
    pages = "052306",
    year = "2000"
}

@article{Peres:1996dw,
    author = "Peres, Asher",
    title = "{Separability criterion for density matrices}",
    eprint = "quant-ph/9604005",
    archivePrefix = "arXiv",
    doi = "10.1103/PhysRevLett.77.1413",
    journal = "Phys. Rev. Lett.",
    volume = "77",
    pages = "1413--1415",
    year = "1996"
}

@article{Vidal:2002zz,
    author = "Vidal, G. and Werner, R. F.",
    title = "{Computable measure of entanglement}",
    eprint = "quant-ph/0102117",
    archivePrefix = "arXiv",
    doi = "10.1103/PhysRevA.65.032314",
    journal = "Phys. Rev. A",
    volume = "65",
    pages = "032314",
    year = "2002"
}

@article{Sabin:2008naa,
    author = "Sab{\'\i}n, Carlos and Garci{\-}a-Alcaine, Guillermo",
    title = "{A classification of entanglement in three-qubit systems}",
    eprint = "0707.1780",
    archivePrefix = "arXiv",
    primaryClass = "quant-ph",
    doi = "10.1140/epjd/e2008-00112-5",
    journal = "Eur. Phys. J. D",
    volume = "48",
    number = "3",
    pages = "435--442",
    year = "2008"
}

@article{Ou:2007ysv,
    author = "Ou, Yong-Cheng and Fan, Heng",
    title = "{Monogamy inequality in terms of negativity for three-qubit states}",
    eprint = "quant-ph/0702127",
    archivePrefix = "arXiv",
    doi = "10.1103/PhysRevA.75.062308",
    journal = "Phys. Rev. A",
    volume = "75",
    number = "6",
    pages = "062308",
    year = "2007"
}

@article{Miranowicz:2004ivl,
    author = "Miranowicz, Adam and Grudka, Andrzej",
    title = "{A comparative study of relative entropy of entanglement, concurrence and negativity}",
    eprint = "quant-ph/0409153",
    archivePrefix = "arXiv",
    doi = "10.1088/1464-4266/6/12/009",
    month = "10",
    journal = "J. Opt. B: Quantum Semiclass. Opt.",
    year = "2004"
}

@article{Xie:2021hsy,
    author = "Xie, Songbo and Eberly, Joseph H.",
    title = "{Triangle Measure of Tripartite Entanglement}",
    eprint = "2101.02260",
    archivePrefix = "arXiv",
    primaryClass = "quant-ph",
    doi = "10.1103/PhysRevLett.127.040403",
    journal = "Phys. Rev. Lett.",
    volume = "127",
    number = "4",
    pages = "040403",
    year = "2021"
}

@article{Baumgratz:2013ecx,
    author = "Baumgratz, T. and Cramer, M. and Plenio, M. {\,}B.",
    title = "{Quantifying Coherence}",
    eprint = "1311.0275",
    archivePrefix = "arXiv",
    primaryClass = "quant-ph",
    doi = "10.1103/physrevlett.113.140401",
    journal = "Phys. Rev. Lett.",
    volume = "113",
    number = "14",
    pages = "140401",
    year = "2014"
}

@article{Singh:2015oby,
    author = "Singh, Uttam and Bera, Manabendra Nath and Dhar, Himadri Shekhar and Pati, Arun Kumar",
    title = "{Maximally coherent mixed states: Complementarity between maximal coherence and mixedness}",
    eprint = "1503.06303",
    archivePrefix = "arXiv",
    primaryClass = "quant-ph",
    reportNumber = "HRI-P-15-05-003",
    doi = "10.1103/PhysRevA.91.052115",
    journal = "Phys. Rev. A",
    volume = "91",
    number = "5",
    pages = "052115",
    year = "2015"
}

@article{Super-Kamiokande:1998kpq,
    author = "Fukuda, Y. and others",
    collaboration = "Super-Kamiokande",
    title = "{Evidence for oscillation of atmospheric neutrinos}",
    eprint = "hep-ex/9807003",
    archivePrefix = "arXiv",
    reportNumber = "BU-98-17, ICRR-REPORT-422-98-18, UCI-98-8, KEK-PREPRINT-98-95, LSU-HEPA-5-98, UMD-98-003, SBHEP-98-5, TKU-PAP-98-06, TIT-HPE-98-09",
    doi = "10.1103/PhysRevLett.81.1562",
    journal = "Phys. Rev. Lett.",
    volume = "81",
    pages = "1562--1567",
    year = "1998"
}

@article{kajita2016nobel,
  author = {{Kajita}, Takaaki},
        title = "{Nobel Lecture: Discovery of atmospheric neutrino oscillations*}",
      journal = {Reviews of Modern Physics},
         year = 2016,
        month = jul,
       volume = {88},
       number = {3},
          eid = {030501},
        pages = {030501},
          doi = {10.1103/RevModPhys.88.030501},
       adsurl = {https://ui.adsabs.harvard.edu/abs/2016RvMP...88c0501K}
}

@article{SNO:2002tuh,
    author = "Ahmad, Q. R. and others",
    collaboration = "SNO",
    title = "{Direct evidence for neutrino flavor transformation from neutral current interactions in the Sudbury Neutrino Observatory}",
    eprint = "nucl-ex/0204008",
    archivePrefix = "arXiv",
    doi = "10.1103/PhysRevLett.89.011301",
    journal = "Phys. Rev. Lett.",
    volume = "89",
    pages = "011301",
    year = "2002"
}

@article{Giunti:2003ax,
    author = "Giunti, C.",
    title = "{Coherence and wave packets in neutrino oscillations}",
    eprint = "hep-ph/0302026",
    archivePrefix = "arXiv",
    doi = "10.1023/B:FOPL.0000019651.53280.31",
    journal = "Found. Phys. Lett.",
    volume = "17",
    pages = "103--124",
    year = "2004"
}

@article{Nussinov:1976uw,
    author = "Nussinov, S.",
    title = "{Solar Neutrinos and Neutrino Mixing}",
    doi = "10.1016/0370-2693(76)90648-1",
    journal = "Phys. Lett. B",
    volume = "63",
    pages = "201--203",
    year = "1976"
}

@article{Wolfenstein:1977ue,
    author = "Wolfenstein, L.",
    title = "{Neutrino Oscillations in Matter}",
    reportNumber = "COO-3066-102",
    doi = "10.1103/PhysRevD.17.2369",
    journal = "Phys. Rev. D",
    volume = "17",
    pages = "2369--2374",
    year = "1978"
}

@article{Mikheyev:1985zog,
    author = "Mikheyev, S. P. and Smirnov, A. Yu.",
    title = "{Resonance Amplification of Oscillations in Matter and Spectroscopy of Solar Neutrinos}",
    journal = "Sov. J. Nucl. Phys.",
    volume = "42",
    pages = "913--917",
    year = "1985"
}

@book{Giunti:2007ry,
    author = {Giunti, Carlo and Kim, Chung W.},
    title = {Fundamentals of Neutrino Physics and Astrophysics},
    publisher = {Oxford University Press},
    address = {Oxford},
    doi = {10.1093/acprof:oso/9780198508717.001.0001},
    isbn = {978-0-19-850871-7},
    year = {2007}
}

@article{SNO:2001kpb,
    author = "Ahmad, Q. R. and others",
    collaboration = "SNO",
    title = "{Measurement of the rate of $\nu_e+d \to p+p+e^-$ interactions produced by $^8$B solar neutrinos at the Sudbury Neutrino Observatory}",
    eprint = "nucl-ex/0106015",
    archivePrefix = "arXiv",
    reportNumber = "UPR-0240E",
    doi = "10.1103/PhysRevLett.87.071301",
    journal = "Phys. Rev. Lett.",
    volume = "87",
    pages = "071301",
    year = "2001"
}

@article{KamLAND:2002uet,
    author = "Eguchi, K. and others",
    collaboration = "KamLAND",
    title = "{First results from KamLAND: Evidence for reactor anti-neutrino disappearance}",
    eprint = "hep-ex/0212021",
    archivePrefix = "arXiv",
    doi = "10.1103/PhysRevLett.90.021802",
    journal = "Phys. Rev. Lett.",
    volume = "90",
    pages = "021802",
    year = "2003"
}

@article{DayaBay:2022eyy,
    author = "An, F. P. and others",
    collaboration = "Daya Bay",
    title = "{First Measurement of High-Energy Reactor Antineutrinos at Daya Bay}",
    eprint = "2203.06686",
    archivePrefix = "arXiv",
    primaryClass = "hep-ex",
    doi = "10.1103/PhysRevLett.129.041801",
    journal = "Phys. Rev. Lett.",
    volume = "129",
    number = "4",
    pages = "041801",
    year = "2022"
}

@article{T2K:2023smv,
    author = "Abe, K. and others",
    collaboration = "T2K",
    title = "{Measurements of neutrino oscillation parameters from the T2K experiment using $3.6\times 10^{21}$ protons on target}",
    eprint = "2303.03222",
    archivePrefix = "arXiv",
    primaryClass = "hep-ex",
    doi = "10.1140/epjc/s10052-023-11819-x",
    journal = "Eur. Phys. J. C",
    volume = "83",
    number = "9",
    pages = "782",
    year = "2023"
}

@article{NOvA:2019cyt,
    author = "Acero, M. A. and others",
    collaboration = "NOvA",
    title = "{First Measurement of Neutrino Oscillation Parameters using Neutrinos and Antineutrinos by NOvA}",
    eprint = "1906.04907",
    archivePrefix = "arXiv",
    primaryClass = "hep-ex",
    reportNumber = "FERMILAB-PUB-19-272-ND",
    doi = "10.1103/PhysRevLett.123.151803",
    journal = "Phys. Rev. Lett.",
    volume = "123",
    number = "15",
    pages = "151803",
    year = "2019"
}

@article{Blasone:1995zc,
    author = "Blasone, M. and Vitiello, Giuseppe",
    title = "{Quantum field theory of fermion mixing}",
    eprint = "hep-ph/9501263",
    archivePrefix = "arXiv",
    reportNumber = "SADF1-1995",
    doi = "10.1006/aphy.1995.1115",
    journal = "Annals Phys.",
    volume = "244",
    pages = "283--311",
    year = "1995",
    note = "[Erratum: Annals Phys. 249, 363--364 (1996)]"
}

@article{Blasone_2009Entanglement,
   title={Entanglement in neutrino oscillations},
   volume={85},
   ISSN={1286-4854},
   url={http://dx.doi.org/10.1209/0295-5075/85/50002},
   DOI={10.1209/0295-5075/85/50002},
   number={5},
   journal={EPL},
   publisher={IOP Publishing},
   author={Blasone, M. and Dell’Anno, F. and De Siena, S. and Illuminati, F.},
   year={2009},
   month=mar, pages={50002} }

@article{Blasone_2008,
    author = "Blasone, M. and Dell'Anno, F. and De Siena, S. and Di Mauro, M. and Illuminati, F.",
    title = "{Multipartite entangled states in particle mixing}",
    eprint = "0711.2268",
    archivePrefix = "arXiv",
    primaryClass = "quant-ph",
    doi = "10.1103/PhysRevD.77.096002",
    journal = "Phys. Rev. D",
    volume = "77",
    pages = "096002",
    year = "2008"
}

@article{Blasone_2010,
   title={On entanglement in neutrino mixing and oscillations},
   volume={237},
   ISSN={1742-6596},
   url={http://dx.doi.org/10.1088/1742-6596/237/1/012007},
   DOI={10.1088/1742-6596/237/1/012007},
  journal={Journal of Physics: Conference Series},
   publisher={IOP Publishing},
   author={Blasone, Massimo and Dell’Anno, Fabio and Siena, Silvio De and Illuminati, Fabrizio},
   year={2010},
   month=jun, pages={012007} }

@article{Konwar:2024nrd,
    author = "Konwar, Lekhashri and Yadav, Bhavna",
    title = "{NSI effects on tripartite entanglement in neutrino oscillations}",
    eprint = "2402.09952",
    archivePrefix = "arXiv",
    primaryClass = "hep-ph",
    doi = "10.1016/j.nuclphysb.2024.116544",
    journal = "Nucl. Phys. B",
    volume = "1002",
    pages = "116544",
    year = "2024"
}

@article{Konwar:2025ipv,
    author = "Konwar, Lekhashri and Panda, Papia and Mohanta, Rukmani",
    title = "{Effect of off-diagonal NSI parameters on entanglement measurements in neutrino oscillations}",
    eprint = "2507.04885",
    archivePrefix = "arXiv",
    primaryClass = "hep-ph",
    doi = "10.1088/1361-6471/ae5076",
    journal = "J. Phys. G",
    volume = "53",
    number = "3",
    pages = "035002",
    year = "2026"
}

@article{Alok_2016,
   title={Quantum correlations in terms of neutrino oscillation probabilities},
   volume={909},
   ISSN={0550-3213},
   url={http://dx.doi.org/10.1016/j.nuclphysb.2016.05.001},
   DOI={10.1016/j.nuclphysb.2016.05.001},
   journal={Nucl. Phys. B},
   publisher={Elsevier BV},
   author={Alok, Ashutosh Kumar and Banerjee, Subhashish and Uma Sankar, S.},
   year={2016},
   month=aug, pages={65–72} }

@article{Banerjee:2024lih,
    author = "Banerjee, Rajrupa and Panda, Papia and Mohanta, Rukmani and Patra, Sudhanwa",
    title = "{Analysis of neutrino oscillation parameters in the light on quantum entanglement}",
    eprint = "2410.05727",
    archivePrefix = "arXiv",
    primaryClass = "hep-ph",
    doi = "10.1016/j.nuclphysb.2025.117263",
    journal = "Nucl. Phys. B",
    volume = "1022",
    pages = "117263",
    year = "2026"
}

@article{Banerjee:2025gau,
    author = "Banerjee, Rajrupa and Panigrahi, Prasanta K. and Mishra, Hiranmaya and Patra, Sudhanwa",
    title = "{Concurrence fill and mode distribution of entanglement in neutrino oscillation}",
    eprint = "2512.11320",
    archivePrefix = "arXiv",
    primaryClass = "hep-ph",
    doi = "10.1103/h72t-ndl2",
    journal = "Phys. Rev. D",
    volume = "113",
    number = "7",
    pages = "073005",
    year = "2026"
}

@article{Li:2021fft,
    author = "Li, Li-Juan and Ming, Fei and Song, Xue-Ke and Ye, Liu and Wang, Dong",
    title = "{Characterizing entanglement and measurement{\textquoteright}s uncertainty in neutrino oscillations}",
    doi = "10.1140/epjc/s10052-021-09503-z",
    journal = "Eur. Phys. J. C",
    volume = "81",
    number = "8",
    pages = "728",
    year = "2021"
}

@article{banerjee2015quantum,
  title={A quantum-information theoretic analysis of three-flavor neutrino oscillations: Quantum entanglement, nonlocal and nonclassical features of neutrinos},
  author={Banerjee, Subhashish and Alok, Ashutosh Kumar and Srikanth, R and Hiesmayr, Beatrix C},
  journal={ Eur. Phys. J. C},
  volume={75},
  number={10},
  pages={487},
  year={2015},
  publisher={Springer},
  url={http://dx.doi.org/10.1140/epjc/s10052-015-3717-x},
   DOI={10.1140/epjc/s10052-015-3717-x}
}

@article{Li:2022mus,
    author = "Li, Yu-Wen and Li, Li-Juan and Song, Xue-Ke and Wang, Dong and Ye, Liu",
    title = "{Geuine tripartite entanglement in three-flavor neutrino oscillations}",
    eprint = "2205.11058",
    archivePrefix = "arXiv",
    primaryClass = "quant-ph",
    doi = "10.1140/epjc/s10052-022-10759-2",
    journal = "Eur. Phys. J. C",
    volume = "82",
    number = "9",
    pages = "799",
    year = "2022"
}

@misc{Quinta:2022sgq,
    author = {Quinta, Gon\c{c}alo M. and Sousa, Alexandre and Omar, Yasser},
    title = {Predicting leptonic CP violation via minimization of neutrino entanglement},
    eprint = {2207.03303},
    archivePrefix = {arXiv},
    primaryClass = {hep-ph},
    year = {2022},
    journal = {}
}

@article{Blasone_2015,
author = {{Blasone}, Massimo and {Dell'Anno}, Fabio and {De Siena}, Silvio and {Illuminati}, Fabrizio},
        title = "{Flavor entanglement in neutrino oscillations in the wave packet description}",
      journal = {EPL (Europhysics Letters)},
         year = 2015,
        month = oct,
       volume = {112},
       number = {2},
          eid = {20007},
        pages = {20007},
          doi = {10.1209/0295-5075/112/20007},
archivePrefix = {arXiv},
       eprint = {1510.06761},
 primaryClass = {quant-ph},
       adsurl = {https://ui.adsabs.harvard.edu/abs/2015EL....11220007B}
}

@article{sarkar2021effects,
    author = "Sarkar, Trisha and Dixit, Khushboo",
    title = "{Effects of nonstandard interaction on temporal and spatial correlations in neutrino oscillations}",
    eprint = "2010.02175",
    archivePrefix = "arXiv",
    primaryClass = "hep-ph",
    doi = "10.1140/epjc/s10052-021-08874-7",
    journal = "Eur. Phys. J. C",
    volume = "81",
    number = "1",
    pages = "88",
    year = "2021"
}

@article{Dixit:2018kev,
    author = "Dixit, Khushboo and Naikoo, Javid and Banerjee, Subhashish and Alok, Ashutosh Kumar",
    title = "{Quantum correlations and the neutrino mass degeneracy problem}",
    eprint = "1807.01546",
    archivePrefix = "arXiv",
    primaryClass = "hep-ph",
    doi = "10.1140/epjc/s10052-018-6376-x",
    journal = "Eur. Phys. J. C",
    volume = "78",
    number = "11",
    pages = "914",
    year = "2018"
}

@article{Dixit:2019swl,
    author = "Dixit, Khushboo and Kumar Alok, Ashutosh",
    title = "{New physics effects on quantum coherence in neutrino oscillations}",
    eprint = "1909.04887",
    archivePrefix = "arXiv",
    primaryClass = "hep-ph",
    doi = "10.1140/epjp/s13360-021-01311-4",
    journal = "Eur. Phys. J. Plus",
    volume = "136",
    number = "3",
    pages = "334",
    year = "2021"
}

@article{Yadav:2022grk,
    author = "Yadav, Bhavna and Sarkar, Trisha and Dixit, Khushboo and Alok, Ashutosh Kumar",
    title = "{Can NSI affect non-local correlations in neutrino oscillations?}",
    eprint = "2201.05580",
    archivePrefix = "arXiv",
    primaryClass = "hep-ph",
    doi = "10.1140/epjc/s10052-022-10373-2",
    journal = "Eur. Phys. J. C",
    volume = "82",
    pages = "446",
    year = "2022"
}

@article{Yadav:2024qav,
    author = "Yadav, Bhavna and Alok, Ashutosh Kumar",
    title = "{Impact of scalar NSI on spatial and temporal correlations in neutrino oscillations}",
    eprint = "2411.17503",
    archivePrefix = "arXiv",
    primaryClass = "hep-ph",
    doi = "10.1088/1361-6471/ae263c",
    journal = "J. Phys. G",
    volume = "52",
    number = "12",
    pages = "125004",
    year = "2025"
}

@article{PhysRevA.98.050302,
 author = {{Song}, Xue-Ke and {Huang}, Yanqi and {Ling}, Jiajie and {Yung}, Man-Hong},
        title = "{Quantifying quantum coherence in experimentally observed neutrino oscillations}",
      journal = {\pra},
         year = 2018,
        month = nov,
       volume = {98},
       number = {5},
          eid = {050302},
        pages = {050302},
          doi = {10.1103/PhysRevA.98.050302},
archivePrefix = {arXiv},
       eprint = {1806.00715},
 primaryClass = {hep-ph},
       adsurl = {https://ui.adsabs.harvard.edu/abs/2018PhRvA..98e0302S}
}

@article{Dong:2022thu,
    author = "Dong, Dong-Dong and Wei, Geng-Biao and Song, Xue-Ke and Wang, Dong and Ye, Liu",
    title = "{Unification of coherence and quantum correlations in tripartite systems}",
    eprint = "2205.07265",
    archivePrefix = "arXiv",
    primaryClass = "quant-ph",
    doi = "10.1103/PhysRevA.106.042415",
    journal = "Phys. Rev. A",
    volume = "106",
    number = "4",
    pages = "042415",
    year = "2022"
}

@article{Dong:2022eys,
    author = "Dong, Dong-Dong and Song, Xue-Ke and Fan, Xiao-Gang and Ye, Liu and Wang, Dong",
    title = "{Complementary relations of entanglement, coherence, steering, and Bell nonlocality inequality violation in three-qubit states}",
    eprint = "2212.09326",
    archivePrefix = "arXiv",
    primaryClass = "quant-ph",
    doi = "10.1103/PhysRevA.107.052403",
    journal = "Phys. Rev. A",
    volume = "107",
    number = "5",
    pages = "052403",
    year = "2023"
}

@article{konwar2024violation,
    author = "Konwar, Lekhashri and Vardani, Juhi and Yadav, Bhavna",
    title = "{Violation of LGtI inequalities in the light of NO$\nu $A and T2K anomaly}",
    eprint = "2401.02886",
    archivePrefix = "arXiv",
    primaryClass = "hep-ph",
    doi = "10.1140/epjc/s10052-024-13370-9",
    journal = "Eur. Phys. J. C",
    volume = "84",
    number = "10",
    pages = "1103",
    year = "2024"
}

@article{Schrodinger:2008pyl,
    author = {Schr{\"o}dinger, E.},
    title = "{Discussion of Probability Relations between Separated Systems}",
    doi = "10.1017/S0305004100013554",
    journal = "Math. Proc. Cambridge Phil. Soc.",
    volume = "31",
    number = "4",
    pages = "555--563",
    year = "2008"
}

@article{Schrodinger:2008opj,
    author = {Schr{\"o}dinger, E.},
    title = "{Probability relations between separated systems}",
    doi = "10.1017/S0305004100019137",
    journal = "Math. Proc. Cambridge Phil. Soc.",
    volume = "32",
    number = "3",
    pages = "446--452",
    year = "2008"
}

@book{Nielsen_Chuang_2010, place={Cambridge}, title={Quantum Computation and Quantum Information: 10th Anniversary Edition}, publisher={Cambridge University Press}, author={Nielsen, Michael A. and Chuang, Isaac L.}, year={2010}}

@article{Sabin2008,
  author       = {C. Sab{\'i}n and G. Garc{\'i}a-Alcaine},
  title        = {A classification of entanglement in three-qubit systems},
  journal      = {Eur. Phys. J. C},
  volume       = {48},
  number       = {3},
  pages        = {435--442},
  year         = {2008},
  doi          = {10.1140/epjc/s10052-006-0041-1},
  publisher    = {Springer}
}

@article{Dur:2000zz,
    author = "Dur, W. and Vidal, G. and Cirac, J. I.",
    title = "{Three qubits can be entangled in two inequivalent ways}",
    eprint = "quant-ph/0005115",
    archivePrefix = "arXiv",
    doi = "10.1103/PhysRevA.62.062314",
    journal = "Phys. Rev. A",
    volume = "62",
    pages = "062314",
    year = "2000"
}

@article{Cunha:2019jex,
    author = "Cunha, M{\'a}rcio M. and Fonseca, Alejandro and Silva, Edilberto O.",
    title = "{Tripartite Entanglement: Foundations and Applications}",
    eprint = "1909.00862",
    archivePrefix = "arXiv",
    primaryClass = "quant-ph",
    doi = "10.3390/universe5100209",
    journal = "Universe",
    volume = "5",
    number = "10",
    pages = "209",
    year = "2019"
}

@article{Bennett:1996gf,
    author = "Bennett, Charles H. and DiVincenzo, David P. and Smolin, John A. and Wootters, William K.",
    title = "{Mixed state entanglement and quantum error correction}",
    eprint = "quant-ph/9604024",
    archivePrefix = "arXiv",
    doi = "10.1103/PhysRevA.54.3824",
    journal = "Phys. Rev. A",
    volume = "54",
    pages = "3824--3851",
    year = "1996"
}

@article{Guo2020,
  author       = {Yong Guo and Lin Zhang},
  title        = {Multipartite entanglement measure and complete monogamy relation},
  journal      = {Phys. Rev. A},
  volume       = {101},
  number       = {3},
  pages        = {032301},
  year         = {2020},
  doi          = {10.1103/PhysRevA.101.032301},
  publisher    = {American Physical Society}
}

@article{Hill:1997pfa,
    author = "Hill, Scott and Wootters, William K.",
    title = "{Entanglement of a pair of quantum bits}",
    eprint = "quant-ph/9703041",
    archivePrefix = "arXiv",
    doi = "10.1103/PhysRevLett.78.5022",
    journal = "Phys. Rev. Lett.",
    volume = "78",
    pages = "5022--5025",
    year = "1997"
}

@article{Wootters:1997id,
    author = "Wootters, William K.",
    title = "{Entanglement of formation of an arbitrary state of two qubits}",
    eprint = "quant-ph/9709029",
    archivePrefix = "arXiv",
    doi = "10.1103/PhysRevLett.80.2245",
    journal = "Phys. Rev. Lett.",
    volume = "80",
    pages = "2245--2248",
    year = "1998"
}

@article{DUNE:2015lol,
    author = "Acciarri, R. and others",

    title = "{Long-Baseline Neutrino Facility (LBNF) and Deep Underground Neutrino Experiment (DUNE)}: {Conceptual Design Report, Volume 2: The Physics Program for DUNE at LBNF}",
    eprint = "1512.06148",
    archivePrefix = "arXiv",
    primaryClass = "physics.ins-det",
    reportNumber = "FERMILAB-DESIGN-2016-02",
    month = "12",
    journal = {arXiv preprint},
    year = "2015"
}

@article{DUNE:2021tad,
    author = "Hewes, V. and others",
    collaboration = "DUNE",
    title = "{Deep Underground Neutrino Experiment (DUNE) Near Detector Conceptual Design Report}",
    eprint = "2103.13910",
    archivePrefix = "arXiv",
    primaryClass = "physics.ins-det",
    reportNumber = "FERMILAB-PUB-21-067-E-LBNF-PPD-SCD-T",
    doi = "10.3390/instruments5040031",
    journal = "Instruments",
    volume = "5",
    number = "4",
    pages = "31",
    year = "2021"
}

@article{DUNE:2020fgq,
    author = "Abi, B. and others",
    collaboration = "DUNE",
    title = "{Prospects for beyond the Standard Model physics searches at the Deep Underground Neutrino Experiment}",
    eprint = "2008.12769",
    archivePrefix = "arXiv",
    primaryClass = "hep-ex",
    reportNumber = "FERMILAB-PUB-20-459-LBNF-ND, FERMILAB-PUB-20-459-LBNF-ND",
    doi = "10.1140/epjc/s10052-021-09007-w",
    journal = "Eur. Phys. J. C",
    volume = "81",
    number = "4",
    pages = "322",
    year = "2021"
}

@article{DUNE:2020lwj,
    author = "Abi, Babak and others",
    collaboration = "DUNE",
    title = "{Deep Underground Neutrino Experiment (DUNE), Far Detector Technical Design Report, Volume I Introduction to DUNE}",
    eprint = "2002.02967",
    archivePrefix = "arXiv",
    primaryClass = "physics.ins-det",
    reportNumber = "FERMILAB-PUB-20-024-ND, FERMILAB-DESIGN-2020-01",
    doi = "10.1088/1748-0221/15/08/T08008",
    journal = "JINST",
    volume = "15",
    number = "08",
    pages = "T08008",
    year = "2020"
}

@article{Bouri:2024kcl,
    author = "Bouri, Subhadip and Jha, Abhishek Kumar and Banerjee, Subhashish",
    title = "{Probing CP violation and mass ordering in neutrino oscillations in matter through quantum speed limits}",
    eprint = "2405.13114",
    archivePrefix = "arXiv",
    primaryClass = "hep-ph",
    doi = "10.1103/w6rq-l3ql",
    journal = "Phys. Rev. D",
    volume = "112",
    number = "3",
    pages = "036007",
    year = "2025"
}

@article{Jha:2021itm,
    author = "Jha, Abhishek Kumar and Chatla, Akshay",
    title = "{Quantum studies of neutrinos on IBMQ processors}",
    doi = "10.1140/epjs/s11734-021-00358-9",
    journal = "Eur. Phys. J. ST",
    volume = "231",
    number = "2",
    pages = "141--149",
    year = "2022"
}

@article{Jha:2022yik,
    author = "Jha, Abhishek Kumar and Chatla, Akshay and Bambah, Bindu A.",
    title = "{Neutrinos as qubits and qutrits}",
    eprint = "2203.13485",
    archivePrefix = "arXiv",
    primaryClass = "hep-ph",
    doi = "10.1140/epjp/s13360-024-04861-5",
    journal = "Eur. Phys. J. Plus",
    volume = "139",
    number = "1",
    pages = "68",
    year = "2024"
}

@article{Dixit:2019lsg,
    author = "Dixit, Khushboo and Naikoo, Javid and Mukhopadhyay, Banibrata and Banerjee, Subhashish",
    title = "{Quantum correlations in neutrino oscillations in curved spacetime}",
    eprint = "1903.05664",
    archivePrefix = "arXiv",
    primaryClass = "hep-ph",
    doi = "10.1103/PhysRevD.100.055021",
    journal = "Phys. Rev. D",
    volume = "100",
    number = "5",
    pages = "055021",
    year = "2019"
}

@article{Jha:2025ekn,
    author = "Jha, Abhishek Kumar and Dutta, Mriganka and Pathak, Mayank and Banerjee, Subhashish and Mukhopadhyay, Banibrata",
    title = "{Probing the quantum speed limit and entanglement in flavor oscillations of neutrino-antineutrino system in curved spacetime}",
    eprint = "2504.20236",
    archivePrefix = "arXiv",
    primaryClass = "gr-qc",
    doi = "10.1103/w9gw-j8zv",
    journal = "Phys. Rev. D",
    volume = "112",
    number = "4",
    pages = "043045",
    year = "2025"
}

@article{Jha:2024gdq,
    author = "Jha, Abhishek Kumar and Dutta, Mriganka and Banerjee, Subhashish and Mukhopadhyay, Banibrata",
    title = "{Influence of Gravity on the Quantum Speed Limit in Neutrino Oscillations}",
    eprint = "2411.18558",
    archivePrefix = "arXiv",
    primaryClass = "gr-qc",
    doi = "10.1007/978-3-031-90186-7_40",
    journal = "Astrophys. Space Sci. Proc.",
    volume = "61",
    pages = "503--514",
    year = "2025"
}

@inproceedings{Jha:2024asl,
    author = "Jha, Abhishek Kumar and Mukhopadhyay, Banibrata and Dutta, Mriganka and Pathak, Mayank and Banerjee, Subhashish",
    title = "{Gravitational Influence on the Quantum Speed Limit in Flavor Oscillations of Neutrino-Antineutrino System}",
    booktitle = "{17th Marcel Grossmann Meeting}: {On Recent Developments in Theoretical and Experimental General Relativity, Gravitation, and Relativistic Field Theories}",
    eprint = "2411.18604",
    archivePrefix = "arXiv",
    primaryClass = "gr-qc",
    month = "11",
    year = "2024"
}

@article{Jha:2026jef,
    author = "Jha, Abhishek Kumar and Konwar, Lekhashri and Mohanta, Rukmani",
    title = "{Quantum speed limit time for bipartite entanglement in neutrino oscillations in matter with nonstandard interactions}",
    eprint = "2602.03748",
    archivePrefix = "arXiv",
    primaryClass = "hep-ph",
    doi = "10.1103/r6t7-wnbz",
    journal = "Phys. Rev. D",
    volume = "114",
    number = "7",
    pages = "076011",
    year = "2026"
}

\end{document}